\pdfoutput=1
\documentclass[12pt]{article}

\catcode`\@=11

\global\arraycolsep=2pt
\usepackage{amsbsy,amssymb,latexsym,amsfonts,amsmath}
\usepackage{graphicx,color}
\usepackage{booktabs}
\usepackage{placeins}
\usepackage{tikz}
\usetikzlibrary{arrows.meta,decorations.pathmorphing,decorations.markings,decorations.pathreplacing,calc}
\tikzset{
  axisline/.style={-{Stealth[length=2mm]},gray!70,thin},
  contour/.style={blue!70!black,very thick,
    decoration={markings,
      mark=at position 0.28 with {\arrow{Stealth[length=2.2mm]}},
      mark=at position 0.75 with {\arrow{Stealth[length=2.2mm]}}},
    postaction={decorate}},
  sdpath/.style={red!75!black,thick,dashed},
  cutline/.style={brown!70!black,line width=1pt,
    decorate,decoration={snake,amplitude=0.7mm,segment length=2.2mm}},
  saddle/.style={circle,fill=black,inner sep=1.6pt},
  ghost/.style={circle,draw=black!45,fill=white,inner sep=1.5pt,line width=0.8pt},
}
\usepackage[numbers]{natbib}

\newcommand{\dd}{\mathrm{d}}
\newcommand{\ee}{\mathrm{e}}

\newcommand{\ii}{\mathrm{i}}
\newcommand{\Om}{\Omega}
\newcommand{\vy}{\vec y}

\usepackage{hyperref}
\hypersetup{
    colorlinks=true,
    linkcolor=blue,
    citecolor=blue,
    urlcolor=blue
}

\begin{document}

\begin{flushright}
YITP-26-173
\end{flushright}

\vspace*{0.7cm}

\begin{center}
{ \Large Lecture Notes on OpenAI paper ``Scale and conformal symmetry in four-dimensional operational quantum field theory''}
\vspace*{1.5cm}\\
{Yu Nakayama}
\end{center}
\vspace*{1.0cm}
\begin{center}

Yukawa Institute for Theoretical Physics,
Kyoto University, Kitashirakawa Oiwakecho, Sakyo-ku, Kyoto 606-8502, Japan

\vspace{3.8cm}
\end{center}

\begin{abstract}
OpenAI's abstract announces a proof that ``scale symmetry implies local conformal symmetry for a class of four-dimensional
unitary, positive-energy quantum field theories with a Poincar\'e- and scale-invariant vacuum. The
class has a discrete, bounded-below spectrum of scaling dimensions with finite multiplicities, and
each original field has finite scaling support.'' The theorem itself assumes in addition that the dilatation acts diagonally with real scaling dimensions, which the abstract does not mention. This assumption is not innocent, since the question that human physicists such as Polchinski, Dorigoni--Rychkov and Fortin--Grinstein--Stergiou actually faced is whether the flavor rotation contained in a given dilatation can be removed, and under this assumption the question cannot even be posed. The theorem therefore cannot replace their work. In these lectures we review the OpenAI approach in the language of physics and follow the ideas that lead toward a complete proof that covers the physically relevant cases, one that does not assume that the scaling dimensions are real or that the dilatation is diagonalizable.
\end{abstract}

\setcounter{page}{0}
\thispagestyle{empty}

\newpage

\section{Introduction}
Conformal invariance is central to our understanding of critical phenomena, in two dimensions and beyond. In three dimensions, for example, the conformal bootstrap \cite{Rattazzi:2008pe,Poland:2018epd} has determined the critical exponents of the Ising model with high precision \cite{ElShowk:2012ht,Kos:2014bka,Kos:2016ysd,Chang:2024whx}. Crossing symmetry of the four-point functions of the spin and energy operators and the energy-momentum tensor, together with unitarity and assumptions on the gaps in the spectrum, confines their scaling dimensions to a tiny island, $\Delta_\sigma=0.518148806(24)$ and $\Delta_\epsilon=1.41262528(29)$ \cite{Chang:2024whx}.

This success rests on an assumption that is far from obvious. The method is built on the conformal block decomposition, so it assumes that a critical system is not only scale invariant but also conformally invariant, that is, described by a conformal field theory (CFT). Is this true? For the Ising model the assumption has been suggested by a renormalization group argument \cite{Delamotte:2015aaa} and is supported by a direct lattice test \cite{Meneses:2018xpu}, which we come back to below. In general, however, it can fail. The free Maxwell theory in $d\neq4$ dimensions is unitary and scale invariant but not conformally invariant \cite{ElShowk:2011gz}, and the theory of elasticity in two dimensions is a non-unitary example \cite{Riva:2005gd} (see \cite{Jackiw:2011vz,Nakayama:2013is,Nakayama:2016cym,Gimenez-Grau:2023lpz,Nakayama:2024psst} for further discussion).

The difference between the two symmetries can be stated precisely with the energy-momentum tensor. Before we get there, let me begin with a fatal flaw in a famous quantum field theory textbook that you probably own, Sec.~19.5 of Peskin and Schroeder \cite{Peskin:1995ev}. They start from the idea that under a dilatation $x\to\ee^{\sigma}x$ of spacetime the effective action responds through the trace of the energy-momentum tensor, $\delta_{\sigma} S = \int \dd^4x\, \sigma\, T^{\mu}{}_{\mu}$. The dilatation current is taken to be $D^\mu=\Theta^{\mu\nu}x_\nu$, where $\Theta^{\mu\nu}$ is their notation for the energy-momentum tensor, so that $\partial_\mu D^\mu=\Theta^\mu{}_\mu$. The change of the Lagrangian under a change of scale is thus identified with the trace. A change of scale also moves the renormalized coupling along its beta function. In a gauge theory such as QCD, with the gauge coupling $g$, this leads to the famous formula
\begin{equation}
\Theta^\mu{}_\mu=\frac{\beta(g)}{2g^3}F^a_{\mu\nu}F^{a\mu\nu}
\end{equation}
(in their normalization of the gauge field), together with the chain of equivalences
\begin{equation}\label{eq:chainPS}
\text{scale invariance}\ \Longleftrightarrow\ T^\mu{}_\mu=0\ \Longleftrightarrow\ \beta=0 .
\end{equation}
For perturbative QCD the formula is correct, but the flaw lies in the chain \eqref{eq:chainPS}. As a general principle neither implication from left to right holds. Let me explain why.

We begin with the first arrow. A dilatation is a global transformation, so with a constant $\sigma$ we only probe $\int\dd^4x\,T^\mu{}_\mu$. For scale invariance we therefore only need $T^\mu{}_\mu$ to be a total divergence. To conclude that $T^\mu{}_\mu=0$ we would have to let $\sigma$ depend on $x$, but invariance under such local rescalings, the Weyl transformations, is what characterizes conformal invariance. The textbook does not discuss conformal invariance, but in this step it treats a global rescaling as if it were a local one. Equivalently, $D^\mu=\Theta^{\mu\nu}x_\nu$ is not the most general dilatation current \cite{Coleman:1970je,Callan:1970ze}.

With several couplings the second arrow can fail as well, since a non-zero beta function can describe a mere rotation of the fields, which leaves the trace zero. We return to this point below. In perturbative QCD, fortunately, neither problem arises. The only gauge-invariant vectors of dimension three are the quark currents, of which the vector currents are conserved, while the divergence of an axial current is a pseudoscalar, which cannot appear in the trace. So there is no candidate for the missing total divergence, and with a single gauge coupling there is no rotation of the couplings either. This is why the textbook formula holds there.

What, then, is the correct relation? For scale invariance we only need the trace to be a total divergence,
\begin{equation}
\Theta\equiv T^{\mu}{}_{\mu}=\partial^\mu V_\mu ,
\end{equation}
where $V_\mu$ is a local operator called the \emph{virial current}. Then $D_\mu=x^\nu T_{\mu\nu}-V_\mu$ is conserved. For conformal invariance we need more. The virial current must be a gradient up to a conserved current, $V_\mu=\partial_\mu L+J_\mu$ with a local scalar $L$ and $\partial^\mu J_\mu=0$.\footnote{In general the improvement can also use a symmetric tensor $L_{\mu\nu}$, and the condition is $V_\mu=\partial^\nu L_{\nu\mu}+J_\mu$ \cite{Polchinski:1987dy,Nakayama:2013is}. In these lectures only the scalar case $L_{\mu\nu}=\eta_{\mu\nu}L$ appears.} The conserved part is harmless, since with it we only redefine the dilatation by the charge of an internal symmetry, as we will see below. The gradient part can be removed by improving the energy-momentum tensor to a traceless one. In four dimensions,
\begin{equation}\label{eq:improve}
T_{\mu\nu}\ \to\ T_{\mu\nu}+\tfrac13(\partial_\mu\partial_\nu-\eta_{\mu\nu}\Box)L,\qquad \Box L=T^\mu{}_\mu .
\end{equation}

In a given theory we can therefore show conformal invariance by showing that there is no candidate for the virial current, as we did for perturbative QCD above. Since the energy-momentum tensor has dimension $d$ in $d$ spacetime dimensions, such a candidate is a vector operator of dimension $d-1$ that is neither conserved nor a gradient. The lattice test of the three-dimensional Ising model mentioned above is a search for such an operator. It bounds the dimension of the lowest candidate from below by about $5$, far above the required value $2$ \cite{Meneses:2018xpu}.

In general, however, we do not know the spectrum of vector operators, so the question ``does scale invariance imply conformal invariance?'' becomes the question whether $L=\Box^{-1}T^\mu{}_\mu$, the scalar that we need for the improvement \eqref{eq:improve}, is a local operator \cite{Polchinski:1987dy,Nakayama:2013is}. Indeed, if it is, $V_\mu-\partial_\mu L$ is a conserved current, which is the harmless case. In four dimensions, the answer is known to be yes in perturbation theory \cite{Luty:2012ww,Fortin:2012hn}. There are also non-perturbative arguments \cite{Dymarsky:2013pqa,Yonekura:2014tha,Dymarsky:2014zja}, but they need additional input, such as the coupling to a dilaton and the behavior of its forward amplitudes. The OpenAI paper \cite{OpenAI:2026} proposes an argument whose main physical inputs are the causality of the energy-momentum tensor and the positivity of energy. It answers the question for unitary theories in which every field has a real scaling dimension and the dilatation acts diagonally.

In these lectures I would first like to explain the strategy of the OpenAI argument. Of the paper itself I review only the first half, which shows that the state created by $L$ is as local as that of a local operator, and not the second half, which constructs $L$ as a local operator. From the second half, Lectures two and three need only correlation functions with a single insertion of $L$, which I explain in Appendix~\ref{sec:insert}. I would then like to remove the two assumptions that the scaling dimensions are real and that the dilatation acts diagonally. For this part I follow a manuscript by the AI system Astra \cite{Astra:2026}, which is released together with these lecture notes. I discuss its status at the end of the lectures.

We will see that neither assumption is needed, although for complex dimensions the last step to conformal invariance uses a technical condition on the fields (Sec.~\ref{sec:conf}). In a unitary theory with an energy-momentum tensor the dilatation cannot act on the fields through a Jordan block. Moreover, the imaginary part of a scaling dimension is the charge of an internal symmetry, so we can remove it by redefining the dilatation. At the end of this introduction, in Sec.~\ref{sec:guide}, I suggest shorter routes through the lectures.

Before we go on, let me fix our conventions. We work in four dimensions unless stated otherwise, with the mostly-minus metric, so that $p^2>0$ for timelike momenta. By a correlation function we always mean a Wightman function, the vacuum expectation value $\langle\Om|O_1(x_1)\cdots O_n(x_n)|\Om\rangle$ of operators in the order written. For a Hermitian scalar of real dimension $\Delta>1$,
\begin{equation}\label{eq:wightman}
\begin{aligned}
\langle\Om|O(x)O(0)|\Om\rangle&=\frac{C}{(-x^2+\ii\epsilon x^0)^{\Delta}},\\
\int\dd^4x\,\ee^{\ii p\cdot x}\,\langle\Om|O(x)O(0)|\Om\rangle&=C_\Delta\,\theta(p^0)\theta(p^2)\,(p^2)^{\Delta-2},
\end{aligned}
\end{equation}
where $C_\Delta$ is $C$ times a positive constant. We call the right-hand side the spectral density. It is the squared norm of the momentum-$p$ component $O(p)|\Om\rangle$ of the state $O(0)|\Om\rangle$, and unitarity requires it to be non-negative. On states created from the vacuum, $\partial_\mu$ acts as $\ii p_\mu$ and $\Box$ as $-p^2$. For a momentum $p$ in the forward cone we call $M=\sqrt{p^2}$ the invariant mass.

Let me also say what we assume about the theory. Throughout these lectures we assume that the energy-momentum tensor $T_{\mu\nu}$ and the virial current $V_\mu$ generate translations, Lorentz transformations and dilatations through their Ward--Takahashi identities, including the mixing of fields under dilatations. Moreover, as in the OpenAI paper \cite{OpenAI:2026} and in Astra's manuscript \cite{Astra:2026}, we assume that a field smeared with a test function supported in a region is an operator localized in that region, so that the states it creates from the vacuum are states prepared in that region. A complete list of the assumptions can be found in Sec.~\ref{sec:conf}.

\subsection{Complex scaling dimensions in scale-invariant theories}\label{sec:complexintro}

How can a field acquire a complex scaling dimension? In a scale-invariant theory this can happen through the virial current, which can contain the current of a flavor rotation. To understand what the virial current does, let us recall angular momentum. The Noether current of Lorentz transformations is
\begin{equation}
M_\mu{}^{\nu\rho}=x^\nu T_\mu{}^\rho-x^\rho T_\mu{}^\nu+S_\mu{}^{\nu\rho}.
\end{equation}
The first two terms are the orbital part and $S_\mu{}^{\nu\rho}$ is the intrinsic spin part. The spin part can always be absorbed by passing to the symmetric Belinfante tensor. The dilatation current has the same structure,
\begin{equation}
D_\mu=\underbrace{x^\nu T_{\mu\nu}}_{\text{orbital}}\ -\ \underbrace{V_\mu}_{\text{intrinsic}} .
\end{equation}
We need the intrinsic part because the canonical energy-momentum tensor need not be traceless, in which case the orbital part alone is not conserved, $\partial^\mu(x^\nu T_{\mu\nu})=T^\mu{}_\mu$.

Take a free massless scalar with the canonical $T_{\mu\nu}$. The trace is $T^\mu{}_\mu=-(\partial\phi)^2=-\tfrac12\Box\phi^2$ on shell, so
\begin{equation}
V_\mu=-\phi\,\partial_\mu\phi=\partial_\mu\big(-\tfrac12\phi^2\big).
\end{equation}
This intrinsic part is a gradient, so the improvement \eqref{eq:improve} with $L=-\tfrac12\phi^2$ absorbs it. The result is the improved tensor of Callan, Coleman and Jackiw \cite{Callan:1970ze}. Our question can therefore be phrased as a question about spin. Can the intrinsic part of the dilatation current always be absorbed into $T_{\mu\nu}$, the way spin can?

With several scalars $\phi^a$ the general form is
\begin{equation}\label{eq:Vpol}
V_\mu=c_{ab}\,\partial_\mu(\phi^a\phi^b)+Q_{ab}\,\phi^a\partial_\mu\phi^b,\qquad c_{ab}=c_{ba},\quad Q_{ab}=-Q_{ba}.
\end{equation}
The first term is again a gradient. The symmetric matrix $c_{ab}$ depends on the improvement term that we include in $T_{\mu\nu}$, and for the canonical tensor $c_{ab}=-\tfrac12\delta_{ab}$. The second term is new. In the free theory it is the conserved current of a rotation of the scalars, so it does not change $\partial^\mu V_\mu$. But it does change the dilatation charge $D=\int\dd^3x\,D_0$, under which the fields now transform as
\begin{equation}
\ii[D,\phi^a(x)]=(x\cdot\partial+1)\phi^a(x)+Q^a{}_b\,\phi^b(x) .
\end{equation}
The fields are rotated while they are rescaled. A real antisymmetric $Q$ has eigenvalues $\pm\ii q$, so the scaling dimensions become $1\pm\ii q$.

In the free theory this is only a redefinition of the dilatation by a symmetry. It becomes physical in an interacting theory whose couplings break the rotation. Then the divergence of the second term no longer vanishes, so it contributes to the trace. Can we still write the trace as $\Box L$ with a local scalar $L$, and so remove the term by an improvement? This is a non-trivial question. If we cannot, the dilatation rotates the fields, and generically their scaling dimensions are complex. So if we assume from the start that the given dilatation has real scaling dimensions, we exclude exactly the cases that we want to understand.

Whether the term can be removed was the question of a series of studies, starting with Polchinski \cite{Polchinski:1987dy}. Following Zamolodchikov \cite{Zamolodchikov:1986gt}, he proved that in two dimensions scale invariance implies conformal invariance for unitary theories with a discrete spectrum, and then turned to four dimensions. Consider the most general renormalizable theory of scalars $\phi^a$ and Weyl fermions $\psi^i$, with couplings $g^I=(\lambda_{abcd},y_{a\,ij},\dots)$. The only candidates for a non-trivial virial current are
\begin{equation}
V_\mu=Q_{ab}\,\phi^a\partial_\mu\phi^b-\ii P_{ij}\,\bar\psi^i\bar\sigma_\mu\psi^j ,
\end{equation}
with $Q$ antisymmetric and $P$ anti-Hermitian. These are the currents of the flavor rotations broken by the couplings. By the equations of motion $\partial^\mu V_\mu=(Qg)^I\mathcal O_I$, where $(Qg)^I$ is the coupling rotated by $Q$ and $P$, and $\mathcal O_I=\partial\mathcal L/\partial g^I$. If we compare this with the naive trace identity $T^\mu{}_\mu=\beta^I\mathcal O_I$, we obtain the naive conditions
\begin{equation}\label{eq:cycle}
\text{scale invariance:}\quad\beta^I=(Qg)^I,\qquad\qquad\text{conformal invariance:}\quad\beta^I=0 .
\end{equation}
Scale without conformal invariance would therefore be a renormalization group \emph{limit cycle}. The couplings do not approach a fixed point, but the theory at different scales is the same theory up to a rotation of the fields.

Does such a cycle exist? Polchinski showed that a scale-invariant theory of scalars in $4-\epsilon$ dimensions has none at one loop. Dorigoni and Rychkov then extended this one-loop argument to scalars together with Weyl fermions \cite{Dorigoni:2009ra}, and Fortin, Grinstein and Stergiou extended the scalar case to two loops \cite{Fortin:2011ks}. Up to this point the answer was no. But in that same paper Fortin, Grinstein and Stergiou also found a candidate cycle in $4-\epsilon$ dimensions, in a theory with scalars and fermions. They studied it further in \cite{Fortin:2011sz,Fortin:2012cq}, and later they found one in a four-dimensional gauge theory as well \cite{Fortin:2012ic}.

So the next question is whether these cycles are conformal. If the trace were simply $\beta^I\mathcal O_I$, they would not be, because on a cycle $\beta^I=(Qg)^I\neq0$. To find the trace on such a cycle, however, we have to keep the divergence of the flavor current in the trace identity. This was done soon after \cite{Luty:2012ww,Fortin:2012hn,Baume:2014rla}, and the correct operator statement is
\begin{equation}\label{eq:B}
T^\mu{}_\mu=B^I\mathcal O_I,\qquad B^I=\beta^I-(Sg)^I ,
\end{equation}
with a calculable antisymmetric $S$ \cite{Jack:1990eb,Osborn:1991gm}. Now, on the cycles that Fortin, Grinstein and Stergiou found, $Q=S$. Then $B^I=\beta^I-(Sg)^I=(Qg)^I-(Sg)^I=0$, and $T^\mu{}_\mu=0$. That is, these cycles are CFTs whose beta functions do not vanish. This is how the second arrow in \eqref{eq:chainPS} fails (see \cite{Nakayama:2013is,Dymarsky:2013pqa,Farnsworth:2013osa,Yonekura:2014tha,Dymarsky:2014zja} for the status in four dimensions).

From this story we learn that in theories of this kind, before we know whether they are conformal, the dilatation defined by a given virial current acts on the fields through a matrix, an anomalous dimension matrix plus the antisymmetric rotation $Q$. We have no reason to expect this matrix to be diagonal with real eigenvalues.

Complex scaling dimensions are known at non-unitary fixed points, for instance in the complex CFTs that underlie walking behavior \cite{Kaplan:2009kr,Gorbenko:2018ncu}. So it is natural to ask whether unitarity alone rules them out. It does not, as the following example shows.

We would like to have a complex scalar $\phi$ of dimension $\alpha+\ii\beta$ in a Hilbert space with a positive norm. To construct it, let us take two independent real generalized free fields $\chi_\pm$ of dimension $\alpha$ \cite{Greenberg:1961mr}, which serve only as auxiliary fields, and define in their Fock space, in momentum space,
\begin{equation}\label{eq:gffbuilt}
\phi(p)=M^{\ii\beta}\,\psi(p),\qquad\psi=\frac{\ee^{\ii\arg\hat A/2}}{\sqrt2}\Big(\sqrt{\hat B+|\hat A|}\,\chi_++\ii\sqrt{\hat B-|\hat A|}\,\chi_-\Big),
\end{equation}
with constants $\hat B$ and $\hat A$ that satisfy $\hat B\ge|\hat A|$, where $M$ is the invariant mass. The field $\psi$ has the real dimension $\alpha$, and the only unusual feature is the phase $M^{\ii\beta}=\ee^{\ii\beta\log M}$. The theory that we consider is the theory of $\phi$ and $\phi^\dagger$, not that of $\chi_\pm$. We need the auxiliary fields only to make sure that the norm is positive, so that the theory is unitary by construction.

The Wightman functions of $\phi$ and $\phi^\dagger$ are given by Wick's theorem in terms of the two-point functions
\begin{equation}\label{eq:gffc}
\begin{aligned}
\langle\Om|\phi(x)\phi^\dagger(0)|\Om\rangle=\langle\Om|\phi^\dagger(x)\phi(0)|\Om\rangle&=\frac{B}{(-x^2+\ii\epsilon x^0)^{\alpha}},\\
\langle\Om|\phi(x)\phi(0)|\Om\rangle&=\frac{A}{(-x^2+\ii\epsilon x^0)^{\alpha+\ii\beta}} ,
\end{aligned}
\end{equation}
where $\hat B$ and $\hat A$ are $B$ and $A$ multiplied by the Fourier constants of the two power laws, as in \eqref{eq:wightman}. Is $\phi$ a local field? The phase $M^{\ii\beta}$ is not a polynomial in the momentum, so this is not obvious. But at spacelike separation the two-point functions depend only on $x^2$, and therefore the commutators of $\phi$ and $\phi^\dagger$ vanish there. So $\phi$ is a local field. In terms of $A$ and $B$ the condition $|\hat A|\le\hat B$ is an upper bound on $|A|/B$ that is smaller than one and decays like $\ee^{-\pi|\beta|}$. Unitarity thus constrains complex dimensions, but it does not forbid them.

Both two-point functions are scale invariant if $\phi$ has the dimension $\alpha+\ii\beta$ and $\phi^\dagger$ the dimension $\alpha-\ii\beta$. Indeed, a dilatation by $s$ rescales the invariant mass, $M\to\ee^{-s}M$, so the factor $M^{\ii\beta}$ produces an extra phase $\ee^{\ii\beta s}$. The unitary dilatation therefore acts as
\begin{equation}
\ee^{\ii sD}\,\phi(x)\,\ee^{-\ii sD}=\ee^{s(\alpha+\ii\beta)}\,\phi(\ee^sx).\label{eq:Ucomplex}
\end{equation}

Can we remove the imaginary part by redefining the dilatation? If $A=0$, the theory has a $U(1)$ charge $F$, and with the normalization $[F,\phi]=\phi$ the dilatation $D'=D-\beta F$ gives $\phi$ the real dimension $\alpha$. The imaginary part then comes from an internal rotation. In a theory with an energy-momentum tensor this is the harmless freedom $V_\mu\to V_\mu+\beta J_\mu$ with a conserved current $J_\mu$, which is the case of \eqref{eq:Vpol} in which the rotation is a symmetry. For $A\neq0$, on the other hand, the second two-point function does not allow a $U(1)$ that rotates $\phi$, and one can check that the only internal symmetry acting linearly on $\phi$ and $\phi^\dagger$ is $\phi\to-\phi$. So we have no symmetry left with which we could absorb the phase. Note, however, that this example has no energy-momentum tensor. In Lecture three we will see that, once an energy-momentum tensor exists, an imaginary part of a scaling dimension must be generated by an internal symmetry, just as in the case $A=0$.

\subsection{Scaling dimensions in Jordan form}\label{sec:jordanintro}
The second assumption of the OpenAI theorem is that the dilatation acts diagonally on the fields. Can unitarity alone guarantee this? We usually associate a Jordan block of the dilatation with a non-unitary logarithmic CFT, so one might think so. But again a generalized free field shows that unitarity does not guarantee it. Take a real generalized free field $\varphi_1$ with spectral density $\rho(M)$ and define a second field in momentum space by
\begin{equation}\label{eq:gffjordan}
\varphi_0(p)=\log(M/\mu)\,\varphi_1(p) ,
\end{equation}
where $\mu$ is an arbitrary reference mass.

Let us check that this is a sensible theory. Both fields are local, since their commutators are superpositions of free massive commutator functions with weights $\rho$, $\rho\log(M/\mu)$ and $\rho\log^2(M/\mu)$, each of which vanishes at spacelike separation. They are different local fields, because $\log M$ is not a polynomial in $p$. The theory is also unitary. As in the complex example, $\varphi_0$ is defined by \eqref{eq:gffjordan} as an operator in the Fock space of $\varphi_1$, whose norm is positive, and it is Hermitian because $\log(M/\mu)$ is real and even in $p$. With $\ell=\log(M/\mu)$ the matrix of spectral densities of $(\varphi_0,\varphi_1)$ is
\[
\rho\begin{pmatrix}\ell^2&\ell\\ \ell&1\end{pmatrix},
\]
which has rank one, since $\varphi_0(p)|\Om\rangle=\ell\,\varphi_1(p)|\Om\rangle$ is a multiple of a state that $\varphi_1$ creates.

If $\rho\propto M^{2\Delta-4}$, the dilatation is unitary and acts as
\begin{equation}\label{eq:jordandil}
\ee^{\ii sD}\varphi_1(x)\ee^{-\ii sD}=\ee^{s\Delta}\varphi_1(\ee^sx),\qquad
\ee^{\ii sD}\varphi_0(x)\ee^{-\ii sD}=\ee^{s\Delta}\big[\varphi_0+s\,\varphi_1\big](\ee^sx) .
\end{equation}
Indeed, a dilatation rescales the invariant mass, $M\to\ee^{-s}M$, and the logarithm turns a rescaling into a shift. In terms of the rescaled mass the logarithm is
\[
\log(M/\mu)=\log(\ee^{-s}M/\mu)+s .
\]
The first term is again the logarithm of the invariant mass and reproduces $\varphi_0$. The second term gives $s\,\varphi_1$.

On the pair $(\varphi_0,\varphi_1)$ the dilatation is a Jordan block, and $\varphi_1$ is the \emph{logarithmic partner} of $\varphi_0$. This is not a logarithmic CFT in the usual sense \cite{Gurarie:1993xq,Hogervorst:2016itc}, where the two-point function of the field that we call the partner vanishes and the theory is not unitary. Note that in that literature the names are often used the other way around, so that the field with the logarithm is called the partner. In our example the Hilbert space is positive and $D$ is self-adjoint on states. The Jordan block appears only in the way $D$ acts on the local fields. On the Hilbert space there is none, since the two fields create the same states up to the factor $\ell$.

The two examples are in fact closely related. If we write the phase $M^{\ii\beta}$ of the complex example as $\ee^{\ii\beta\log(M/\mu)}$, up to a constant, and differentiate it with respect to $\beta$ at $\beta=0$, we obtain $\ii\log(M/\mu)$, the factor that defines the Jordan pair. In a generalized free theory we may multiply the fields by a real function of $M$, or rotate a Hermitian multiplet by an angle depending on $\log M$, without losing locality or unitarity. Neither example, however, has an energy-momentum tensor.

Now, can the energy-momentum tensor itself sit in a Jordan block? At first sight this sounds strange. The energy-momentum tensor is conserved, and we usually say that a conserved current is not renormalized. But what is protected is only the charge, here the momentum $P_\mu$. The tensor itself is defined only up to improvement terms, which do not contribute to $P_\mu$, and nothing protects their coefficients.

Let us see this in the simplest example, the $\lambda\phi^4$ theory. We can take its energy-momentum tensor to be
\begin{equation}
T_{\mu\nu}=T^{\rm can}_{\mu\nu}-\xi\,(\partial_\mu\partial_\nu-\eta_{\mu\nu}\Box)\phi^2 ,
\end{equation}
where $\xi$ is a parameter. In curved space it is the coefficient of the coupling $\tfrac12\xi R\phi^2$. Classically the tensor with $\xi=\tfrac16$ is traceless. In the interacting theory, however, $\xi$ is a coupling like any other, and it runs:
\begin{equation}\label{eq:betaxi}
\mu\frac{\dd\xi}{\dd\mu}=\big(\xi-\tfrac16\big)\gamma_{\phi^2}(\lambda)+\hat\eta(\lambda).
\end{equation}
Here $\gamma_{\phi^2}$ is the anomalous dimension of $\phi^2$, and $\hat\eta(\lambda)$ is an inhomogeneous term, which does not depend on $\xi$. In minimal subtraction it first appears at four loops \cite{Collins:1976vm,Hathrell:1981zb}. So even if we start with $\xi=\tfrac16$, the renormalization group moves us away from it.

What does this mean for the dilatation? A change of scale changes $\xi$, and hence adds to $T_{\mu\nu}$ a multiple of the improvement term $I_{\mu\nu}=(\partial_\mu\partial_\nu-\eta_{\mu\nu}\Box)\phi^2$. That is, the dilatation mixes $T_{\mu\nu}$ with $I_{\mu\nu}$. On the pair $(T_{\mu\nu},I_{\mu\nu})$, where we take $T_{\mu\nu}$ with $\xi=\tfrac16$, the matrix of scaling dimensions is
\begin{equation}
\begin{pmatrix}4&-\hat\eta\\0&4+\gamma_{\phi^2}\end{pmatrix}.
\end{equation}
Usually this causes no problem. At a fixed point with $\gamma_{\phi^2}\neq0$ the two diagonal entries are different, so we can diagonalize the matrix. Concretely, we shift $\xi$ to $\xi_*=\tfrac16-\hat\eta/\gamma_{\phi^2}$, and the energy-momentum tensor with this $\xi$ has a definite scaling dimension. But if $\gamma_{\phi^2}=0$ while $\hat\eta\neq0$, no shift of $\xi$ helps. Then
\begin{equation}
\ee^{\ii sD}\,T_{\mu\nu}(x)\,\ee^{-\ii sD}=\ee^{4s}\big[T_{\mu\nu}-s\,\hat\eta\,I_{\mu\nu}\big](\ee^sx),
\end{equation}
which is a Jordan block of the same form as for $(\varphi_0,\varphi_1)$, with the improvement term $I_{\mu\nu}$ as the partner.

This happens only for a scalar of dimension exactly two, since only then is its improvement term degenerate with $T_{\mu\nu}$. Two is also the dimension of the would-be $L=\Box^{-1}T^\mu{}_\mu$, and to see that this is no accident we take the trace of the dilatation law. With $O=\phi^2$ the trace of $I_{\mu\nu}$ is $-3\Box O$, so
\begin{equation}
\ii[D,\Theta(x)]=(x\cdot\partial+4)\,\Theta(x)+3\hat\eta\,\Box O(x) .
\end{equation}
If $\Theta=\Box L$, then $L$ must transform as $L\to L+3\hat\eta\,s\,O$. So $L$ itself sits in a Jordan block with $O$. If we now improve with this $L$, $T'_{\mu\nu}=T_{\mu\nu}+\tfrac13(\partial_\mu\partial_\nu-\eta_{\mu\nu}\Box)L$, we obtain a traceless tensor without any Jordan block. That is, a Jordan block of $T_{\mu\nu}$ means that $L=\Box^{-1}\Theta$, if it is local, is a \emph{logarithmic} scalar. In that case the theory is conformal, with the improved tensor $T'_{\mu\nu}$. So the question is again whether $L$ is local.

So far we have looked at one example. The conclusion of the general analysis will be that we can always remove such a Jordan block by an improvement if the theory is conformal, so that it can survive only in the theories that we want to understand. The systematic framework is the local renormalization group of Jack and Osborn \cite{Jack:1990eb,Osborn:1991gm,Baume:2014rla}. The idea is to promote not only the metric but also the couplings to functions, $g^I\to g^I(x)$, and to ask how the theory responds to a \emph{local} rescaling $\sigma(x)$. This is, so to speak, the correct version of the textbook argument. With a local rescaling we really do measure $T^\mu{}_\mu(x)$ point by point. But we then also pick up the beta functions and every total-derivative operator that we cannot probe with a constant $\sigma$. Schematically, for constant couplings on a curved background and up to the c-number terms of the trace anomaly,
\begin{equation}\label{eq:localRG}
T^\mu{}_\mu=\beta^I\mathcal O_I-\partial_\mu\big(S^AJ^\mu_A\big)+\Box\big(\tau^a\mathcal O_a\big)+\hat\eta^a R\,\mathcal O_a .
\end{equation}
Here $J_A$ are the currents of the flavor symmetries broken by the couplings, $\mathcal O_a$ are the scalars of dimension two, and $S^A$, $\tau^a$ and $\hat\eta^a$ are coefficients that depend on the couplings.

Let us go through the four terms one by one. The first is the term of the textbook formula. The second has the form of Polchinski's virial current, and this is where complex scaling dimensions come from. Since $\partial_\mu J^\mu_A=(\mathsf t_Ag)^I\mathcal O_I$, where $\mathsf t_A$ are the generators of the flavor rotations, it combines with the first term into $B^I$ of \eqref{eq:B} with $(Sg)^I=S^A(\mathsf t_Ag)^I$, and only this combination is meaningful. The third term is harmless. It corresponds to the conformal case $V_\mu=\partial_\mu(\tau\mathcal O)$, and we can remove it by an improvement.

The dangerous one is the last term. On a flat background it seems to disappear. But recall from our example that the coefficient of $R\phi^2$ is also the coefficient of the improvement term. In the same way the last term remains in flat space, as the logarithmic renormalization of $T_{\mu\nu}$ itself,
\begin{equation}\label{eq:Trun}
\mu\frac{\dd}{\dd\mu}T_{\mu\nu}\ \propto\ \hat\eta^a\,(\partial_\mu\partial_\nu-\eta_{\mu\nu}\Box)\mathcal O_a ,
\end{equation}
that is, up to normalization, as the term $\hat\eta$ of our example \eqref{eq:betaxi}. As in the example, we can shift the coefficients of the improvement terms, and this removes $\hat\eta$ whenever it lies in the image of the anomalous dimension matrix $\gamma$ of the $\mathcal O_a$, as Polchinski already noted \cite{Polchinski:1987dy}. If it does not, the theory is scale invariant while $T_{\mu\nu}$ is still logarithmically renormalized, and $T_{\mu\nu}$ forms a Jordan block with an improvement term \cite{Nakayama:2013is,Nakayama:2013d3}.

Now, the Wess--Zumino consistency condition relates the last two terms of \eqref{eq:localRG}. Where the theory is conformal, $B^I=0$, it gives $\hat\eta=\gamma\tau$ \cite{Jack:2013sha,Nakayama:2013d3}. So there $\hat\eta$ lies in the image of $\gamma$, and we can always remove it. A Jordan block of $T_{\mu\nu}$ that we cannot remove by an improvement can therefore occur only in a theory that is scale invariant but not conformal, that is, in the theories that we want to understand, so we cannot assume it away. The two-point functions cannot exclude it either, as the generalized free field \eqref{eq:gffjordan} shows. In Lecture two we will rule it out with the energy-momentum tensor itself, by measuring the energy flux of the state that it creates through a null plane.

The Wess--Zumino consistency conditions have another famous consequence, the perturbative $a$-theorem \cite{Osborn:1989td,Jack:1990eb,Osborn:1991gm}. There is a function $\tilde a$ of the couplings that equals the $a$-anomaly at fixed points, and along the flow it changes at the rate $\dd\tilde a/\dd\log\mu\propto\chi_{IJ}B^IB^J$. Here $\chi_{IJ}$ is a positive metric on the space of couplings, at leading order the Zamolodchikov metric given by the two-point functions of the $\mathcal O_I$. In a scale-invariant theory the couplings move only by a rotation of the fields, and a rotation does not change $\tilde a$. So $\chi_{IJ}B^IB^J=0$, and hence $B^I=0$. This is the perturbative proof that scale invariance implies conformal invariance \cite{Luty:2012ww,Fortin:2012hn}. Beyond perturbation theory, the $a$-theorem was proved by Komargodski and Schwimmer, who coupled the theory to a background dilaton, a spurion for the Weyl factor of the metric that couples to $\Theta$, and studied its low-energy effective action \cite{Komargodski:2011vj,Komargodski:2011xv}. We return to this in Sec.~\ref{sec:lookback} and Appendix~\ref{app:lookback}.

\subsection{How to read these lectures}\label{sec:guide}
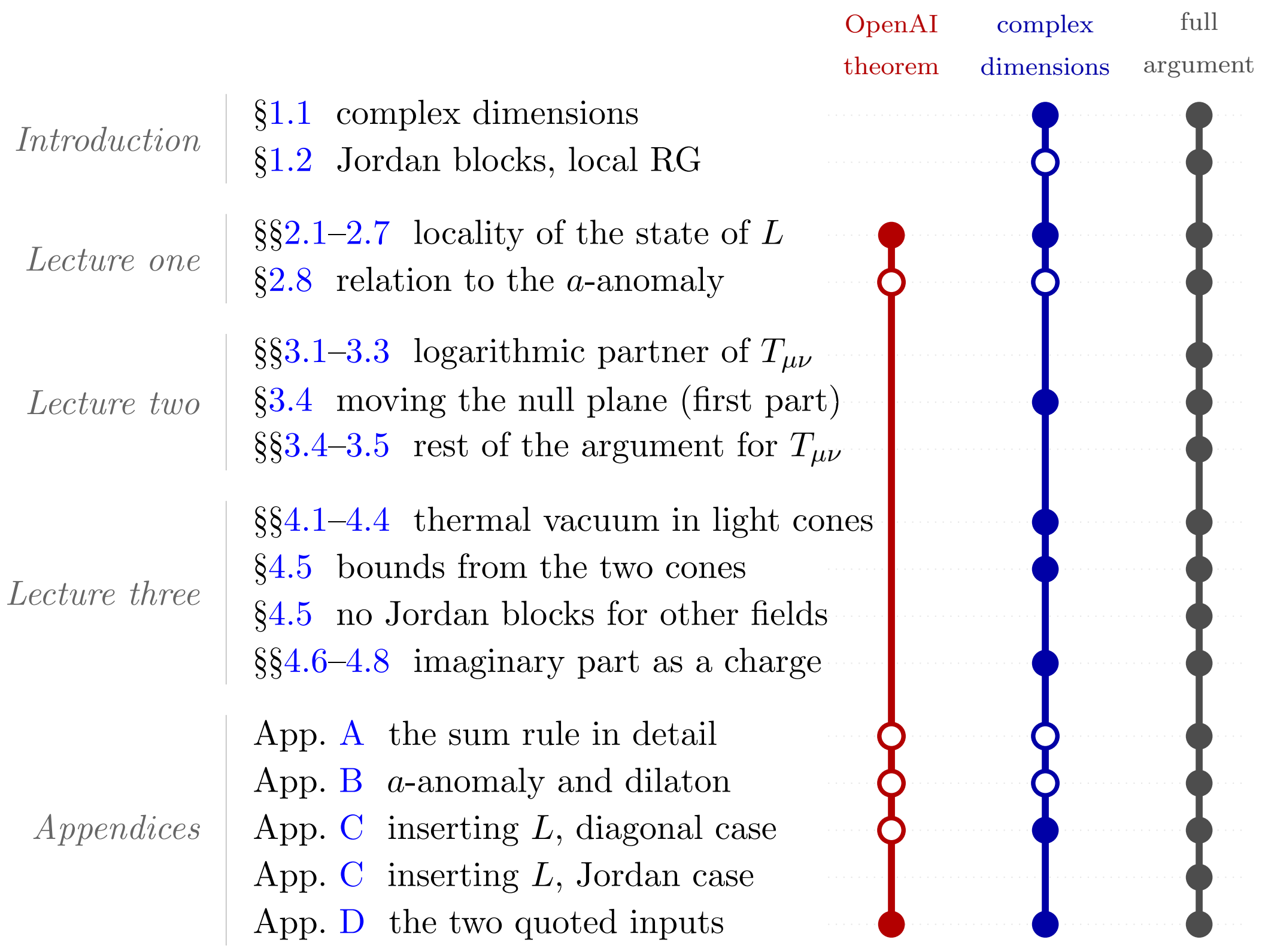
\begin{figure}[t!]
\centering
\begin{tikzpicture}[x=1cm,y=0.52cm,font=\small]
\node[anchor=east,font=\small\itshape,gray!80!black] at (2.0,-0.500) {Introduction};
\draw[gray!40] (2.15,0.450)--(2.15,-1.450);
\node[anchor=east,font=\small\itshape,gray!80!black] at (2.0,-3.050) {Lecture one};
\draw[gray!40] (2.15,-2.100)--(2.15,-4.000);
\node[anchor=east,font=\small\itshape,gray!80!black] at (2.0,-6.100) {Lecture two};
\draw[gray!40] (2.15,-4.650)--(2.15,-7.550);
\node[anchor=east,font=\small\itshape,gray!80!black] at (2.0,-10.150) {Lecture three};
\draw[gray!40] (2.15,-8.200)--(2.15,-12.100);
\node[anchor=east,font=\small\itshape,gray!80!black] at (2.0,-15.200) {Appendices};
\draw[gray!40] (2.15,-12.750)--(2.15,-17.650);
\node[anchor=west] at (2.3,0.000) {\S\ref{sec:complexintro}\ \ complex dimensions};
\draw[gray!20,dotted] (8.8,0.000)--(13.4,0.000);
\node[anchor=west] at (2.3,-1.000) {\S\ref{sec:jordanintro}\ \ Jordan blocks, local RG};
\draw[gray!20,dotted] (8.8,-1.000)--(13.4,-1.000);
\node[anchor=west] at (2.3,-2.550) {\S\S\ref{sec:twopt}--\ref{sec:sumrule}\ \ locality of the state of $L$};
\draw[gray!20,dotted] (8.8,-2.550)--(13.4,-2.550);
\node[anchor=west] at (2.3,-3.550) {\S\ref{sec:lookback}\ \ relation to the $a$-anomaly};
\draw[gray!20,dotted] (8.8,-3.550)--(13.4,-3.550);
\node[anchor=west] at (2.3,-5.100) {\S\S\ref{sec:plan2}--\ref{sec:step2}\ \ logarithmic partner of $T_{\mu\nu}$};
\draw[gray!20,dotted] (8.8,-5.100)--(13.4,-5.100);
\node[anchor=west] at (2.3,-6.100) {\S\ref{sec:step3}\ \ moving the null plane (first part)};
\draw[gray!20,dotted] (8.8,-6.100)--(13.4,-6.100);
\node[anchor=west] at (2.3,-7.100) {\S\S\ref{sec:step3}--\ref{sec:noassume}\ \ rest of the argument for $T_{\mu\nu}$};
\draw[gray!20,dotted] (8.8,-7.100)--(13.4,-7.100);
\node[anchor=west] at (2.3,-8.650) {\S\S\ref{sec:thermal}--\ref{sec:cones}\ \ thermal vacuum in light cones};
\draw[gray!20,dotted] (8.8,-8.650)--(13.4,-8.650);
\node[anchor=west] at (2.3,-9.650) {\S\ref{sec:logconv}\ \ bounds from the two cones};
\draw[gray!20,dotted] (8.8,-9.650)--(13.4,-9.650);
\node[anchor=west] at (2.3,-10.650) {\S\ref{sec:logconv}\ \ no Jordan blocks for other fields};
\draw[gray!20,dotted] (8.8,-10.650)--(13.4,-10.650);
\node[anchor=west] at (2.3,-11.650) {\S\S\ref{sec:charge}--\ref{sec:examples}\ \ imaginary part as a charge};
\draw[gray!20,dotted] (8.8,-11.650)--(13.4,-11.650);
\node[anchor=west] at (2.3,-13.200) {App.~\ref{app:sumrule}\ \ the sum rule in detail};
\draw[gray!20,dotted] (8.8,-13.200)--(13.4,-13.200);
\node[anchor=west] at (2.3,-14.200) {App.~\ref{app:lookback}\ \ $a$-anomaly and dilaton};
\draw[gray!20,dotted] (8.8,-14.200)--(13.4,-14.200);
\node[anchor=west] at (2.3,-15.200) {App.~\ref{sec:insert}\ \ inserting $L$, diagonal case};
\draw[gray!20,dotted] (8.8,-15.200)--(13.4,-15.200);
\node[anchor=west] at (2.3,-16.200) {App.~\ref{sec:insert}\ \ inserting $L$, Jordan case};
\draw[gray!20,dotted] (8.8,-16.200)--(13.4,-16.200);
\node[anchor=west] at (2.3,-17.200) {App.~\ref{sec:entered}\ \ the two quoted inputs};
\draw[gray!20,dotted] (8.8,-17.200)--(13.4,-17.200);
\draw[red!70!black,line width=2.2pt] (9.5,-2.550)--(9.5,-17.200);
\node[align=center,font=\scriptsize,text=red!70!black] at (9.5,1.500) {OpenAI\\theorem};
\fill[red!70!black] (9.5,-2.550) circle (4.2pt);
\filldraw[draw=red!70!black,fill=white,line width=1.4pt] (9.5,-3.550) circle (3.8pt);
\filldraw[draw=red!70!black,fill=white,line width=1.4pt] (9.5,-14.200) circle (3.8pt);
\filldraw[draw=red!70!black,fill=white,line width=1.4pt] (9.5,-15.200) circle (3.8pt);
\fill[red!70!black] (9.5,-17.200) circle (4.2pt);
\draw[blue!65!black,line width=2.2pt] (11.2,0.000)--(11.2,-17.200);
\node[align=center,font=\scriptsize,text=blue!65!black] at (11.2,1.500) {complex\\dimensions};
\fill[blue!65!black] (11.2,0.000) circle (4.2pt);
\filldraw[draw=blue!65!black,fill=white,line width=1.4pt] (11.2,-1.000) circle (3.8pt);
\fill[blue!65!black] (11.2,-2.550) circle (4.2pt);
\filldraw[draw=blue!65!black,fill=white,line width=1.4pt] (11.2,-3.550) circle (3.8pt);
\fill[blue!65!black] (11.2,-6.100) circle (4.2pt);
\fill[blue!65!black] (11.2,-8.650) circle (4.2pt);
\fill[blue!65!black] (11.2,-9.650) circle (4.2pt);
\fill[blue!65!black] (11.2,-11.650) circle (4.2pt);
\filldraw[draw=blue!65!black,fill=white,line width=1.4pt] (11.2,-14.200) circle (3.8pt);
\fill[blue!65!black] (11.2,-15.200) circle (4.2pt);
\fill[blue!65!black] (11.2,-17.200) circle (4.2pt);
\draw[black!70,line width=2.2pt] (12.9,0.000)--(12.9,-17.200);
\node[align=center,font=\scriptsize,text=black!70] at (12.9,1.500) {full\\argument};
\fill[black!70] (12.9,0.000) circle (4.2pt);
\fill[black!70] (12.9,-1.000) circle (4.2pt);
\fill[black!70] (12.9,-2.550) circle (4.2pt);
\fill[black!70] (12.9,-3.550) circle (4.2pt);
\fill[black!70] (12.9,-5.100) circle (4.2pt);
\fill[black!70] (12.9,-6.100) circle (4.2pt);
\fill[black!70] (12.9,-7.100) circle (4.2pt);
\fill[black!70] (12.9,-8.650) circle (4.2pt);
\fill[black!70] (12.9,-9.650) circle (4.2pt);
\fill[black!70] (12.9,-10.650) circle (4.2pt);
\fill[black!70] (12.9,-11.650) circle (4.2pt);
\fill[black!70] (12.9,-14.200) circle (4.2pt);
\fill[black!70] (12.9,-15.200) circle (4.2pt);
\fill[black!70] (12.9,-16.200) circle (4.2pt);
\fill[black!70] (12.9,-17.200) circle (4.2pt);
\filldraw[draw=red!70!black,fill=white,line width=1.4pt] (9.5,-13.200) circle (3.8pt);
\filldraw[draw=blue!65!black,fill=white,line width=1.4pt] (11.2,-13.200) circle (3.8pt);
\fill[black!70] (12.9,-13.200) circle (4.2pt);
\end{tikzpicture}
\caption{Three routes through the lectures (Sec.~\ref{sec:guide}). A filled circle marks a part that the route needs, and an open circle a part that is optional or needed only in part. The lines pass by the parts that a route skips.}\label{fig:routes}
\end{figure}

Lecture one is the core of the OpenAI argument, and the other two lectures build on it. Depending on what you want, you can take one of three routes, which Fig.~\ref{fig:routes} shows section by section.

\begin{itemize}
\item \emph{Only the OpenAI theorem.} Read Lecture one. Appendix~\ref{sec:entered} derives the main input of Lecture one, which Lecture one quotes without proof, Appendix~\ref{app:sumrule} gives the details of the sum rule, and Appendix~\ref{app:lookback} relates Lecture one to the $a$-anomaly.
\item \emph{Complex dimensions, but no Jordan blocks.} Assume that the dilatation acts diagonally on every field. Then Lecture one applies as it stands, and you can go on to Lecture three and skip its arguments against Jordan blocks. From Lecture two you need only the two derivatives \eqref{eq:Kp} and \eqref{eq:Kpp} of the second moment $K^+$ of the energy flux in Step 3, and from the appendices the beginning of Appendix~\ref{sec:insert} and Appendix~\ref{sec:entered}.
\item \emph{The full argument.} Read the lectures in order.
\end{itemize}
In each lecture I have boxed the important formulas, and a footnote in Sec.~\ref{sec:summary} lists the inputs that I quote without proof.

\FloatBarrier

\section{Lecture one: A physicist's way into the first half of the OpenAI paper}

\subsection{Why two-point functions are not enough}\label{sec:twopt}
In this lecture I would like to explain the idea of the first half of the OpenAI paper. As we saw in the Introduction, the question is whether $L=\Box^{-1}\Theta$ is a local operator. The first half of the paper does not yet construct $L$ as an operator. It looks only at the state that $L$ would create from the vacuum, and asks whether this state looks like one created by a local operator. I will not follow the paper line by line. Rather, I would like to explain why each step is natural for a physicist, leaving the precise statements to the paper. Throughout this lecture we assume that $T_{\mu\nu}$ and $V_\mu$ have definite scaling dimensions. We will come back to this assumption at the end of the lecture, and in Lecture two we will see that it is in fact not needed.

Now, the first thing one would look at is the two-point function. Since $\Theta$ is a Hermitian scalar of dimension four, \eqref{eq:wightman} gives
\begin{equation}
\int\dd^4x\,\ee^{\ii p\cdot x}\,\langle\Om|\Theta(x)\Theta(0)|\Om\rangle=C_4\,\theta(p^0)\theta(p^2)\,(p^2)^2 ,\qquad C_4\ge0 .
\end{equation}
The spectral density vanishes like $(p^2)^2$ at the boundary $p^2=0$ of the forward cone, so nothing prevents us from dividing the state $\Theta(p)|\Om\rangle$ by $p^2$. Since $\Box$ acts as $-p^2$ on states, we define
\[
L(p)|\Om\rangle\equiv-\frac1{p^2}\,\Theta(p)|\Om\rangle .
\]
In the squared norm of this state the two factors of $1/p^2$ cancel the factor $(p^2)^2$, so the spectral density of $L$ is $C_4\,\theta(p^0)\theta(p^2)$, which by \eqref{eq:wightman} is that of a scalar of dimension two. In other words, as far as the two-point function is concerned, $L$ is a perfectly good local scalar of dimension two.

This is not surprising, because a two-point function gives only the inner products of states, and from inner products alone we cannot see where a state was created. Here is an example, which we will use throughout this lecture. In the theory of a free massless scalar $\phi$, both $\Box\phi^2$ and $\phi^4$ are scalars of dimension four. If we apply $\Box^{-1}$ to the first, we get back the local field $\phi^2$. If we apply it to the second, we get $\Box^{-1}\phi^4$, which is not a local field, since $\phi^4$ is a primary field and hence not $\Box$ of any local field. Nevertheless, by the argument above, both have the spectral density of a scalar of dimension two, so up to normalization their two-point functions are the same.

How, then, can we see that one of them is local and the other is not? Suppose we smear both fields with a test function $h$ supported in a bounded region. The field $\phi^2$ smeared with $h$ creates a state in this region. The field $\Box^{-1}\phi^4$ smeared with $h$, on the other hand, is the same as $\phi^4$ smeared with $\Box^{-1}h$. This function is not supported in a bounded region but extends over the whole light cone of the region. The state is thus created by a source that reaches out to infinity, so we expect its energy to be spread more widely than that of a state created in the bounded region. The same would be true for any field of dimension four, unless the field is itself $\Box$ of a local field, as $\Box\phi^2$ is.

The OpenAI paper makes this difference quantitative by looking at the energy that the state carries through a null plane. For a generic test function, the energy of the state of $\Box^{-1}\phi^4$ turns out to be spread so widely in the transverse directions that its mean squared width is infinite, whereas for $\phi^2$ it is finite. For $L=\Box^{-1}\Theta$ we will then ask which of the two happens.

\subsection{Energy flux through a null plane}
We begin by defining the energy flux through a null plane and its width. We use the light-cone coordinates
\[
u=\frac{x^0-x^3}{\sqrt2},\qquad v=\frac{x^0+x^3}{\sqrt2},\qquad \dd s^2=2\,\dd u\,\dd v-|\dd\vy|^2 ,
\]
with the transverse coordinates $\vy=(x^1,x^2)$, and take the null plane $u=0$, which is spanned by the null direction $v$ and the two transverse directions. We write $p^+\equiv p_v$ and $p^-\equiv p_u$ for the momenta conjugate to $v$ and $u$, so that $p^2=2p^+p^--|\vec p|^2$ with the transverse momentum $\vec p$.

The energy that passes through the plane at the transverse position $\vy$ is obtained by integrating $T_{vv}$ along the light ray at $\vy$:
\begin{equation}
E(\vy)=\int_{-\infty}^{\infty}\dd v\,T_{vv}(u=0,v,\vy)\ \ge0 .
\end{equation}
This is the light-ray operator familiar from conformal collider physics \cite{Hofman:2008ar,Kravchuk:2018htv}, and its positivity is the averaged null energy condition \cite{Klinkhammer:1991ki,Faulkner:2016mzt,Hartman:2016lgu}. It also annihilates the vacuum, since it is positive and its vacuum expectation value vanishes due to the Lorentz symmetry.

Two light rays at different transverse positions are spacelike separated, so their light-ray operators commute. We can therefore measure the energy at all transverse positions at once and consider the moments of its distribution,
\begin{equation}\label{eq:moments}
P^+=\int\dd^2y\,E(\vy),\qquad P^+Y^i=\int\dd^2y\,y^iE(\vy),\qquad K^+=\int\dd^2y\,|\vy|^2E(\vy) ,
\end{equation}
which are commuting operators. Here $P^+$ is the total null momentum, the generator of translations in $v$, and $Y^i$ is the centroid of the energy. Note that in a CFT the second moment $K^+$ is a component of the special conformal generator and generates a symmetry, whereas in a theory that is only scale invariant it does not.

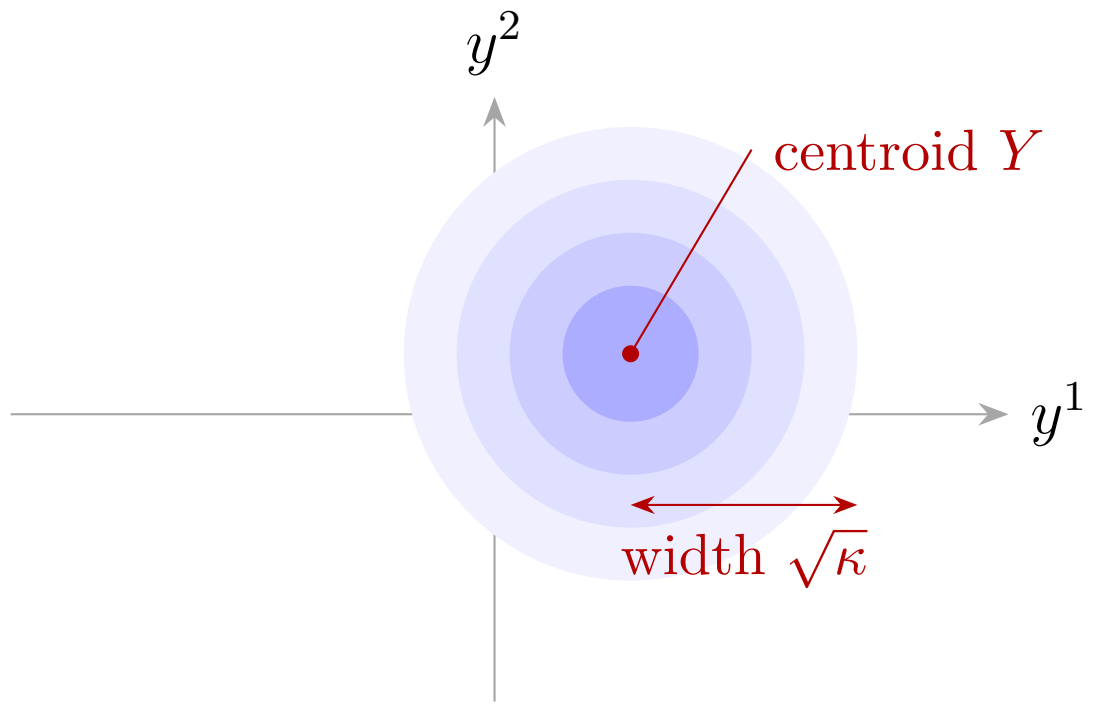
\begin{figure}[htb]
\centering
\begin{tikzpicture}[scale=1.0]
\draw[axisline] (-3.2,0)--(3.4,0) node[right,black] {$y^1$};
\draw[axisline] (0,-1.9)--(0,2.1) node[above,black] {$y^2$};
\foreach \r/\c in {1.5/6,1.15/12,0.8/20,0.45/32} {\fill[blue!\c] (0.9,0.4) circle (\r);}
\fill[red!70!black] (0.9,0.4) circle (1.6pt);
\draw[red!70!black,thin] (0.9,0.4)--(1.7,1.75) node[right,font=\small] {centroid $Y$};
\draw[{Stealth[length=1.6mm]}-{Stealth[length=1.6mm]},red!70!black] (0.9,-0.6)--(2.4,-0.6);
\node[red!70!black,font=\small] at (1.65,-0.95) {width $\sqrt{\kappa}$};
\end{tikzpicture}
\caption{Energy flux through the null plane, shown in the transverse plane. The arrow marks the root-mean-square distance of the energy from its centroid.}\label{fig:kappa}
\end{figure}

The total energy $P^+$ and the centroid $Y$ describe the state as a whole. What we are interested in is how the energy is spread around the centroid, and the simplest measure of this is the squared width
\begin{equation}\label{eq:kappa}
\boxed{\ \kappa=\frac{K^+}{P^+}-|Y|^2\ \ge0\ }
\end{equation}
that is, the variance of the transverse position weighted with the energy (Fig.~\ref{fig:kappa}). It is non-negative because the weights $E(\vy)$ are. Let me emphasize that $\kappa$ is an operator. If we prepare a state $|\psi\rangle$ and measure the energy flux, we obtain some value of $\kappa$, and if we repeat the experiment many times, we find that $\kappa$ is larger than $t$ with the probability
\begin{equation}
\mathrm{Prob}_\psi(\kappa>t)=\frac{\langle\psi|\mathbf 1(\kappa>t)|\psi\rangle}{\langle\psi|\psi\rangle} ,
\end{equation}
where $\mathbf 1(\kappa>t)$ is the spectral projection of $\kappa$. We omit the subscript $\psi$ when the state is clear from the context. In what follows we call $\sqrt\kappa$ the width, and the mean value of $\kappa$,
\[
\langle\kappa\rangle=\int_0^\infty\dd t\,\mathrm{Prob}(\kappa>t)
\]
the mean squared width.

\subsection{The width in a CFT and the Claim}\label{sec:claim}
How does the width depend on the way we prepare the state? Let us first answer this question in a CFT, where conformal symmetry allows us to compute the width, and then ask what remains of the answer in a theory that is only scale invariant.

We prepare a state by smearing a field $O$ with a test function $h(p)$ in momentum space,
\[
O(h)|\Om\rangle=\int\dd^4p\,h(p)\,O(p)|\Om\rangle .
\]
Only momenta in the forward cone contribute, and there we use the coordinates $(M,p^+,\vec p)$, with $p^-$ determined by $M^2=2p^+p^--|\vec p|^2$. For a field of dimension $\Delta$ it is convenient to take
\[
h(p)=M^{2-\Delta}f(M)\,\zeta(p^+,\vec p),
\]
and we write $|O(f)\rangle$ for the resulting state. The function $\zeta$ describes how the state moves as a whole, and $f$ how it depends on the invariant mass. The factor $M^{2-\Delta}$ has been inserted for the following reason. By \eqref{eq:wightman} the spectral density is proportional to $M^{2\Delta-4}$, and at fixed $p^+$ and $\vec p$ we have $\dd p^-=M\,\dd M/p^+$, so the squared norm of $|O(f)\rangle$ is proportional to
\[
\int_0^\infty|f(M)|^2M\,\dd M
\]
times a norm of $\zeta$, whatever $\Delta$ is.

This looks just like the squared norm of a function on a plane, written in polar coordinates with the radius $M$. So let us introduce an auxiliary two-dimensional plane with a vector $\vec q$ of length $|\vec q|=M$, and regard $f(M)$ as a radial function $f(|\vec q|)$ on this plane, which I will call the \emph{wave function} of the state. Note that the auxiliary plane is not a plane in spacetime. We denote by $\hat f(\vec\rho)$ the two-dimensional Fourier transform of $f(|\vec q|)$, which is a function on the conjugate plane, and write $r=|\vec\rho|$.

Now, in a CFT, $K^+$ generates a symmetry, so the conformal algebra determines how $K^+$, and hence $\kappa$, acts on the states of a scalar primary field $O$ of dimension $\Delta$. The result, which I quote without derivation, is surprisingly simple: for every $\zeta$, $\kappa$ acts on the wave function $f$ as
\begin{equation}\label{eq:kappaCFT}
\boxed{\ \kappa=-\frac{\partial^2}{\partial M^2}-\frac1M\frac{\partial}{\partial M}+\frac{\nu^2}{M^2},\qquad\nu=\Delta-2\ }
\end{equation}
This is nothing but the Laplacian $-\nabla^2$ of the auxiliary plane in polar coordinates, acting on the function $f(M)\,\ee^{\ii\nu\vartheta}$, where $\vartheta$ is the polar angle of $\vec q$. Since the Laplacian becomes the multiplication by $r^2$ in the conjugate plane, in a CFT the width of the energy flux is distributed as the radius $r$ is distributed in the Fourier transform of $f(M)\,\ee^{\ii\nu\vartheta}$. The invariant mass is not the momentum conjugate to the transverse position, and yet, to put it as a slogan, ``in a CFT the width is conjugate to the invariant mass.''

For $\Delta=2$ the angular momentum $\nu$ vanishes, so $\kappa$ is distributed as $r^2$ is distributed in $|\hat f(\vec\rho)|^2\,\dd^2\rho$. In particular, if $\hat f$ vanishes outside the disk of radius $R$, no measurement ever gives a width larger than $R$. For $\Delta\neq2$, on the other hand, the function $f(M)\,\ee^{\ii\nu\vartheta}$ carries the angular momentum $\nu$, and $\nu^2/M^2$ is the familiar centrifugal term. Integrating by parts, we find the mean squared width in the state $|O(f)\rangle$:
\begin{equation}\label{eq:kapparef}
\frac{\langle O(f)|\kappa|O(f)\rangle}{\langle O(f)|O(f)\rangle}
=\frac{\int_0^\infty\big(|M\partial_Mf|^2+\nu^2|f|^2\big)\frac{\dd M}{M}}{\int_0^\infty|f|^2M\,\dd M} .
\end{equation}
We write $\kappa_0[f]$ for the right-hand side with $\nu=0$, which is the mean value of $r^2$ in $|\hat f(\vec\rho)|^2$.

What does this mean for our two examples, $L_1=\phi^2$ and $L_2=\Box^{-1}\phi^4$? The free theory is a CFT, so we can use \eqref{eq:kappaCFT}. Both fields have dimension two, so we prepare their states with the same test function $h=f(M)\zeta$, that is, with the same wave function $f$. Since $L_1=\phi^2$ is a primary field of dimension two, $\kappa$ acts on its state with $\nu=0$. For $L_2$, we recall that $\Box$ acts as $-M^2$ on states, so that
\[
L_2(h)|\Om\rangle=-\int\dd^4p\,M^{-2}f(M)\,\zeta\,\phi^4(p)|\Om\rangle=-|\phi^4(f)\rangle .
\]
In the last step we noticed that $M^{-2}f\zeta$ is the test function $M^{2-\Delta}f\zeta$ with $\Delta=4$, which gives $\phi^4$ the wave function $f$. The state of $L_2$ is therefore the state of the primary field $\phi^4$ of dimension four, on which $\kappa$ acts with $\nu=2$.

With \eqref{eq:kapparef} we can now compare the two. For $L_1$ we get $\kappa_0[f]$, while for $L_2$ we get in addition the centrifugal term
\[
\frac{\int_0^\infty4|f|^2\,\dd M/M}{\int_0^\infty|f|^2M\,\dd M} ,
\]
which diverges logarithmically at $M=0$ unless $f(0)=0$. Thus, for a smooth $f$ with $f(0)\neq0$, the mean squared width is finite for the local field and infinite for the non-local one, just as we anticipated in Sec.~\ref{sec:twopt}. This is the difference that we will look for in the state of $L$.

So far we have been in a CFT. In a theory that is only scale invariant, $K^+$ does not generate a symmetry, so there is no reason for \eqref{eq:kappaCFT} to hold. Still, we can compute with \eqref{eq:kappaCFT} as if $O$ were a primary field of a CFT with the same two-point function. We denote the probability computed in this way by $\mathrm{Prob}_{\rm CFT}$. The OpenAI paper shows that the true probability is never larger:
\begin{equation}\label{eq:compare}
\mathrm{Prob}(\kappa>t)\ \le\ \mathrm{Prob}_{\rm CFT}(\kappa>t)\qquad\text{for every }t\ge0 .
\end{equation}
I will call this inequality the \emph{comparison with the CFT}. It says that a local field can spread its energy at most as widely as the corresponding field of a CFT does.

The comparison holds for a local scalar field $O$ of definite dimension $\Delta$ and for every wave function $f$, with $\nu=\Delta-2$. Note that $O$ must have a definite dimension here, since otherwise its two-point function is not that of a CFT and there is nothing to compare with. This will become important in Lecture two. In these lectures I use the comparison without proof. It comes from the thermal property of the vacuum that we discuss in Lecture three, and I sketch the argument in Appendix~\ref{sec:entered}.

For a local scalar of dimension two, the comparison gives the central input of this lecture, which I call the Claim.

\begin{center}
\setlength{\fboxsep}{8pt}\fbox{\begin{minipage}{0.9\textwidth}
\textbf{Claim} \cite{OpenAI:2026}.\quad Let $O$ be a local scalar of dimension two. If the Fourier transform $\hat f(\vec\rho)$ of the wave function $f(|\vec q|)$ is supported in a disk of radius $R$, then in the state $|O(f)\rangle$
\begin{equation}\label{eq:claim}
\mathrm{Prob}(\kappa>R^2)=0 .
\end{equation}
\end{minipage}}
\end{center}

\noindent That is, if we prepare the state with such a wave function, no measurement ever gives a width larger than $R$, just as in a CFT. In a CFT this followed from the symmetry generated by $K^+$. In a theory that is only scale invariant there is no such symmetry, and yet the width is still controlled by the way the state depends on the invariant mass.

Finally, if we integrate \eqref{eq:compare} over $t$, we also obtain a bound on the mean value. For a local scalar of dimension two,
\begin{equation}\label{eq:claimineq}
\langle\kappa\rangle\ \le\ \kappa_0[f]
\end{equation}
in every state $|O(f)\rangle$, which I will call the mean-value form of the comparison. For a field of another dimension $\Delta$, the right-hand side is replaced by \eqref{eq:kapparef} with $\nu=\Delta-2$.

\subsection{The tail of the state of $L$}\label{sec:tailL}
Now we come back to our question: is the state of $L$ like the state of a local field of dimension two, or like the state of $L_2=\Box^{-1}\phi^4$?

We cannot use the Claim, since we do not yet know whether $L$ is local. But $\Theta$ is local, and on states $L$ is just $\Theta$ divided by $-p^2$. With the test function $h=f(M)\zeta$ we therefore have
\[
|l(f)\rangle\equiv L(h)|\Om\rangle=-|\Theta(f)\rangle ,
\]
that is, the state of $L$ with the wave function $f$ is, up to sign, the state of $\Theta$ with the same wave function. This is what happened for $L_2$ in the free theory.

Since $\Theta$ is a local field of dimension four, we can use the comparison \eqref{eq:compare} with $\nu=2$:
\[
\mathrm{Prob}_{l(f)}(\kappa>t)\ \le\ \mathrm{Prob}_{\rm CFT}(\kappa>t) .
\]
Here $\mathrm{Prob}_{\rm CFT}$ is computed with the function $f(M)\,\ee^{2\ii\vartheta}$ on the auxiliary plane, as for the state of $L_2$. So the energy of the state of $L$ is spread at most as widely as that of the state of $L_2$. It is therefore worth looking at the latter more closely.

What does the function $f(M)\,\ee^{2\ii\vartheta}$ look like in the conjugate plane? We call its Fourier transform $\psi_{\rm CFT}(\vec\rho)$. In general, the behavior of a Fourier transform at large $r$ is determined by the singularities of the function. For $f(0)\neq0$ our function is indeed singular at $\vec q=0$. Near the origin it is approximately $f(0)\,\ee^{2\ii\vartheta}$, whose value depends on the direction from which we approach the origin. Since this is a function of the angle alone, its Fourier transform carries angular momentum two and falls off like $1/r^2$. The precise statement, which I quote from the OpenAI paper \cite{OpenAI:2026}, is that if $\hat f$ is supported in the disk of radius $R$, as in the Claim, then outside the disk
\begin{equation}\label{eq:psitail}
\psi_{\rm CFT}(\vec\rho)=-\frac{2f(0)\,\ee^{2\ii\varphi}}{r^2}\qquad(r>R) ,
\end{equation}
where $\varphi$ is the polar angle of $\vec\rho$.

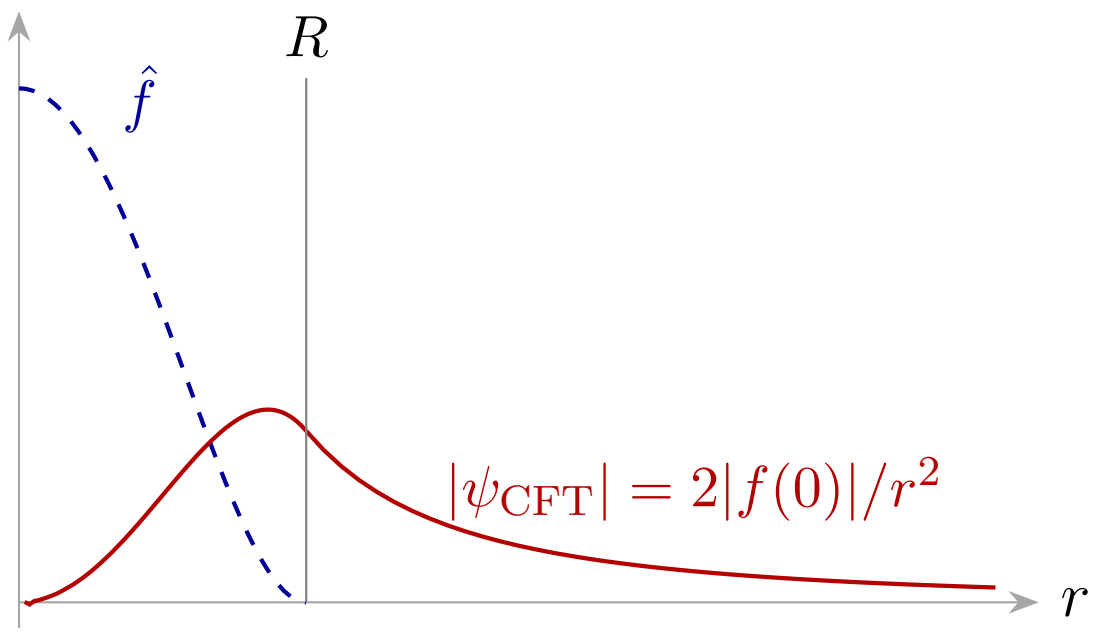
\begin{figure}[htb]
\centering
\begin{tikzpicture}[x=1.9cm,y=3.4cm]
\draw[axisline] (0,0)--(3.55,0) node[right,black] {$r$};
\draw[axisline] (0,-0.05)--(0,1.15);
\draw[thick,dashed,blue!60!black] plot[domain=0:1,samples=40] (\x,{(1-\x*\x)^2});
\draw[thick,red!70!black] plot[domain=0.02:1,samples=60] (\x,{(1-(1-\x*\x)^3)/(3*\x*\x)-(1-\x*\x)^2});
\draw[thick,red!70!black] plot[domain=1:3.4,samples=40] (\x,{1/(3*\x*\x)});
\draw[gray] (1,0)--(1,1.02) node[above,font=\small,black] {$R$};
\node[blue!60!black,font=\small] at (0.42,0.98) {$\hat f$};
\node[red!70!black,font=\small] at (2.35,0.22) {$|\psi_{\rm CFT}|=2|f(0)|/r^2$};
\end{tikzpicture}
\caption{The Fourier transform $\hat f$ (dashed) of a radial wave function $f$ with $f(0)\neq0$ is supported in the disk $r\le R$. The CFT wave function $\psi_{\rm CFT}$ of the state with $\nu=2$, the Fourier transform of $f\,\ee^{2\ii\vartheta}$, vanishes at $r=0$ since it carries angular momentum two, and outside the disk its modulus is the single tail $2|f(0)|/r^2$. The curves are drawn for $\hat f\propto(1-r^2/R^2)^2$.}\label{fig:tail}
\end{figure}

Thus, for $\nu=2$, the function $\psi_{\rm CFT}$ does not stay inside the disk. It has a tail $1/r^2$, whose size is fixed by a single number, the value $f(0)$ of the wave function at zero invariant mass (Fig.~\ref{fig:tail}). Since the width is distributed as $r$ in $|\psi_{\rm CFT}|^2$, the CFT state can give widths larger than $R$. Moreover, with the density $|\psi_{\rm CFT}|^2\,\dd^2\rho\propto\dd r/r^3$ the mean value of $r^2$ diverges logarithmically, which is the divergence that we found for $L_2$.

Note that the tail is the only difference from a local field of dimension two. If $f(0)=0$, the tail \eqref{eq:psitail} is absent and the CFT state stays inside the disk, and by the comparison so does the state of $L$. It never gives a width larger than $R$, as the Claim says for a local field. The question is what happens for $f(0)\neq0$.

For $f(0)\neq0$ the comparison gives only an upper bound on the tail of $|l(f)\rangle$, but we can still determine its form. Take two wave functions $f_1$ and $f_2$, both with $\hat f_i$ supported in the disk, and with the same value at $M=0$. Their difference vanishes at $M=0$, so the state $|l(f_1)\rangle-|l(f_2)\rangle$ never gives a width larger than $R$. The part of $|l(f)\rangle$ with a width larger than $R$ therefore depends on $f$ only through $f(0)$. Writing $\mathcal P_t=\mathbf 1(\kappa>t)$ for the spectral projection, we find by linearity
\[
\big\|\mathcal P_t|l(f)\rangle\big\|^2=|f(0)|^2\,C(t)\qquad(t>R^2)
\]
with a function $C(t)$ that does not depend on $f$.

Scale invariance then fixes $C(t)$. Indeed, $|f(0)|^2C(t)$ is a squared norm, so it has the same dimension as $\int|f|^2M\,\dd M$, that is, the dimension of $|f(0)|^2$ times a mass squared. Since $t$ has the dimension of a length squared and the theory has no scale, we must have $C(t)=C(1)/t$. Following the OpenAI paper, we write $C(1)=\tfrac12\|\mathbf d_l\|^2$ and call $\mathbf d_l$ the \emph{tail amplitude}. (In the OpenAI paper $\mathbf d_l$ is a vector, but we need only its norm.) Dividing by the norm of the state, we obtain
\begin{equation}\label{eq:tail}
\mathrm{Prob}(\kappa>t)=\frac{|f(0)|^2\,\|\mathbf d_l\|^2}{2\,\langle l(f)|l(f)\rangle\,t}\qquad(t>R^2) .
\end{equation}

A tail $1/t$ makes the mean squared width $\langle\kappa\rangle=\int_0^\infty\dd t\,\mathrm{Prob}(\kappa>t)$ diverge logarithmically, as for $L_2$. Our question has thus been reduced to a question about a single number. If $\mathbf d_l\neq0$, the state of $L$ has an infinite mean squared width, like the state of $L_2$. If $\mathbf d_l=0$, on the other hand, it never gives a width larger than $R$, like the state of a local field of dimension two. The goal of the rest of this lecture is to show that $\mathbf d_l=0$.

\subsection{A key identity from causality}\label{sec:key}
So far we have used only the comparison, which holds for $\phi^4$ just as well as for $\Theta$, so it cannot be enough to show $\mathbf d_l=0$. We need a property that distinguishes $\Theta$ from $\phi^4$. This property is that $\Theta$ is the trace of the energy-momentum tensor, the same operator with which we measure the energy flux. Consider two energy-momentum tensors, one of which creates a state while the other measures its energy flux. Causality then allows us to exchange the roles of the two.

For this we need a light ray that creates a state. A complete light-ray integral annihilates the vacuum, so it creates nothing, and we therefore take only half of it. We define the half light-ray integrals
\begin{equation}
E_\pm(\vy)=\int_{v\gtrless0}\dd v\,T_{vv}(u=0,v,\vy),\qquad
\bar E_\pm(\vy)=\int_{u\gtrless0}\dd u\,T_{uu}(u,v=0,\vy),
\end{equation}
so that $E=E_++E_-$ and $\bar E=\bar E_++\bar E_-$. Then we have the \emph{key identity}
\begin{equation}\label{eq:key}
\boxed{\ E(\vy_1)\,\bar E_+(\vy_2)|\Om\rangle=\bar E(\vy_2)\,E_+(\vy_1)|\Om\rangle\ }
\end{equation}
On the left-hand side, $\bar E_+(\vy_2)$ creates a state and $E(\vy_1)$ measures its energy flux through the plane $u=0$. On the right-hand side, it is $E_+(\vy_1)$ that creates the state, and $\bar E(\vy_2)$ measures its energy flux through the plane $v=0$.

The proof is almost disappointingly easy. Since complete light-ray integrals annihilate the vacuum, $E_-|\Om\rangle=-E_+|\Om\rangle$ and $\bar E_-|\Om\rangle=-\bar E_+|\Om\rangle$. Moreover, the interval between $(0,v,\vy_1)$ and $(u,0,\vy_2)$ is $-2uv-|\vy_1-\vy_2|^2$, so half rays of the same sign are spacelike separated and commute. Hence
\[
E_-\bar E_+|\Om\rangle=-E_-\bar E_-|\Om\rangle=-\bar E_-E_-|\Om\rangle=\bar E_-E_+|\Om\rangle ,
\]
and adding $E_+\bar E_+|\Om\rangle=\bar E_+E_+|\Om\rangle$ gives \eqref{eq:key}.

\subsection{Where the tail shows up}\label{sec:spin}
In the key identity the state is created by a half light ray of $T$, whereas the tail is a property of the state $|l\rangle$ of $L$. To use the identity, we therefore have to find the tail of $|l\rangle$ in the states created by $T$. The two are connected by the relation $\Theta=\partial^\mu V_\mu=T^\mu{}_\mu$, because of which the state of $L$ appears in the states created by $V$ and by $T$. We will see that $V$ and $T$ can be local with a tail of $|l\rangle$ only if the other parts of their states have the opposite tail, and this is where we can look for it.

To make this explicit, let us decompose these states into spin zero, one and two in the rest frame of $p$, as we would do for a massive particle. In momentum space we define
\begin{equation}\label{eq:spin}
\begin{gathered}
|l\rangle=\Box^{-1}\Theta|\Om\rangle,\qquad |j_\mu\rangle=\frac{1}{\ii\sqrt{p^2}}\big(V_\mu|\Om\rangle-\ii p_\mu|l\rangle\big),\\
|s_{\mu\nu}\rangle=\frac{1}{p^2}\Big(T_{\mu\nu}|\Om\rangle-\tfrac13(p_\mu p_\nu-p^2\eta_{\mu\nu})\,|l\rangle\Big).
\end{gathered}
\end{equation}
One can check that $|j_\mu\rangle$ is transverse to $p$, and that $|s_{\mu\nu}\rangle$ is transverse and traceless. So $|l\rangle$ is the spin-zero part of both states, $|j_\mu\rangle$ is the spin-one part of the state of $V$, and $|s_{\mu\nu}\rangle$ is the spin-two part of the state of $T$.

Now, let $n$ and $\bar n$ be the null vectors along the $v$ and $u$ directions, so that $T_{vv}=n^\mu n^\nu T_{\mu\nu}$. In the rest frame of $p$ the vector $n$ points in some spatial direction, which we call $z$, and $n^\mu=(p^+/M)(1,e_z)$. Contracting \eqref{eq:spin} with $n$, we find
\begin{equation}\label{eq:cVcT}
V_v|\Om\rangle=\ii p^+\big(|l\rangle+|j_z\rangle\big)\equiv\ii p^+|c_V\rangle,\qquad
T_{vv}|\Om\rangle=(p^+)^2\big(\tfrac13|l\rangle+|s_{zz}\rangle\big)\equiv(p^+)^2|c_T\rangle ,
\end{equation}
where $|j_z\rangle$ and $|s_{zz}\rangle$ are the components along $z$.

We do not know whether $|l\rangle$, $|j_z\rangle$ and $|s_{zz}\rangle$ are states of local fields. But $|c_V\rangle$ and $|c_T\rangle$ are, up to powers of $p^+$ that we absorb into $\zeta$, the states of the local fields $V_v$ and $T_{vv}$. These fields have dimensions three and four, but once the factors $p^+$ and $(p^+)^2$ have been removed, $|c_V\rangle$ and $|c_T\rangle$ have the spectral density of a scalar of dimension two, and the OpenAI paper shows that the Claim holds for them as well \cite{OpenAI:2026}. As for $|l(f)\rangle$, we write $|c_V(f)\rangle$, $|j_z(f)\rangle$ and so on for these states with the test function $h=f(M)\zeta$.

So let $\hat f$ be supported in the disk of radius $R$. The Claim then says that $\mathcal P_t|c_V(f)\rangle=\mathcal P_t|c_T(f)\rangle=0$ for $t>R^2$, that is,
\[
\mathcal P_t|j_z(f)\rangle=-\mathcal P_t|l(f)\rangle,\qquad\mathcal P_t|s_{zz}(f)\rangle=-\tfrac13\,\mathcal P_t|l(f)\rangle\qquad(t>R^2) .
\]
That is, the spin-one and the spin-two parts must have tails that cancel the tail of $|l\rangle$. But then $|j_z\rangle$ and $|s_{zz}\rangle$ themselves have tails, so the matrix element of $\kappa$ between them diverges just as that of $|l\rangle$ does.

To make this quantitative, let us cut off the width $\sqrt\kappa$ at $\Lambda$ and write $[\,\cdot\,]_{\log\Lambda}$ for the coefficient of $\log\Lambda$. Since $\kappa=\int_0^\infty\dd t\,\mathcal P_t$ and $\mathcal P_t^2=\mathcal P_t$, the two relations above and the tail \eqref{eq:tail} give
\begin{equation}\label{eq:dd}
\big[\langle j_z(f)|\kappa|s_{zz}(f)\rangle\big]_{\log\Lambda}=\tfrac13\big[\langle l(f)|\kappa|l(f)\rangle\big]_{\log\Lambda}=\tfrac13\,|f(0)|^2\,\|\mathbf d_l\|^2 .
\end{equation}
This means that the matrix element $\langle j_z(f)|\kappa|s_{zz}(f)\rangle$ grows like $\log\Lambda$ if the state of $L$ has a tail.\footnote{States of different spin are orthogonal, but this matrix element need not vanish, because $\kappa$ singles out the null direction $n$, and both $|j_z\rangle$ and $|s_{zz}\rangle$ have helicity zero about this direction.}

In this matrix element the bra is the spin-one part of the state of $V$, the ket is the spin-two part of the state of $T$, and $\kappa$ is measured with a second $T$. A matrix element of this kind appears if we take the matrix element of the key identity with a bra created by $V$, as we do next.

\subsection{The sum rule}\label{sec:sumrule}
We now take the matrix element of the key identity \eqref{eq:key} with the bra $\langle V_\mu(p)|$ of the state $V_\mu(p)|\Om\rangle$. In terms of the Fourier transforms $E(\vec k_1)=\int\dd^2y\,\ee^{\ii\vec k_1\cdot\vy}E(\vy)$ and so on, it reads
\[
A_\mu(\vec k_1,\vec k_2)\equiv\langle V_\mu(p)|E(\vec k_1)\,\bar E_+(\vec k_2)|\Om\rangle=\langle V_\mu(p)|\bar E(\vec k_2)\,E_+(\vec k_1)|\Om\rangle\equiv A'_\mu(\vec k_1,\vec k_2),
\]
where the transverse momentum of the bra is $\vec p=\vec k_1+\vec k_2$. In $A$, a half light ray on the plane $v=0$ creates the state and a light ray on the plane $u=0$ measures its energy flux, while in $A'$ the two planes exchange their roles.

Now, there is another way to exchange the two planes, namely a rotation by $\pi$ about the $y^2$ axis. It exchanges $u$ and $v$, and with them $p^+$ and $p^-$ of the bra. A boost along $z$ brings $p^+$ and $p^-$ back to their original values. It rescales $E$ and $\bar E_+$ by inverse factors, so it does not change their product, and it does not change the ratios $V_v/p^+$ and $V_u/p^-$ either. Combining the rotation and the boost, we obtain
\[
p^-A_v(\vec k_1,\vec k_2)=p^+A'_u(S\vec k_2,S\vec k_1),\qquad S=\mathrm{diag}(-1,1) .
\]
So far this is only Lorentz invariance. If we now use the key identity, $A=A'$, it becomes a relation for $A$ alone,
\[
p^-A_v(\vec k_1,\vec k_2)=p^+A_u(S\vec k_2,S\vec k_1) .
\]
In $A$ the two light rays play different roles, and this relation says what happens when we exchange them. In particular, since $S^2=1$, the combination $\mathcal A=p^-A_v-p^+A_u$ changes sign under the exchange,
\[
\mathcal A(S\vec k_2,S\vec k_1)=-\mathcal A(\vec k_1,\vec k_2) .
\]
Therefore, if we weight the two light rays with the same radial function of the transverse position and integrate, $\mathcal A$ gives zero. In momentum space such a weight is the same function of $\vec k_1$ and of $\vec k_2$, so it does not change under the exchange. This is the \emph{sum rule}.

To use the sum rule we have to choose the weight. What we want on the measuring light ray is $\kappa$. By \eqref{eq:moments} and \eqref{eq:kappa},
\[
P^+\kappa=\int\dd^2y\,|\vy-Y|^2\,E(\vy) ,
\]
so the weight that gives $\kappa$ is the squared distance of the light ray from the centroid. We therefore weight both light rays with $|\vy-Y|^2$. What the sum rule then says has to be worked out, and the computation is not short. I do it in Appendix~\ref{app:sumrule}, and here I only quote the result,
\begin{equation}\label{eq:sumrule0}
M\langle j_z|\kappa|s_{zz}(f)\rangle=2\ii\,\langle j_i|\mathcal D_i|s_{zz}(f)\rangle+\tfrac23\,\langle j_i|\kappa|s_{zi}(Mf)\rangle ,
\end{equation}
where $\mathcal D_i$ is the first moment of the energy about its centroid,
\[
\mathcal D_i=\frac1{P^+}\int\dd^2y\,(y^i-Y^i)\,E(\vy) .
\]
In \eqref{eq:sumrule0} $f$ is a non-zero constant, the bra has the definite invariant mass $M$, and $i=1,2$ labels the two directions transverse to $z$ in the rest frame of $p$ and is summed over \cite{OpenAI:2026}.

Let us look at the two sides of this equation. On the left-hand side, $\kappa$ is taken between the spin-one state $\langle j_z|$ and the spin-two state $|s_{zz}(f)\rangle$ with a constant wave function. A constant wave function has $f(0)\neq0$, so up to the wave function of the bra this is the matrix element of \eqref{eq:dd}. If we cut off the width at $\Lambda$, it grows like $\log\Lambda$, with a coefficient that is a positive multiple of $\|\mathbf d_l\|^2$. On the right-hand side, the first term vanishes, because the first moment $\mathcal D_i$ is zero by the definition of the centroid, and the second term does not grow like $\log\Lambda$, because its wave function $Mf$ vanishes at $M=0$.\footnote{A constant $f$ is not normalizable, so to make this precise we cut off the weight to the annulus $1<|\vy-Y|<\Lambda$ around the centroid and give the bra a wave function as well. With the cutoff the first term no longer vanishes exactly, which is why we have kept it in \eqref{eq:sumrule0}, but both terms on the right-hand side grow more slowly than $\log\Lambda$ (Appendix~\ref{app:sumrule}).} The two sides can agree only if
\begin{equation}\label{eq:dl0}
\boxed{\ \mathbf d_l=0\ }
\end{equation}
The state of $L$ has the property that the Claim \eqref{eq:claim} asserts for a local scalar of dimension two: for a wave function whose Fourier transform is supported in a disk of radius $R$, the state of $L$ never gives a width larger than $R$.\footnote{This concerns only the vector $|l(f)\rangle=L(h)|\Om\rangle$. To improve the energy-momentum tensor with $L$, we need $L$ as a local operator, that is, also its matrix elements between other states, such as $\langle\Om|O_2^\dagger\,L(h)\,O_1|\Om\rangle$ with other fields $O_1$ and $O_2$. This is the subject of the second half of the OpenAI paper. In Lecture two we need this single insertion of $L$, which I explain in Appendix~\ref{sec:insert}.}

Why does this argument not apply to $L_2=\Box^{-1}\phi^4$, whose state does have a tail? For $\Theta$ we used that its state is the spin-zero part of the state of $T_{\mu\nu}$, the same field whose light rays appear in the key identity. The field $\phi^4$ is not the trace of the energy-momentum tensor of the free theory, so we cannot repeat the argument for it.

\subsection{What the tail means for the $a$-anomaly}\label{sec:lookback}
Before we leave this lecture, let me explain what the number $\|\mathbf d_l\|^2$ means physically. It turns out to be proportional to the rate at which the $a$-anomaly would keep flowing in a theory that is scale invariant but not conformal. Here I only describe the idea. The details are in Appendix~\ref{app:lookback}.

In perturbation theory this can be checked directly. Near a CFT the trace is $\Theta=B^I\mathcal O_I$ with nearly marginal scalars $\mathcal O_I$, as in \eqref{eq:B}. As for $L_2$, the state of $L$ is then, up to sign, $B^I|\mathcal O_I(f)\rangle$, and at lowest order $\kappa$ acts on it with $\nu=2$. If we cut off the width at $\Lambda$, the tail \eqref{eq:psitail} gives $\langle\kappa\rangle$ the logarithm $4|f(0)|^2\log\Lambda/\int_0^\infty|f|^2M\,\dd M$, while by \eqref{eq:tail} the same logarithm is $|f(0)|^2\|\mathbf d_l\|^2\log\Lambda/\langle l(f)|l(f)\rangle$. Comparing the two, we find
\[
\|\mathbf d_l\|^2=\frac{4\,\langle l(f)|l(f)\rangle}{\int_0^\infty|f|^2M\,\dd M}\ \propto\ \chi_{IJ}B^IB^J ,
\]
since the squared norm of $B^I|\mathcal O_I(f)\rangle$ is given by the Zamolodchikov metric $\chi_{IJ}$. This is the quantity that determines how fast $\tilde a$ changes in the perturbative argument of the Introduction.

Beyond perturbation theory, Hartman, Mathys and Stergiou have recently related the flow of $a$ to the energy flux \cite{Hartman:2026pos}. For a renormalization group flow between two CFTs they found
\[
\Delta a=\frac1{32}\int\dd^2b\,|\vec b|^2\mathcal E(\vec b) ,
\]
where $\mathcal E(\vec b)\ge0$ is the energy flux at the transverse distance $\vec b$ from the centroid, in the state created by $\Theta$ integrated over the half line $u<0$, with no transverse momentum. The half-line integral divides $\Theta(p)$ by $p^-=M^2/2p^+$, so this state is the state of $L$ with a wave function that does not depend on $M$, and the integral is its second moment $\langle P^+\kappa\rangle$. That is, up to normalization, the change of $a$ along the flow is the mean squared width of the state of $L$.

Hartman, Mathys and Stergiou read $1/b$ as the renormalization scale, so that the energy flux at the distance $b$ tells us how fast $a$ changes at the scale $1/b$. In a theory that is scale invariant but not conformal there is no scale, so every logarithmic interval of $b$ must contribute the same amount. This means $\mathcal E\propto1/b^4$, and $a$ would keep flowing at a constant rate. This profile is the tail \eqref{eq:tail} seen in the energy flux, and the rate is proportional to $\|\mathbf d_l\|^2$. So by \eqref{eq:dl0} $a$ does not keep flowing, which is the non-perturbative counterpart of $\chi_{IJ}B^IB^J=0$.

The earlier non-perturbative arguments \cite{Luty:2012ww,Dymarsky:2013pqa,Yonekura:2014tha,Dymarsky:2014zja} obtain the same rate from the forward scattering of a background dilaton, following Komargodski and Schwimmer \cite{Komargodski:2011vj}. Under three assumptions, which I state in Appendix~\ref{app:lookback}, the two rates agree. But to show that the rate vanishes, those arguments need further properties of the dilaton amplitudes. In this lecture we needed no dilaton, and our main physical inputs were the causality of the energy-momentum tensor and the positivity of energy.

\subsection{Summary of Lecture one}\label{sec:defdim}
Let me summarize what we have done. The two-point function of $L$ is that of a local scalar of dimension two, so it cannot show whether $L$ is local. We looked instead at the energy flux of the state that $L$ creates, and asked how widely it is spread in the transverse directions of a null plane. The comparison with a CFT leaves only one way in which this state can differ from the state of a local scalar, namely a tail of large widths, whose size is a single number $\|\mathbf d_l\|^2$. Causality then gives $\mathbf d_l=0$, because it allows us to exchange the light ray that creates the state with the light ray that measures its energy flux. By the formula of Hartman, Mathys and Stergiou, the same number is proportional to the rate at which the $a$-anomaly would keep flowing at a scale-invariant point. So we have also seen, without a dilaton, that it does not flow there.

Finally, let me come back to the assumption that I made at the beginning of this lecture. We took $T_{\mu\nu}$ and $V_\mu$, and hence $\Theta$ and $L$, to have definite scaling dimensions, because the comparison holds only for fields of definite dimension. If the dilatation mixes $T_{\mu\nu}$ with another operator through a Jordan block, $\Theta$ and $T_{vv}$ have no definite dimension, so we can apply neither the comparison to $\Theta$ nor the Claim to $|c_T\rangle$. In the next lecture I would like to explain how we can deal with this problem.

\section{Lecture two: How can we exclude the Jordan form?}

\subsection{The problem and two assumptions}\label{sec:plan2}
In Lecture one we assumed that $T_{\mu\nu}$ and $V_\mu$ have definite scaling dimensions. In this lecture I would like to show that, in a unitary theory, they cannot sit in a Jordan block, so that this assumption is not needed.

First, recall how a Jordan block could arise. In the $\lambda\phi^4$ example of the Introduction, $T_{\mu\nu}$ mixes under renormalization with the improvement term of $\phi^2$, and if $\gamma_{\phi^2}=0$ while $\hat\eta\neq0$ at a fixed point, this mixing becomes a Jordan block. In general we write
\begin{equation}\label{eq:chain}
\ee^{\ii sD}\,T_{\mu\nu}(x)\,\ee^{-\ii sD}=\ee^{4s}\big[T_{\mu\nu}+s\,A_{\mu\nu}\big](\ee^sx),\qquad
\ee^{\ii sD}\,V_\mu(x)\,\ee^{-\ii sD}=\ee^{3s}\big[V_\mu+s\,NV_\mu\big](\ee^sx),
\end{equation}
where $s$ is the logarithm of the scale factor and $N$ denotes the nilpotent part of the dilatation. We call $A_{\mu\nu}=NT_{\mu\nu}$ the logarithmic partner of $T_{\mu\nu}$, in the same sense as $\varphi_1$ is the partner of $\varphi_0$ in \eqref{eq:jordandil}. We assume that $A$ itself has a definite dimension, $NA=0$.\footnote{We display the case $N^2T=N^2V=0$. The argument for a longer finite chain is given in Astra's manuscript \cite{Astra:2026}.} What we want to show is that $A=0$ and $NV=0$.

Why is this not easy? The first thing one would try is the conservation law. The momentum $P_\mu=\int\dd^3x\,T_{0\mu}$ generates translations, and a dilatation turns a translation into a rescaled translation, so the momentum scales without a logarithm:
\[
\ee^{\ii sD}P_\mu\ee^{-\ii sD}=\ee^sP_\mu .
\]
Integrating \eqref{eq:chain} over space, we then find $\int\dd^3x\,A_{0\mu}=0$. However, as we saw in the Introduction, an improvement term satisfies this identically, so the conservation law does not exclude a partner. Nor do the two-point functions, since the generalized free field \eqref{eq:gffjordan} is local and unitary and yet has a logarithmic partner.

In our case, however, the field with a partner is the energy-momentum tensor itself, and with the energy-momentum tensor we can measure the energy flux of a state through a null plane. Let me outline the argument. If $T_{\mu\nu}$ had a partner, the state that $T_{vv}$ creates from a source in a bounded region would contain $\log M$ times the state of a scalar of dimension two, as we will see in Step 1. Now, in Lecture one we learned that the width of the energy flux is conjugate to the invariant mass, and that a singularity at $M=0$ gives a tail of large widths. A logarithm is such a singularity, so it would make the second moment of the energy flux infinite. On the other hand, for a state created in a bounded region the second moment must be finite. So there can be no partner.

To keep the argument as simple as possible, I make two assumptions in the main part of this lecture, both of which concern the conserved local operators of the theory.
\begin{itemize}
\item[(A)] Every local, symmetric and conserved tensor of dimension four has the form $c\,T_{\mu\nu}+(\partial_\mu\partial_\nu-\eta_{\mu\nu}\Box)X$ with a constant $c$ and a local scalar $X$ of dimension two.
\item[(B)] Every local conserved vector of dimension three whose charge $\int\dd^3x\,J_0$ vanishes is zero.
\end{itemize}
Assumption (A) says that $T_{\mu\nu}$ is, up to normalization, the only conserved spin-two operator of dimension four, and that the rest of any such tensor is the improvement term of a \emph{local} scalar. Note that we cannot exclude the improvement terms themselves, since every local scalar of dimension two gives one, and in the $\lambda\phi^4$ example they do mix with $T_{\mu\nu}$. We need assumption (B) for the virial current, whose partner could contain a conserved current of dimension three. It still allows several conserved currents, as long as no combination of them has zero charge. In the language of the local renormalization group \eqref{eq:localRG}, the main part of this lecture treats the mixing through $\hat\eta$. In Sec.~\ref{sec:noassume} I explain what changes without (A) and (B).

\subsection{Step 1: the partner is an improvement term}\label{sec:step1}
We first collect what we know about the partner $A_{\mu\nu}=NT_{\mu\nu}$. Since $T_{\mu\nu}$ is conserved at every scale, $A_{\mu\nu}$ is a local, symmetric and conserved tensor of dimension four, and as we have just seen, its charges vanish:
\begin{equation}\label{eq:Acharge}
\int\dd^3x\,A_{0\nu}=0 .
\end{equation}
By assumption (A) we can write $A_{\mu\nu}=c\,T_{\mu\nu}+(\partial_\mu\partial_\nu-\eta_{\mu\nu}\Box)X$. The improvement term does not contribute to the charges, because its $0\nu$ components are total spatial derivatives,
\[
(\partial_0\partial_0-\eta_{00}\Box)X=\nabla^2X,\qquad(\partial_0\partial_i-\eta_{0i}\Box)X=\partial_i\partial_0X ,
\]
so \eqref{eq:Acharge} gives $c\,P_\nu=0$, and hence $c=0$. The partner is thus a pure improvement term. The scalar $X$ itself has a definite dimension, which we will need below.\footnote{Differentiation commutes with $N$, so $NA_{\mu\nu}=(\partial_\mu\partial_\nu-\eta_{\mu\nu}\Box)NX$, and $NA=0$ says that the improvement term of $NX$ vanishes. Then $NX$ is a multiple of the identity operator. The identity has dimension zero, while $NX$ has dimension two, so $NX=0$. The scalar $X$ is Hermitian, since $T_{\mu\nu}$ and hence its partner are Hermitian.}

The same $X$ also determines the partner of $V$. Since $\Theta=\partial^\mu V_\mu$ holds at every scale, we can compare the terms linear in $s$ on both sides and find $A^\mu{}_\mu=\partial^\mu(NV)_\mu$. The trace of the improvement term is $\eta^{\mu\nu}(\partial_\mu\partial_\nu-\eta_{\mu\nu}\Box)X=(1-4)\Box X=-3\Box X$, so the vector
\[
J_\mu=NV_\mu+3\,\partial_\mu X
\]
is conserved. It is local and has dimension three. Moreover, its charge vanishes. To see this, note that the dilatation does not change its own charge, $\ee^{\ii sD}D\,\ee^{-\ii sD}=D$. So in $D=\int\dd^3x\,(x^\nu T_{0\nu}-V_0)$ the term linear in $s$ that \eqref{eq:chain} produces must vanish,
\[
0=\int\dd^3x\,(x^\nu A_{0\nu}-NV_0)=-\int\dd^3x\,(NV_0+3\,\partial_0X)=-\int\dd^3x\,J_0 ,
\]
where we used $A_{\mu\nu}=(\partial_\mu\partial_\nu-\eta_{\mu\nu}\Box)X$ and integrated by parts. By assumption (B), then, $J=0$. We have found
\begin{equation}\label{eq:NV}
\boxed{\ NT_{\mu\nu}=(\partial_\mu\partial_\nu-\eta_{\mu\nu}\Box)X,\qquad NV_\mu=-3\,\partial_\mu X\ }
\end{equation}
with a local scalar $X$ of dimension two. What remains to be shown is that $X=0$.

What does this mean for the states of Lecture one? Let $|X\rangle=X(p)|\Om\rangle$ be the state of $X$ in momentum space, and $|X(f)\rangle$ the state with the wave function $f$, prepared as in Sec.~\ref{sec:claim} with the test function $h=f(M)\,\zeta(p^+,\vec p)$. Recall that $\zeta$ describes how the state moves as a whole. We call the closed subspace spanned by the states $|X(f)\rangle$, for all $f$ and $\zeta$, the states of $X$.

Recall from \eqref{eq:cVcT} that $T_{vv}$ creates the state $(p^+)^2(\tfrac13|l\rangle+|s_{zz}\rangle)$. The partner $(\partial_\mu\partial_\nu-\eta_{\mu\nu}\Box)X$ has no spin-two part, so only the spin-zero part $|l\rangle$ has a partner. Since the partner of $\Theta$ is $-3\Box X$, the partner of $L$ is $-3X$, so a dilatation by $s$ shifts $|l\rangle$ by $-3s\,|X\rangle$. Now, a dilatation rescales $M$, and a function that is shifted when its argument is rescaled is a logarithm, as we saw for the pair \eqref{eq:gffjordan}. So we can write
\[
|l\rangle=|l_0\rangle-3\log(M/\mu_0)\,|X\rangle ,
\]
where $|l_0\rangle$ transforms under the dilatation like a state of dimension two, without a shift, and $\mu_0$ is an arbitrary reference mass. That is, a logarithmic partner of $T_{\mu\nu}$ means that the state of $L$ contains $\log M$ times the state of $X$. We now want to show that this is impossible.

\subsection{Step 2: a logarithm makes the mean squared width infinite}\label{sec:step2}
At first sight, the partner seems to drop out of an energy measurement. The partner of $T_{vv}$ is $\partial_v^2X$, since $\eta_{vv}=0$, and it integrates to zero along a complete light ray, so the light-ray operator $E(\vy)$ with which we measure the energy flux has no partner at all.

The way out is that we use $T_{vv}$ twice. We integrate it along complete light rays to measure the energy flux, and there the partner drops out. But we also smear it over a bounded region and use it as the source that creates the state, and there the partner survives. So let us create the state $|c_T\rangle$ of $T_{vv}$ with a smooth wave function $f_T$ with $f_T(0)\neq0$. We will see below that a generic source in a bounded region gives such a wave function. By the last equation of Step 1, the state $\tfrac13|l\rangle+|s_{zz}\rangle$ of $T_{vv}$ contains $-\log(M/\mu_0)\,|X\rangle$, so up to sign it contains the state $|X(\tilde f)\rangle$ with the wave function
\[
\tilde f(M)=\log(M/\mu_0)\,f_T(M) .
\]
That is, although the wave function $f_T$ of the source is smooth, the mixing gives the state of $X$ that the source creates a wave function with a logarithmic singularity at $M=0$.

Why is such a singularity dangerous? Recall what we found in Lecture one. For a scalar of dimension two in a CFT, the width is distributed as the radius $r$ in the Fourier transform of the wave function, and a wave function that is singular at $M=0$ has a Fourier transform with a tail at large $r$. For $L_2$ the singularity was the phase $\ee^{2\ii\vartheta}$, which gave the tail $1/r^2$. Now, in two dimensions the Fourier transform of $\log|\vec q|$ is also proportional to $1/r^2$ away from $r=0$. So the logarithm gives the same tail, and with it an infinite mean squared width.

We can see the same thing from the conformal value $\kappa_0[\tilde f]$ of the mean squared width. The derivative of $\tilde f$ is
\[
\partial_M\tilde f=\frac{f_T(M)}{M}+\log(M/\mu_0)\,\partial_Mf_T ,
\]
whose first term behaves as $f_T(0)/M$ near $M=0$, so that
\begin{equation}\label{eq:div}
\boxed{\ \int_0|\tilde f|^2M\,\dd M<\infty,\qquad\int_0|\partial_M\tilde f|^2M\,\dd M\ \supset\ |f_T(0)|^2\int_0\frac{\dd M}{M}=\infty\ }
\end{equation}
The norm is finite, but $\kappa_0[\tilde f]$ is infinite. So if the state of $T_{vv}$ really contains $|X(\tilde f)\rangle$, and if the states of $X$ behave as in a CFT, the second moment of the energy flux of this state is infinite.

On the other hand, the second moment of the energy flux of this state must be finite. More precisely, let $h(x)$ be a real smooth test function supported in a bounded region strictly on one side of the null plane, say $u<0$, and let $\psi=T_{vv}(h)|\Om\rangle$. Then
\begin{equation}\label{eq:Kfinite}
\langle\psi|K^+|\psi\rangle<\infty .
\end{equation}
The reason is causality. Light from a bounded source can reach the null plane at every transverse position $\vy$, but the farther away, the later. If we choose the origin of $v$ suitably, it arrives only at $v>c\,|\vy|^2$ with some constant $c>0$. If the energy density were a positive classical quantity, we would therefore have
\[
c\,K^+=\int\dd v\,\dd^2y\,c\,|\vy|^2\,T_{vv}\ \le\ \int\dd v\,\dd^2y\,v\,T_{vv}=B ,
\]
where $B$ is the boost, whose expectation value is finite. In Appendix~\ref{sec:entered} I explain how the thermal property of the vacuum turns this into an inequality of the quantum theory. Note that no scaling dimension enters this argument, so \eqref{eq:Kfinite} holds for the energy-momentum tensor even if it has a logarithmic partner.

Since $K^+=P^+(|Y|^2+\kappa)\ge P^+\kappa$, it follows from \eqref{eq:Kfinite} that the mean squared width is finite on the part of the state with $p^+$ bounded away from zero. This applies to the state $|c_T\rangle$, which $T_{vv}(h)$ creates. After we absorb the factor $(p^+)^2$ into $\zeta$, its wave function $f_T$ is the $M$-dependence of the Fourier transform of $h$ at fixed $p^+$ and $\vec p$. It is smooth, and $f_T(0)$, the Fourier transform of $h$ at a null momentum, does not vanish for a generic $h$. The divergence \eqref{eq:div} comes from small $M$ at fixed $p^+>0$, so it contradicts \eqref{eq:Kfinite}.\footnote{In the generalized free field \eqref{eq:gffjordan} the same logarithm appears, but there is no energy-momentum tensor, so we cannot define $K^+$ and the argument does not apply.} We would therefore like to conclude that $|X\rangle=0$.

To be honest, there are still two gaps in this argument. First, we computed the mean squared width of $|X(\tilde f)\rangle$ as in a CFT. In a theory that is only scale invariant, the mean-value form \eqref{eq:claimineq} of the comparison says only that the true value is not larger than the conformal one, and an upper bound that is infinite is of no use. What we need is that on the states of $X$ the mean squared width is equal to its conformal value. Second, the other components of the state of $T_{vv}$ might cancel the divergence, just as the tail of $|s_{zz}\rangle$ cancelled that of $|l\rangle$ in Lecture one. We close both gaps in Step 3.

\subsection{Step 3: moving the null plane}\label{sec:step3}
We begin with the first gap. We want to show that in every state $|X(f)\rangle$ the mean squared width is equal to its conformal value:
\begin{equation}\label{eq:equal}
\boxed{\ \langle\kappa\rangle\ =\ \kappa_0[f]\ =\ \frac{\int_0^\infty|\partial_Mf|^2M\,\dd M}{\int_0^\infty|f|^2M\,\dd M}\ }
\end{equation}
Since $X$ has a definite dimension, the mean-value form \eqref{eq:claimineq} of the comparison already gives $\langle\kappa\rangle\le\kappa_0[f]$. We call the difference between the conformal and the true quadratic forms $U[f]$,
\[
\langle\kappa\rangle\,\|f\|^2=\int_0^\infty|\partial_Mf|^2M\,\dd M-U[f],\qquad U[f]\ge0,\qquad \|f\|^2=\int_0^\infty|f|^2M\,\dd M ,
\]
so that $U$ is a non-negative quadratic form in $f$, and what we have to show is $U=0$.

How can we learn anything about $U$? In a theory that is only scale invariant, $K^+$ is not a symmetry generator, so we cannot compute its matrix elements. But we can compute how $K^+$ changes when we move the null plane on which we measure it, and it changes as in a CFT. So the deficit $U$ is not affected by moving the plane, and we will see that a deficit with this property must vanish.

So let us move the null plane. The following computation does not use the Jordan block, so we can use it again in Lecture three. We work at fixed $p^+$ and $\vec p$ throughout this step, and we write
\[
K^+(u_0)=\int_{u=u_0}\dd v\,\dd^2y\,|\vy|^2\,T_{vv}
\]
for $K^+$ evaluated on the plane $u=u_0$. In the following computation every integral is over this plane with the measure $\dd v\,\dd^2y$. In light-cone coordinates the conservation law $\partial^\mu T_{\mu\nu}=0$ reads
\[
\partial_uT_{vv}=\partial_iT_{iv}-\partial_vT_{uv},\qquad
\partial_uT_{vi}=\partial_jT_{ji}-\partial_vT_{ui} .
\]
We differentiate $K^+(u_0)$ and use the first equation. The $\partial_v$ term integrates to zero along each light ray, and integrating the $\partial_i$ term by parts with $\partial_i|\vy|^2=2y^i$, we get
\begin{equation}\label{eq:Kp}
\frac{\dd K^+}{\dd u_0}=\int|\vy|^2\big(\partial_iT_{iv}-\partial_vT_{uv}\big)=-2\int y^iT_{iv} .
\end{equation}
If we differentiate once more and use the second equation in the same way, we obtain $2\int T_{ii}$. With the trace $\Theta=2T_{uv}-T_{ii}$ and $\int T_{uv}=P^-$, this is
\[
\frac{\dd^2K^+}{\dd u_0^2}=2\int T_{ii}=4P^--2\int\Theta .
\]

In a CFT the last term is absent. In our case we need the integral of $\Theta$ over the null plane between states created by $X$. Here we use one more input. Inside these matrix elements we may write $\Theta=\Box L=\partial_v(2\partial_uL)-\partial_i^2L$, as I explain in Appendix~\ref{sec:insert}. Both terms integrate to zero over the null plane, so we find
\begin{equation}\label{eq:Kpp}
\frac{\dd^2K^+}{\dd u_0^2}=4P^- ,
\end{equation}
just as in a CFT.

Now we come back to $U$. Moving the plane by $u_0$ is the same as translating the state by $u_0$ in the $u$ direction, which multiplies it by $\ee^{\ii p^-u_0}$. Since $p^-=(M^2+|\vec p|^2)/2p^+$, the translation multiplies the wave function, up to a constant phase that we absorb into $\zeta$, by a phase that depends on $M$:
\[
f(M)\ \to\ f_{u_0}(M)=\ee^{\ii u_0M^2/2p^+}f(M) .
\]
By \eqref{eq:Kpp}, the true $\langle K^+\rangle$ in the translated state is a quadratic function of $u_0$ with the second derivative $4\langle P^-\rangle$, and the same holds for the value that $\langle K^+\rangle$ would have in a CFT. The centroid contributes equally to both, since the $P^+Y^i$ are generators of the Poincar\'e group. So the difference of the two, which is $p^+U[f_{u_0}]$, is a linear function of $u_0$. But a linear function that is non-negative for all real $u_0$ must be a constant, and therefore
\[
U[f_{u_0}]=U[f]\qquad\text{for every }u_0 .
\]

This is a strong constraint. The phases $\ee^{\ii u_0M^2/2p^+}$ with all real $u_0$ are the Fourier modes in the variable $M^2$. A quadratic form that does not change when we multiply $f$ by any of them cannot depend on how $f$ varies with $M$. It can only depend on $|f(M)|^2$ at each $M$, that is,
\[
U[f]=\int_0^\infty|f(M)|^2\,\sigma(M)\,\dd M
\]
with some weight $\sigma(M)\ge0$.

Now let us compare this with the conformal quadratic form $\int_0^\infty|\partial_Mf|^2M\,\dd M$. It bounds $U[f]$ from above because $\kappa\ge0$, and it depends only on the derivative of $f$. On a two-dimensional plane, such as our auxiliary plane, this form has a remarkable property. A function can be equal to one on any bounded range of $M$ and still make the form arbitrarily small, provided that it falls to zero slowly enough in $\log M$.\footnote{This is the property behind a well-known fact of quantum mechanics: in two dimensions an attractive potential has a bound state however shallow it is. The form is the kinetic energy of a rotationally symmetric wave function in a plane with the radius $M$. Since a wave function such as $\chi_H$ makes it arbitrarily small, the potential energy always wins.} For example, take
\[
\chi_H(M)=\begin{cases}1 & (M\le M_1)\\ 1-\log(M/M_1)/H & (M_1\le M\le M_1\ee^H)\\ 0 & (M\ge M_1\ee^H)\end{cases}
\]
with a large number $H$. In the variable $x=\log M$ the form is $\int|\partial_x\chi_H|^2\dd x$, and $\chi_H$ falls linearly from one to zero over the length $H$, so
\[
\int_0^\infty|\partial_M\chi_H|^2M\,\dd M=\frac1H .
\]
Therefore
\[
\int_0^{M_1}\sigma(M)\,\dd M\ \le\ U[\chi_H]\ \le\ \frac1H .
\]
Letting $H\to\infty$, we find that $\sigma$ vanishes below $M_1$, and since $M_1$ was arbitrary, $U=0$. This proves \eqref{eq:equal}, that is, on the states of $X$ the mean squared width is the conformal one.

Now we turn to the second gap. We split the state $|c_T(f_T)\rangle=\tfrac13|l(f_T)\rangle+|s_{zz}(f_T)\rangle$ of $T_{vv}$ into its component $|\psi_X\rangle$ along the states of $X$ and the rest $|\psi_\perp\rangle$, which is orthogonal to them. The probability of a width larger than $\sqrt t$ in $|c_T\rangle$ is proportional to
\[
\big\|\mathcal P_t|\psi_X\rangle+\mathcal P_t|\psi_\perp\rangle\big\|^2 ,
\]
and the two vectors could interfere destructively. Such a cancellation did happen in Lecture one, where $\mathcal P_t|c_T\rangle=0$ at large $t$ although $\mathcal P_t|l\rangle$ and $\mathcal P_t|s_{zz}\rangle$ did not vanish.

Here this cannot happen, because on the states of $X$ a measurement of $\kappa$ gives what it gives in a CFT. Indeed, the true probability of $\kappa>t$ is bounded by the conformal one at every $t$, while by \eqref{eq:equal} the two mean values agree. So the two probabilities must agree at every $t$. But then the projection $\mathcal P_t=\mathbf 1(\kappa>t)$ must act on the states of $X$ as it does in the CFT, where it removes the part of $\hat f$ inside the disk $r^2\le t$. In particular, it maps the states of $X$ to states of $X$ \cite{Astra:2026}.

Since $\mathcal P_t$ is self-adjoint, it also maps the states orthogonal to those of $X$ to states orthogonal to them. The two vectors $\mathcal P_t|\psi_X\rangle$ and $\mathcal P_t|\psi_\perp\rangle$ therefore remain orthogonal, and
\[
\|\mathcal P_t|c_T\rangle\|^2=\|\mathcal P_t|\psi_X\rangle\|^2+\|\mathcal P_t|\psi_\perp\rangle\|^2\ \ge\ \|\mathcal P_t|\psi_X\rangle\|^2 .
\]
The probabilities of large widths add, so the rest of the state cannot cancel the contribution of the states of $X$ at any $t$. Integrating over $t$, we obtain $\langle c_T|\kappa|c_T\rangle\ge\langle\psi_X|\kappa|\psi_X\rangle$.

We can now complete the argument. With $|l\rangle=|l_0\rangle-3\log(M/\mu_0)\,|X\rangle$, the component $|\psi_X\rangle$ consists of $-|X(\tilde f)\rangle$ and the part of $\tfrac13|l_0(f_T)\rangle$ along the states of $X$, while $|s_{zz}(f_T)\rangle$, which has spin two, lies in $|\psi_\perp\rangle$. Up to sign, the wave function of $|\psi_X\rangle$ is therefore $\log(M/\mu_0)\,f_T(M)$ plus a constant multiple of $f_T(M)$. The multiple is a constant because $|l_0\rangle$ and $|X\rangle$ both scale as states of dimension two without a logarithm, so the ratio of their wave functions cannot depend on $M$. The multiple of $f_T$ is smooth and cannot cancel the term $f_T(0)/M$ in the derivative.

So suppose that $|X\rangle\neq0$. Then by \eqref{eq:div} and \eqref{eq:equal} the mean squared width of the state of $T_{vv}$ diverges at $p^+$ bounded away from zero, whereas by \eqref{eq:Kfinite} it is finite. We must therefore have $|X\rangle=0$, that is, the local field $X$ annihilates the vacuum. By the Reeh--Schlieder theorem a local field that annihilates the vacuum vanishes, so $X=0$, and by \eqref{eq:NV} we conclude
\[
NT_{\mu\nu}=0,\qquad NV_\mu=0 .
\]

\subsection{Without the two assumptions}\label{sec:noassume}
What changes if we drop the assumptions (A) and (B)? The ideas of Steps 2 and 3 stay the same, but Astra's manuscript \cite{Astra:2026} needs two more steps. Here I only indicate what they are.

Without (A), the partner of $T_{\mu\nu}$ need not be an improvement term, and it may contain a spin-two part. Astra's manuscript first shows that the partner carries no null energy flux. The manuscript derives this from the Ward--Takahashi identity of the boost on the null plane, but there is a short way to see why it must be true. Under a dilatation by $s$, the null energy $E_{\mathcal R}=\int_{\mathcal R}\dd^2y\,E(\vy)$ through a region $\mathcal R$ of the null plane acquires $s$ times the corresponding flux of the partner. But $0\le E_{\mathcal R}\le P^+$, and $P^+$ scales without a logarithm, so this flux must vanish. For an improvement term this holds automatically, but for a spin-two part it is a non-trivial condition, and together with the translation Ward--Takahashi identity it excludes the spin-two part.

Without (B), the partner of $V_\mu$ may contain a conserved current of zero charge. We used (B) in Step 1 to obtain $NV_\mu=-3\,\partial_\mu X$. This relation entered once more in Step 3, in the single insertion of $L$ behind \eqref{eq:Kpp}, where Appendix~\ref{sec:insert} uses it to show that the spin-two component $|s_{zz}\rangle$ has no tail. Without (B), Astra's manuscript shows the latter by a longer computation, and then excludes the current by applying the argument of Steps 2 and 3 to the state created by $V_v$ \cite{Astra:2026}.

\subsection{Summary of Lecture two}
Let me summarize. Under our two assumptions, a logarithmic partner of the energy-momentum tensor is the improvement term of a scalar $X$ of dimension two. It integrates to zero along every light ray, so we do not see it in the light-ray operator $E(\vy)$. But we see it in the state that $T_{vv}$ creates from a source in a bounded region, as a component $\log M$ times a state of $X$, and a logarithm at $M=0$ makes the second moment of the energy flux infinite. Since causality requires this second moment to be finite, there is no partner. The hard step was to show that on the states of $X$ the width is distributed as in a CFT, and not only bounded by it. For this we moved the null plane. As in Lecture one, we did not use the two-point functions, but the energy flux, which is given by three-point functions with one insertion of $T_{\mu\nu}$.

\section{Lecture three: Modular theory and complex scaling dimensions}
In this last lecture we turn to the remaining fields. In Lecture two we saw that the energy-momentum tensor and the virial current have definite scaling dimensions, under the assumptions (A) and (B), and without them in the argument of Astra's manuscript \cite{Astra:2026} that I sketched. Other fields, however, may still sit in a Jordan block or have a complex dimension, and from the generalized free fields of the Introduction we know that locality and unitarity alone cannot exclude this. What I would like to show in this lecture is that no field has a logarithmic partner, and that the imaginary part of every scaling dimension is the charge of an internal symmetry.

Our tool will be the fact that the vacuum, restricted to certain regions, is a thermal state. In Lectures one and two we relied on this fact only through results that I quoted without proof (I explain them in Appendix~\ref{sec:entered}), but in this lecture we use it directly. We will see that a dilatation by the imaginary amount $\ii\pi$ is nothing but a reflection through the origin.

\subsection{The vacuum is thermal}\label{sec:thermal}
Let us start with the Unruh effect \cite{Unruh:1976db}, probably the best known example of a thermal vacuum. An observer with constant acceleration has access only to a Rindler wedge $W$, and this observer sees the vacuum as a thermal state with respect to the boost that preserves $W$. Let $B$ be the generator of this boost. A boost by the imaginary angle $\ii\pi$ maps the wedge to the opposite wedge, and the thermal property can be written as
\[
\ee^{-\pi B}A|\Om\rangle=J\,A^\dagger|\Om\rangle
\]
for every operator $A$ localized in $W$. Here $J$ is the antiunitary operator that reflects $x^0$ and $x^3$, that is, the product of PCT and the rotation by $\pi$ in the $(x^1,x^2)$ plane, and it exchanges the two wedges.

This is the Bisognano--Wichmann theorem \cite{Bisognano:1976za}. It is equivalent to the Kubo--Martin--Schwinger condition, the periodicity in imaginary time that characterizes thermal equilibrium, here in imaginary boost time and at the temperature $1/2\pi$ \cite{Haag:1967sg}. In the form above, the theorem says that a boost by half of the imaginary period turns the state created by $A$ into the reflected state created by $A^\dagger$. In particular, the boosted state has the norm of the state created by $A^\dagger$,
\[
\|\ee^{-\pi B}A|\Om\rangle\|=\|A^\dagger|\Om\rangle\| .
\]

The same structure exists for other regions. For a region $\mathcal U$ with its algebra of observables $\mathcal M(\mathcal U)$, there is a positive operator $\Delta_{\mathcal U}$ and an antiunitary $J_{\mathcal U}$ such that $\Delta_{\mathcal U}^{1/2}A|\Om\rangle=J_{\mathcal U}A^\dagger|\Om\rangle$ for all $A$ in $\mathcal M(\mathcal U)$, and hence
\begin{equation}\label{eq:tomita}
\boxed{\ \|\Delta_{\mathcal U}^{1/2}A|\Om\rangle\|=\|A^\dagger|\Om\rangle\|\qquad(A\in\mathcal M(\mathcal U))\ }
\end{equation}
This is Tomita--Takesaki theory \cite{Takesaki:1970,Haag:1992hx,Borchers:2000pv,Witten:2018zxz}, but in these lectures we need only \eqref{eq:tomita} and a few facts that I state as we go. We call $K_{\mathcal U}=-\log\Delta_{\mathcal U}$ the \emph{modular Hamiltonian} of the region, and we say that the vacuum restricted to $\mathcal U$ is thermal with respect to it. (The symbol $\Delta_{\mathcal U}$ is standard, but it has nothing to do with a scaling dimension.)

Why does \eqref{eq:tomita} express thermal equilibrium? To see this, let us look at a quantum system with finitely many states and Hamiltonian $H$, in the thermal state $\rho=\ee^{-H}/Z$, where we measure energies in units of the temperature. We can write $\rho$ as a pure state of two copies of the system,
\[
|\Om\rangle=\frac{1}{\sqrt Z}\sum_n\ee^{-E_n/2}\,|n\rangle\otimes|n\rangle ,
\]
and let the operators $A$ act on the first copy. Then $\|A^\dagger|\Om\rangle\|^2=\mathrm{Tr}(\rho AA^\dagger)$. Now take $\Delta=\rho\otimes\rho^{-1}$. Since $\rho^{-1}|n\rangle=Z\ee^{E_n}|n\rangle$, we find
\[
\|\Delta^{1/2}A|\Om\rangle\|^2=\frac1Z\sum_n\ee^{-E_n}\,\langle n|A^\dagger\rho A|n\rangle\,Z\ee^{E_n}=\mathrm{Tr}(A^\dagger\rho A)=\|A^\dagger|\Om\rangle\|^2 ,
\]
which is \eqref{eq:tomita}. The modular Hamiltonian is
\[
K=-\log\Delta=H\otimes1-1\otimes H ,
\]
the Hamiltonian of the first copy minus that of the second, in units of the temperature. In this language the Bisognano--Wichmann theorem reads $K_W=2\pi B$. The wedge plays the role of the first copy and the opposite wedge that of the second, and the boost $B$ plays the role of the Hamiltonian at the temperature $1/2\pi$.

In general $K_{\mathcal U}$ is not the integral of a local density, and we know it explicitly only in special cases. One of them is a light cone in a CFT. A conformal transformation maps the wedge to the future light cone of a point and the boost to the dilatation about that point, so the modular Hamiltonian of the future cone is $2\pi D$ \cite{Hislop:1981uh,Brunetti:1992zf}. That is, in a CFT the vacuum inside a light cone is thermal with respect to the dilatation. In a theory that is only scale invariant we cannot use this argument, so we will have to find out to what extent the statement still holds.

\subsection{If the future cone were thermal}\label{sec:ideal}
Before we deal with this, let us see what we could conclude if the vacuum in the future light cone were thermal with respect to $D$, as in a CFT. We compute the action of the imaginary dilatation $\ee^{-\pi D}$ on a state in two ways. On the one hand, by the thermal property \eqref{eq:tomita} it turns the state created by a field into a state with the norm of the state created by the adjoint field, without any extra factor. On the other hand, we know how the dilatation acts on a field: a dilatation by $s$ maps $x$ to $\ee^sx$, so a dilatation by $\ii\pi$ maps $x$ to $-x$, and it multiplies a field of dimension $\Delta=\alpha+\ii\beta$ by $\ee^{\ii\pi\Delta}$, whose modulus is $\ee^{-\pi\beta}$. The two are compatible only if $\beta=0$. A logarithmic partner will be excluded in the same way, because $\log x$ becomes $\log x+\ii\pi$ under the reflection.

Let us now carry this out. We consider a general field, which may sit in a Jordan block and have a complex dimension $\Delta=\alpha+\ii\beta$. The fields of a Jordan block span a vector space, and we write a general element as $O_a$ with a vector $a$ in this space. For the Jordan pair $(\varphi_0,\varphi_1)$ of the generalized free field \eqref{eq:gffjordan}, for example, $O_a=a_0\varphi_0+a_1\varphi_1$. The dilatation acts on the vector by $\ee^{sN}$,
\begin{equation}\label{eq:lawa}
\ee^{\ii sD}\,O_a(x)\,\ee^{-\ii sD}=\ee^{s\Delta}\,O_{\ee^{sN}a}(\ee^sx),
\end{equation}
and for the pair $N(a_0,a_1)=(0,a_0)$. We take a real test function $h(x)$ supported in the future cone, and let $|\Psi_a\rangle=O_a(h)|\Om\rangle$. For these states we will use the following three relations.

The first is the Tomita relation \eqref{eq:tomita} for the future cone. If the future cone is thermal with respect to $D$, that is, if its modular Hamiltonian is $2\pi D$, the operator $\Delta^{1/2}$ in \eqref{eq:tomita} is $\ee^{-\pi D}$, and \eqref{eq:tomita} reads
\[
\|\ee^{-\pi D}|\Psi_a\rangle\|=\|O_a(h)^\dagger|\Om\rangle\| .
\]

The second relation follows from PCT. Since PCT maps $O(x)$ to $O^\dagger(-x)$ and leaves the vacuum invariant, the state created by $O_a^\dagger$ with the test function $h$ has the same norm as the state created by $O_a$ with the reflected test function $h^-(x)=h(-x)$,
\[
\|O_a(h)^\dagger|\Om\rangle\|=\|O_a(h^-)|\Om\rangle\| .
\]

The third relation is the dilatation law \eqref{eq:lawa} itself, continued to the imaginary value $s=\ii\pi$. Since $D$ annihilates the vacuum, \eqref{eq:lawa} gives
\[
\ee^{\ii sD}\,O_a(x)|\Om\rangle=\ee^{s\Delta}\,O_{\ee^{sN}a}(\ee^sx)|\Om\rangle .
\]
We can continue both sides to $s=\ii\theta$ with $0<\theta<\pi$, and this is why we took $h$ supported in the future cone. Indeed, for a point $x$ in the future cone, the complex point $\ee^{\ii\theta}x$ has its imaginary part in the forward cone, where the state $O(z)|\Om\rangle$ is well defined because the energy is positive.\footnote{Explicitly, $O(z)|\Om\rangle=\ee^{\ii P\cdot z}O(0)|\Om\rangle$ with $|\ee^{\ii P\cdot z}|=\ee^{-P\cdot{\rm Im}\,z}$, and ${\rm Im}\,z=\sin\theta\,x$ for $z=\ee^{\ii\theta}x$. The factor $\ee^{s\Delta}$ and the label $\ee^{sN}a$, which is a polynomial in $s$ since $N$ is nilpotent, cause no problem.}

At the endpoint $\theta=\pi$ we have $\ee^{\ii\pi}x=-x$, as promised. If we smear with $h$, the field at $-x$ becomes the field smeared with the reflected test function $h^-$, the label becomes $\ee^{\ii\pi N}a$, and we obtain the factor $\ee^{\ii\pi\Delta}$,
\[
\ee^{-\pi D}\,O_a(h)|\Om\rangle=\ee^{\ii\pi\Delta}\,O_{\ee^{\ii\pi N}a}(h^-)|\Om\rangle .
\]
Here the factor $\ee^{\ii\pi N}$ is the shift of the logarithm. For the pair of \eqref{eq:gffjordan}, the continuation turns $\varphi_0$ into $\varphi_0+\ii\pi\varphi_1$. To compensate for this shift, let us start instead from the label $\Sigma a$ with $\Sigma=\ee^{-\ii\pi N}$, which for the pair is the field $\varphi_0-\ii\pi\varphi_1$. Then the label after the continuation is $\ee^{\ii\pi N}\Sigma a=a$, and we have
\begin{equation}\label{eq:strip}
\ee^{-\pi D}|\Psi_{\Sigma a}\rangle=\ee^{\ii\pi\Delta}\,O_a(h^-)|\Om\rangle,\qquad|\ee^{\ii\pi\Delta}|=\ee^{-\pi\beta} .
\end{equation}
Note that for this relation we need neither the energy-momentum tensor nor the thermal property, so it holds in the generalized free field examples as well.

Now we combine the three relations. With the first two we compute the norm of $\ee^{-\pi D}|\Psi_a\rangle$, and with the third the norm of $\ee^{-\pi D}|\Psi_{\Sigma a}\rangle$,
\begin{gather*}
\big\|\ee^{-\pi D}|\Psi_a\rangle\big\|\ \overset{\rm Tomita}{=}\ \big\|O_a(h)^\dagger|\Om\rangle\big\|\ \overset{\rm PCT}{=}\ \big\|O_a(h^-)|\Om\rangle\big\|,\\
\big\|\ee^{-\pi D}|\Psi_{\Sigma a}\rangle\big\|\ \overset{\eqref{eq:strip}}{=}\ \ee^{-\pi\beta}\,\big\|O_a(h^-)|\Om\rangle\big\| .
\end{gather*}
The same norm $\|O_a(h^-)|\Om\rangle\|$ appears in both lines. Therefore
\begin{equation}\label{eq:ideal}
\boxed{\ \big\|\ee^{-\pi D}|\Psi_{\Sigma a}\rangle\big\|=\ee^{-\pi\beta}\,\big\|\ee^{-\pi D}|\Psi_a\rangle\big\|\ }
\end{equation}
We have reached the state of the reflected field in two ways, once with the thermal property and PCT, and once by continuing the dilatation. The two results agree only if replacing $a$ by $\Sigma a$ multiplies the norm by $\ee^{-\pi\beta}$. This is the most important relation of this lecture. In the rest of the lecture we ask how much of it remains true in a theory that is only scale invariant.

What does \eqref{eq:ideal} tell us? If the dilatation is diagonal, $N=0$ and $\Sigma=1$, and \eqref{eq:ideal} gives $\beta=0$ at once. With a Jordan block, let us apply \eqref{eq:ideal} repeatedly, to the labels $\Sigma^ka$ with all integers $k$. Each step multiplies the norm by the same factor, so
\[
\big\|\ee^{-\pi D}|\Psi_{\Sigma^ka}\rangle\big\|^2=\ee^{-2\pi\beta k}\,\big\|\ee^{-\pi D}|\Psi_a\rangle\big\|^2 ,
\]
that is, the squared norms form a geometric sequence in $k$.

On the other hand, we can compute these norms directly. Every Jordan block contains a label $b\neq0$ with $Nb=0$, and if $N$ does not vanish on the block, we can choose $b$ at the end of a chain, so that there is a label $a$ with
\begin{equation}\label{eq:poly}
Na=b,\qquad Nb=0,\qquad b\neq0 ,
\end{equation}
as for the Jordan pair of \eqref{eq:gffjordan}, where $a$ and $b$ correspond to $\varphi_0$ and its partner $\varphi_1$. Then $\Sigma^ka=a-\ii\pi k\,b$, that is, each $\Sigma$ subtracts $\ii\pi$ times the partner. Writing $|\tilde\Psi_a\rangle=\ee^{-\pi D}|\Psi_a\rangle$ and $|\tilde\Psi_b\rangle=\ee^{-\pi D}|\Psi_b\rangle$, we find
\[
\big\|\ee^{-\pi D}|\Psi_{\Sigma^ka}\rangle\big\|^2=\big\||\tilde\Psi_a\rangle-\ii\pi k|\tilde\Psi_b\rangle\big\|^2=\langle\tilde\Psi_a|\tilde\Psi_a\rangle+2\pi k\,\mathrm{Im}\langle\tilde\Psi_a|\tilde\Psi_b\rangle+\pi^2k^2\langle\tilde\Psi_b|\tilde\Psi_b\rangle ,
\]
a quadratic polynomial in $k$. But a polynomial in $k$ can be a geometric sequence only if it is a constant, so $\langle\tilde\Psi_b|\tilde\Psi_b\rangle=0$, and since $\ee^{-\pi D}$ has no kernel, $O_b(h)|\Om\rangle=0$. This holds for every real test function $h$ supported in the future cone, which is impossible for a non-zero field $O_b$. Hence $N=0$ on every field, and then $\beta=0$, as we have just seen.

We conclude that if the future cone is thermal with respect to $D$, every field has a real dimension and no logarithmic partner. In the generalized free fields of the Introduction, the field \eqref{eq:gffc} with a complex dimension and the Jordan pair \eqref{eq:gffjordan}, this conclusion fails, so in those theories the future cone is not thermal with respect to $D$.

There is also a simple example that shows what we should expect. Take a CFT with a flavor symmetry of charge $F$, and redefine its conformal dilatation $D_{\rm c}$ by this charge, $D=D_{\rm c}+F$. Then a field of charge $e_F$ has the complex dimension $\Delta_{\rm c}+\ii e_F$. Of course, the cone is thermal with respect to $D_{\rm c}$, not with respect to $D$, and the two differ by the charge $F$. The most we can expect is therefore thermality up to a charge.

\subsection{Moving the edge of a wedge}\label{sec:edge}
In Sec.~\ref{sec:ideal} we used that the modular Hamiltonian of the future light cone is $2\pi D$. This is true in a CFT, but in a theory that is only scale invariant we do not know the modular Hamiltonian of the future cone. What we do know, in any quantum field theory, is the modular Hamiltonian of the wedge, $2\pi B$. So let us start from the wedge and deform it toward the cone, keeping track of how the modular Hamiltonian changes. Fortunately, in every theory with an energy-momentum tensor this is possible for one family of deformations, in which we move the edge of the wedge along a null plane.

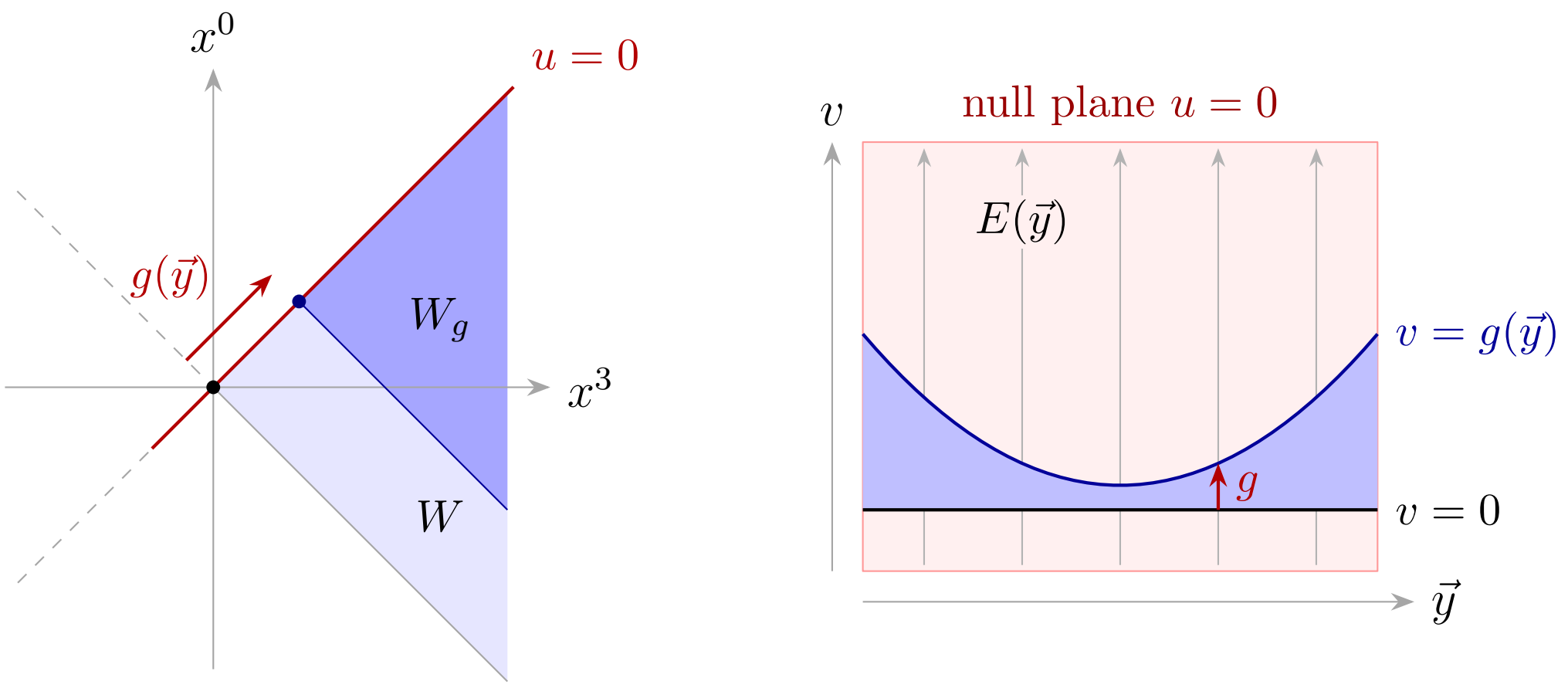
\begin{figure}[htb]
\centering
\begin{tikzpicture}[scale=1.05,line join=round]
\fill[blue!10] (0,0)--(2.4,2.4)--(2.4,-2.4)--cycle;
\fill[blue!35] (0.7,0.7)--(2.4,2.4)--(2.4,-1.0)--cycle;
\draw[axisline] (-1.7,0)--(2.75,0) node[right,black] {$x^3$};
\draw[axisline] (0,-2.3)--(0,2.6) node[above,black] {$x^0$};
\draw[gray!70,dashed] (-1.6,1.6)--(0,0)--(-1.6,-1.6);
\draw[gray!70] (0,0)--(2.4,-2.4);
\draw[red!70!black,thick] (-0.5,-0.5)--(2.45,2.45) node[above right,font=\small] {$u=0$};
\draw[blue!60!black] (0.7,0.7)--(2.4,-1.0);
\fill (0,0) circle (1.6pt);
\fill[blue!50!black] (0.7,0.7) circle (1.6pt);
\draw[-{Stealth[length=2mm]},thick,red!70!black] (-0.22,0.22)--(0.48,0.92) node[midway,above left,font=\small] {$g(\vy)$};
\node[font=\small] at (1.85,-1.05) {$W$};
\node[font=\small] at (1.85,0.55) {$W_g$};
\begin{scope}[xshift=7.4cm,yshift=-1.0cm]
\fill[red!6] (-2.1,-0.5) rectangle (2.1,3.0);
\draw[red!40] (-2.1,-0.5) rectangle (2.1,3.0);
\node[red!60!black,font=\small,above] at (0,3.0) {null plane $u=0$};
\foreach \yy in {-1.6,-0.8,0,0.8,1.6} {\draw[gray!60,-{Stealth[length=1.6mm]}] (\yy,-0.45)--(\yy,2.95);}
\fill[blue!25] plot[domain=-2.1:2.1,samples=40] (\x,{0.2+0.28*\x*\x}) -- (2.1,0) -- (-2.1,0) -- cycle;
\draw[thick] (-2.1,0)--(2.1,0) node[right,font=\small] {$v=0$};
\draw[thick,blue!60!black] plot[domain=-2.1:2.1,samples=40] (\x,{0.2+0.28*\x*\x}) node[right,font=\small] {$v=g(\vy)$};
\draw[-{Stealth[length=2mm]},thick,red!70!black] (0.8,0)--(0.8,{0.2+0.28*0.64}) node[midway,right,font=\small] {$g$};
\node[gray!50!black,font=\small] at (-1.6,3.25) {};
\node[font=\small,fill=red!6,inner sep=1pt] at (-0.8,2.35) {$E(\vy)$};
\draw[axisline] (-2.1,-0.75)--(2.4,-0.75) node[right,black] {$\vy$};
\draw[axisline] (-2.35,-0.5)--(-2.35,3.0) node[above,black] {$v$};
\end{scope}
\end{tikzpicture}
\caption{Moving the edge of a wedge. The left panel shows the wedge $W$ (light) and the wedge $W_g$ (dark) in the $(x^3,x^0)$ plane at a fixed transverse position $\vy$. We move the edge along the null plane $u=0$ by $g(\vy)$. The right panel shows the null plane $u=0$ seen from the front, with the old edge $v=0$ and the new edge $v=g(\vy)$. The vertical lines are the light rays along which $E(\vy)$ integrates $T_{vv}$. In \eqref{eq:Hg} we weight the null energy on each light ray with the displacement $g(\vy)$. For a non-constant $g$ the left panel is schematic, since the boundary of $W_g$ at fixed $\vy$ also depends on $g$ at neighboring positions.}\label{fig:edge}
\end{figure}

To be concrete, let $W$ be the wedge $u<0<v$, whose future boundary lies on the null plane $u=0$ and whose edge is at $v=0$. We move the edge along this plane to $v=g(\vy)$, by an amount that may depend on the transverse position, and call the resulting region $W_g$. More precisely, $W_g$ consists of the points on the side $u<0$ from which no light signal can reach the part $v<g(\vy)$ of the null plane. For $g\ge0$ it is smaller than $W$. Then, for a smooth bounded function $g$, the modular Hamiltonian changes as
\begin{equation}\label{eq:Hg}
\boxed{\ K_W-K_{W_g}=2\pi H_g,\qquad H_g=\int\dd^2y\,g(\vy)\,E(\vy)\ }
\end{equation}
where $E(\vy)=\int\dd v\,T_{vv}(u=0,v,\vy)$ is the light-ray operator of Lecture one \cite{Wiesbrock:1992mg,Faulkner:2016mzt,Casini:2017roe}. Astra's manuscript shows that this formula remains valid when $T_{\mu\nu}$ sits in a Jordan block \cite[Proposition~4.1]{Astra:2026}, which we needed in Lecture two (Appendix~\ref{sec:entered}). In words, the modular Hamiltonian changes by the null energy on each light ray, weighted with the distance $g(\vy)$ by which we moved the edge on that light ray (Fig.~\ref{fig:edge}).

To get a feeling for \eqref{eq:Hg}, let us check it in the simplest case, a constant $g=v_0>0$. Then $W_g$ is just the wedge $W$ translated by $v_0$ in the $v$ direction, so its modular Hamiltonian is $2\pi$ times the boost about the translated edge. On the null plane the boost is $B=\int\dd v\,\dd^2y\,v\,T_{vv}$, and the boost about $v=v_0$ is
\[
\int\dd v\,\dd^2y\,(v-v_0)\,T_{vv}=B-v_0P^+ ,
\]
which agrees with \eqref{eq:Hg} because $\int\dd^2y\,E(\vy)=P^+$. For a general $g$ we can write \eqref{eq:Hg} in the same form,
\[
K_{W_g}=2\pi\int\dd v\,\dd^2y\,\big(v-g(\vy)\big)\,T_{vv}(u=0,v,\vy) ,
\]
that is, on each light ray the modular Hamiltonian is the boost about the point where the light ray crosses the new edge. This is what one would naively guess. The non-trivial content of \eqref{eq:Hg} is that the guess is correct light ray by light ray, even when the edge is not straight.

Incidentally, \eqref{eq:Hg} is also how one proves the averaged null energy condition. Modular Hamiltonians are monotonic: if one region is contained in another, the modular Hamiltonian of the smaller region is the smaller of the two \cite{Borchers:2000pv}. Since $W_g$ is contained in $W$ for $g\ge0$, we find $H_g\ge0$ for every positive $g$, and hence $E(\vy)\ge0$.

\subsection{Light cones cut by a null plane}\label{sec:cones}
We now know the modular Hamiltonian of the regions $W_g$, but what we need is that of the future cone. A light cone is not bounded by a null plane, so it is not of the kind $W_g$. If we cut the cone with a null plane, however, we do obtain a region of this kind. At the end we will move the cut away.

\begin{figure}[htb]
\centering
\includegraphics[trim=20.9 59.7 43.6 66.4,clip,width=0.62\textwidth]{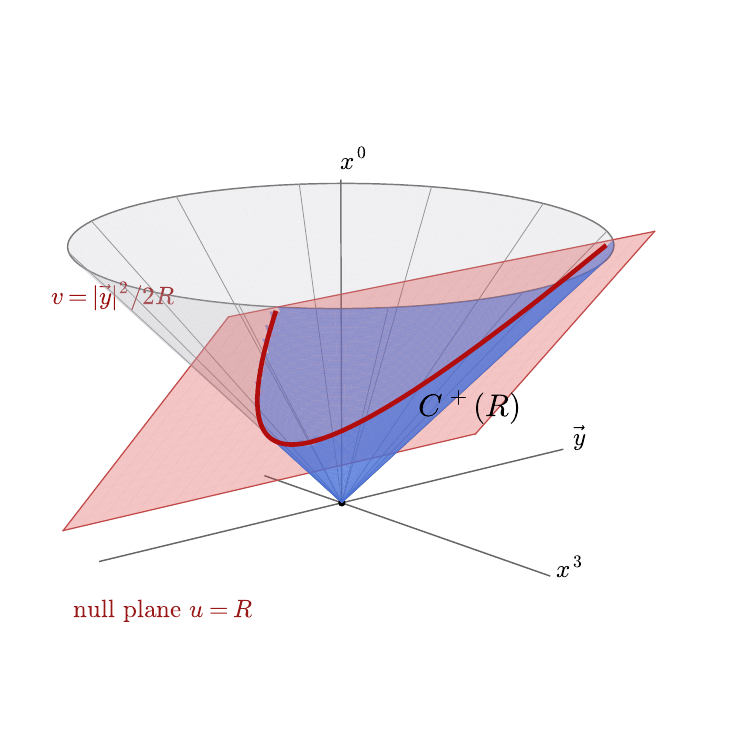}
\caption{The future light cone of the origin and the null plane $u=R$. They intersect along the parabola $v=|\vy|^2/2R$. The cut cone $C^+(R)$ is the part of the interior of the cone that lies below the plane, $u<R$. Its boundary on the cone is shown in blue.}\label{fig:cone}
\end{figure}

So consider the future cone of the origin cut by the null plane $u=R$,
\[
C^+(R)=\{\,2uv>|\vy|^2,\ 0<u<R\,\} ,
\]
which is shown in Fig.~\ref{fig:cone} and which becomes the whole future cone as $R\to\infty$. The cone meets the plane $u=R$ in the parabola $v=|\vy|^2/2R$, and the part of the plane above the parabola, $v>|\vy|^2/2R$, lies inside the cone.

Why is $C^+(R)$ a region of the kind $W_g$? Recall that $W_g$ consists of the points on one side of a null plane from which no light signal can reach the part of the plane below the edge. The cut cone has this property, with the plane $u=R$ and with the parabola as the edge. Indeed, a light signal from a point inside the future cone stays inside the cone, so it can reach the plane $u=R$ only above the parabola, whereas from a point outside the cone a light signal can reach the plane below the parabola. Thus $C^+(R)$ is the region $W_g$ with
\[
g(\vy)=\frac{|\vy|^2}{2R} ,
\]
translated by $R$ in the $u$ direction.

With this $g$, the weighted energy flux $\int\dd^2y\,g(\vy)E(\vy)$ of \eqref{eq:Hg} is $K^+/2R$, so here we meet the second moment $K^+$ of the energy flux again. In Lectures one and two it measured the width of the energy flux, whereas here it appears as the operator that bends the straight edge of a wedge into the parabola in which a light cone meets a null plane. This is the reason why the energy flux of Lectures one and two and the thermal property of this lecture have anything to do with each other.

We now apply \eqref{eq:Hg} to the cut cone. Strictly speaking, we are not allowed to do this, since $g$ is not bounded, but let us first see what we get. Together with the Bisognano--Wichmann theorem $K_W=2\pi B$, \eqref{eq:Hg} reads $K_{W_g}=2\pi(B-H_g)$. If we translate this formula by $R$ in the $u$ direction, $E(\vy)$ becomes the light-ray integral on the plane $u=R$, and the boost $B$ becomes the boost about the translated edge,
\[
B(R)=\ee^{\ii RP^-}B\,\ee^{-\ii RP^-}=B+RP^- .
\]
With $g(\vy)=|\vy|^2/2R$ we thus obtain $2\pi H^+(R)$ with
\[
H^+(R)=B(R)-\frac1{2R}\int_{u=R}\dd v\,\dd^2y\,|\vy|^2\,T_{vv}=B(R)-\frac{K^+(R)}{2R} ,
\]
where $K^+(R)$ is the second moment of the energy flux through the plane $u=R$, that is, $K^+(u_0)$ of Step 3 of Lecture two at $u_0=R$.

In a CFT, $2\pi H^+(R)$ is indeed the modular Hamiltonian of $C^+(R)$, because the conformal transformation generated by $K^+$ maps the wedge onto the cut cone (Appendix~\ref{sec:entered}). In a general theory with an energy-momentum tensor we cannot say this, since $g$ is not bounded and \eqref{eq:Hg} does not apply as it stands. But Astra's manuscript shows that the Tomita relation with $H^+(R)$ still holds for the states created in $C^+(R)$ \cite{Astra:2026}. For an operator $A$ localized in $C^+(R)$ we have
\begin{equation}\label{eq:cutnorm}
\big\|\ee^{-\pi H^+(R)}A|\Om\rangle\big\|=\big\|A^\dagger|\Om\rangle\big\| .
\end{equation}
This relation is all we will use about $H^+(R)$.

Now, when $R\to\infty$ the cut cone becomes the future cone, and in Sec.~\ref{sec:ideal} the operator in the Tomita relation of the future cone was $D$. How is $H^+(R)$ related to $D$? To find out, we have to express $H^+(R)$, which is defined on the plane $u=R$, on the plane $u=0$. For this we need to know how $K^+(u_0)$ depends on $u_0$. We computed this in Step 3 of Lecture two, \eqref{eq:Kp} and \eqref{eq:Kpp}. Since the second derivative $4P^-$ does not depend on $u_0$, the Taylor series stops at the second order,\footnote{In Lecture two we needed \eqref{eq:Kpp} only between the states created by $X$. Astra's manuscript obtains it, with one insertion of $L$, on a dense set of states \cite{Astra:2026}.}
\[
K^+(R)=K^+(0)-2R\int_{u=0}y^iT_{iv}+2R^2P^- .
\]

We insert this into $H^+(R)=B+RP^--K^+(R)/2R$. The term $2R^2P^-$ gives $-RP^-$, which cancels the term $RP^-$ of $B(R)$, and the first-order term gives $\int y^iT_{iv}$. Writing $K^+\equiv K^+(0)$, we find
\[
H^+(R)=\int_{u=0}\big(v\,T_{vv}+y^iT_{iv}\big)-\frac{K^+}{2R} .
\]
The integrand should look familiar. It is almost the dilatation current. Indeed, on the null plane the dilatation charge is
\[
D=\int_{u=0}(x^\nu T_{v\nu}-V_v)=\int_{u=0}(v\,T_{vv}+y^iT_{iv}-V_v) ,
\]
so that
\begin{equation}\label{eq:HRD}
\boxed{\ H^\pm(R)=D+Q_0\mp\frac{K^+}{2R},\qquad Q_0=\int_{u=0}V_v\ }
\end{equation}
Here the lower sign is for the past cone $C^-(R)$ cut at $u=-R$, which we treat in the same way. For a state created in the past cut cone the norm identity has the opposite sign in the exponent, just as the modular Hamiltonian of the past cone in a CFT is $-2\pi D$,
\begin{equation}\label{eq:cutnormpast}
\big\|\ee^{+\pi H^-(R)}A|\Om\rangle\big\|=\big\|A^\dagger|\Om\rangle\big\| .
\end{equation}

The term $K^+/2R$ comes from the cut, and it disappears when we move the cut away, $R\to\infty$. What remains, however, is not $D$ but $D+Q_0$. To put it as a slogan, ``in a theory that is only scale invariant, the light cone is thermal with respect to the dilatation shifted by $Q_0$.''

What is $Q_0$, then? It is the flux of the virial current through the null plane, and since $\int\partial_vL=0$ along each light ray, it is also the charge of $V_\mu-\partial_\mu L$, the part of the virial current that an improvement cannot remove. Note that $Q_0$ annihilates the vacuum, since $H^\pm(R)$, $D$ and $K^+$ all do. In a CFT, of course, $Q_0=0$ and we recover $2\pi D$. In the CFT whose dilatation has been redefined by a flavor charge, $D=D_{\rm c}+F$, we have $Q_0=-F$, so the cone is thermal with respect to $D+Q_0=D_{\rm c}$, as it should be. In that example a field with $[F,O]=e_FO$ has $\beta=e_F$, so it creates a state with $Q_0=-\beta$. We will find that this always holds.

I should add, however, that the cones are thermal with respect to $D+Q_0$ only in a limited sense. We have the norm identities \eqref{eq:cutnorm} and \eqref{eq:cutnormpast} only for states created in the cut cones, and only at finite $R$, where the two generators differ by
\[
H^+(R)-H^-(R)=-\frac{K^+}R .
\]
On the states that we use below, $\langle K^+\rangle$ is finite, since the bound \eqref{eq:Kfinite} of Lecture two holds for every local field (Appendix~\ref{sec:entered}). So this difference disappears as $R\to\infty$.

\subsection{The argument with the cut cones}\label{sec:logconv}
We would now like to repeat the argument of Sec.~\ref{sec:ideal} with $D+Q_0$ in place of $D$. The difficulty is that we know nothing about $Q_0$. We therefore use both the future and the past cone. They enter with $\ee^{-\pi H^+(R)}$ and $\ee^{+\pi H^-(R)}$, that is, with opposite signs, so that the unknown $Q_0$ drops out when we combine them.

As in Sec.~\ref{sec:ideal}, we fix a field label $a$ and a real test function $h$, now supported in the future cut cone, and consider the states with the labels $\Sigma^ka$,
\[
|\Psi_k\rangle=O_{\Sigma^ka}(h)|\Om\rangle,\qquad c_k=\langle\Psi_k|\Psi_k\rangle .
\]
Note that $c_k$ is the squared norm of the state itself, without the factor $\ee^{-\pi D}$. For the pair of \eqref{eq:poly}, $Na=b$ and $Nb=0$, we have $\Sigma^ka=a-\ii\pi kb$, so
\[
|\Psi_k\rangle=O_a(h)|\Om\rangle-\ii\pi k\,O_b(h)|\Om\rangle ,
\]
and $c_k$ is again a quadratic polynomial in $k$.

Let us first make two simplifications. We set $R=\infty$ from the start, and we assume that $Q_0$ commutes with $D$. Neither is justified, but the argument is easiest to see in this case, which I call the simplest situation. We follow the chain of equalities of Sec.~\ref{sec:ideal}. The Tomita relation \eqref{eq:cutnorm} of the future cone now has $D+Q_0$ in place of $D$, while PCT and the continued dilatation law \eqref{eq:strip}, which does not involve the modular Hamiltonian, are used as before. Combining the three relations, we find
\begin{equation}\label{eq:finiteR}
\big\|\ee^{-\pi(D+Q_0)}|\Psi_0\rangle\big\|
\ \overset{\rm Tomita}{=}\ \big\|O_a(h)^\dagger|\Om\rangle\big\|
\ \overset{\rm PCT}{=}\ \big\|O_a(h^-)|\Om\rangle\big\|
\ \overset{\eqref{eq:strip}}{=}\ \ee^{\pi\beta}\,\big\|\ee^{-\pi D}|\Psi_1\rangle\big\| .
\end{equation}
Compared with \eqref{eq:ideal}, the only change is the factor $\ee^{-\pi Q_0}$ acting on $|\Psi_0\rangle$.

In Sec.~\ref{sec:ideal} we worked with the norms of the states $\ee^{-\pi D}|\Psi\rangle$. Here this is not convenient. The past cone will give a second relation, in which $\ee^{-\pi D}$ appears in a different place, and we want to combine the two. So we first remove the factor $\ee^{-\pi D}$ from both sides of \eqref{eq:finiteR}.

Neither $|\Psi_0\rangle$ nor $|\Psi_1\rangle$ is an eigenstate of $D$, so we cannot simply replace $\ee^{-\pi D}$ by a number. But $D$ is a self-adjoint operator on the states, even when it acts on the fields through a Jordan block. We can therefore decompose each state into components with a definite value $\delta$ of $D$, just as we decompose a wave packet into plane waves,
\[
|\Psi_k\rangle=\int\dd\delta\,|\Psi_k(\delta)\rangle .
\]
In the simplest situation $Q_0$ acts on each component separately, so the square of \eqref{eq:finiteR} reads
\[
\int\dd\delta\,\ee^{-2\pi\delta}\,\big\|\ee^{-\pi Q_0}|\Psi_0(\delta)\rangle\big\|^2=\ee^{2\pi\beta}\int\dd\delta\,\ee^{-2\pi\delta}\,\big\||\Psi_1(\delta)\rangle\big\|^2 .
\]
The relation \eqref{eq:finiteR} holds for every label and every real test function supported in the cone, and a dilatation by $s$ maps $h$ to another such test function and the label $a$ to $\ee^{sN}a$. So \eqref{eq:finiteR} also holds if we apply $\ee^{\ii sD}$ to both $|\Psi_0\rangle$ and $|\Psi_1\rangle$, and Astra's manuscript shows that the same is true for any superposition $\gamma(D)=\sum_jw_j\,\ee^{\ii s_jD}$ of dilatations \cite{Astra:2026}. This multiplies the component with the value $\delta$ by the number $\gamma(\delta)$, so it puts the factor $|\gamma(\delta)|^2$ under both integrals. Since we can choose $|\gamma(\delta)|^2$ as we like, the two integrands must agree at each $\delta$. There $\ee^{-2\pi\delta}$ is a number, and it cancels. Integrating over $\delta$ again, we are left with $\|\ee^{-\pi Q_0}|\Psi_0\rangle\|=\ee^{\pi\beta}\,\||\Psi_1\rangle\|$.

Squaring and dividing by $c_0$, we find
\[
\frac{\langle\Psi_0|\,\ee^{-2\pi(Q_0+\beta)}\,|\Psi_0\rangle}{\langle\Psi_0|\Psi_0\rangle}=\frac{c_1}{c_0} .
\]
The left-hand side is the expectation value of $\ee^{-2\pi(Q_0+\beta)}$ in the state $|\Psi_0\rangle$. Let us think of it as we did for $\kappa$ in Lecture one. If we measure $Q_0$ in the state $|\Psi_0\rangle$, we find a value $q$ with some probability $\mathrm{Prob}_0(q)$, and the expectation value of a function of $Q_0$ is the average of the same function of $q$ with these probabilities. We write $\langle\,\cdot\,\rangle_0$ for this average, where the subscript reminds us that it is taken in the state $|\Psi_0\rangle$. So we have
\[
\big\langle\ee^{-2\pi(q+\beta)}\big\rangle_0=\frac{c_1}{c_0} .
\]
That is, the ratio of the squared norms of the shifted state $|\Psi_1\rangle$ and the original state $|\Psi_0\rangle$ measures the average of $\ee^{-2\pi(q+\beta)}$. If $|\Psi_0\rangle$ had a definite value $q$ of $Q_0$, this would say $c_1/c_0=\ee^{-2\pi(q+\beta)}$. Of course, we do not know the probabilities $\mathrm{Prob}_0(q)$.

The past cone gives a second relation of this kind. By \eqref{eq:strip}, the state $\ee^{-\pi D}|\Psi_1\rangle=\ee^{\ii\pi\Delta}O_a(h^-)|\Om\rangle$ is created in the past cone, since $h^-$ is supported there. We may therefore apply to it the Tomita relation \eqref{eq:cutnormpast} of the past cone, and with PCT we find
\begin{equation}\label{eq:pastnorm}
\big\|\ee^{\pi(D+Q_0)}\,\ee^{-\pi D}|\Psi_1\rangle\big\|=\ee^{-\pi\beta}\,\big\|O_a(h^-)^\dagger|\Om\rangle\big\|=\ee^{-\pi\beta}\,\big\||\Psi_0\rangle\big\| .
\end{equation}
Here the operator $\ee^{\pi(D+Q_0)}\ee^{-\pi D}$ is simply $\ee^{+\pi Q_0}$, now with the opposite sign in the exponent, and it acts on $|\Psi_1\rangle$. So \eqref{eq:pastnorm} says that the expectation value of $\ee^{+2\pi(Q_0+\beta)}$ in the state $|\Psi_1\rangle$ is $c_0/c_1$, that is, $\langle\ee^{+2\pi(q+\beta)}\rangle_1=c_0/c_1$. Since this holds for any label, we may apply it to the label $\Sigma^{-1}a$ instead of $a$, which lowers each $k$ by one. Then both averages are taken in the same state $|\Psi_0\rangle$:
\[
\big\langle\ee^{-2\pi(q+\beta)}\big\rangle_0=\frac{c_1}{c_0},\qquad
\big\langle\ee^{+2\pi(q+\beta)}\big\rangle_0=\frac{c_{-1}}{c_0} .
\]

We can now eliminate the unknown probabilities. Since the exponential is convex, every average satisfies Jensen's inequality $\langle\ee^y\rangle\ge\ee^{\langle y\rangle}$. With the mean $m=\langle q\rangle_0$ it gives
\[
\frac{c_1}{c_0}\ge\ee^{-2\pi(m+\beta)},\qquad\frac{c_{-1}}{c_0}\ge\ee^{+2\pi(m+\beta)} ,
\]
and if we multiply the two inequalities, both $m$ and $\beta$ drop out, leaving $c_1c_{-1}\ge c_0^2$. With the future cone alone the unknown $m$ would remain. Notice how little we have used. Of the probabilities $\mathrm{Prob}_0(q)$ only the mean $m$ entered, and we needed the two relations only as upper bounds on the two averages.

So much for the simplest situation. Without the two simplifications we cannot argue in this way, but two inequalities and a common mean survive, and we have just seen that this is enough. Let me explain what changes. We have the Tomita relations only at finite $R$, where the operator is $H^\pm(R)=D+Q_0\mp K^+/2R$, and we do not know that $Q_0$ and $K^+$ commute with $D$. So we cannot split $\ee^{-\pi H^\pm(R)}$ into $\ee^{-\pi D}$ and a factor that contains only $Q_0$, and we cannot speak of the probability of a value of $Q_0$ as we did above.\footnote{So far $Q_0$ is given only by the formal expression in \eqref{eq:HRD}. Nor can we repeat the step with $\gamma(D)$ at a fixed $R$, because a dilatation does not map the cut cone to itself.}

The way out is, once again, a dilatation. A dilatation maps the cone cut at $R$ to the cone cut at $\ee^sR$,
\[
\ee^{\ii sD}\,H^\pm(R)\,\ee^{-\ii sD}=H^\pm(\ee^sR) ,
\]
so all it does is change $R$. Now take the limit $R\to\infty$ as an average over $\log R$. A dilatation shifts $\log R$, and an average over a long range of $\log R$ does not notice the shift. This is only the idea. The result, which I quote from Astra's manuscript \cite{Astra:2026}, is that in this limit $H^\pm(R)$ behaves like an operator that commutes with $D$, so that we can measure the two together. Let $\lambda$ be the value of $H^\pm(R)$ that we find in this limit and $\delta$ the value of $D$, and put $q=\lambda-\delta$. By \eqref{eq:HRD}, it is this $q$ that takes the place of the value of $Q_0$. In the state $|\Psi_0\rangle$ we find it with some probability, $\mathrm{Prob}_0^+(q)$ if we use the future cones and $\mathrm{Prob}_0^-(q)$ if we use the past cones. We write $\langle\,\cdot\,\rangle_0^+$ and $\langle\,\cdot\,\rangle_0^-$ for the averages with these probabilities.

One would expect that $\mathrm{Prob}_0^+=\mathrm{Prob}_0^-$, since $H^+(R)$ and $H^-(R)$ differ only by $K^+/R$. This has not been shown. What Astra's manuscript shows is the following \cite{Astra:2026}. First, the two relations of the simplest situation survive as inequalities,
\begin{equation}\label{eq:twobounds}
\big\langle\ee^{-2\pi(q+\beta)}\big\rangle_0^+\le\frac{c_1}{c_0},\qquad
\big\langle\ee^{+2\pi(q+\beta)}\big\rangle_0^-\le\frac{c_{-1}}{c_0} .
\end{equation}
Second, $\mathrm{Prob}_0^+$ and $\mathrm{Prob}_0^-$ have the same mean $m$, because the expectation values of $H^+(R)$ and $H^-(R)$ differ only by $\langle K^+\rangle/R$, which tends to zero.\footnote{Why only inequalities? In the limit a small part of the probability can move out to large $|q|$, where the exponential is large, and its contribution to the average is lost. For the same reason the mean of a limit of probability distributions need not be the limit of their means, so even the second statement is not automatic.}

But an upper bound on each of the two averages and a common mean are all that we used in the simplest situation. Jensen's inequality therefore gives $c_1c_{-1}\ge c_0^2$ as before. Starting from the label $\Sigma^ka$ instead of $a$, we find in the same way
\begin{equation}\label{eq:logconvex}
\boxed{\ c_k^2\le c_{k-1}\,c_{k+1}\qquad\text{for every integer }k\ }
\end{equation}
This is what remains of \eqref{eq:ideal} in a theory that is only scale invariant. In the ideal case of Sec.~\ref{sec:ideal}, where $Q_0=0$, the same steps make $c_k$ a geometric sequence, which satisfies \eqref{eq:logconvex} with equality. Now we know only that $\log c_k$ is a convex function of $k$.

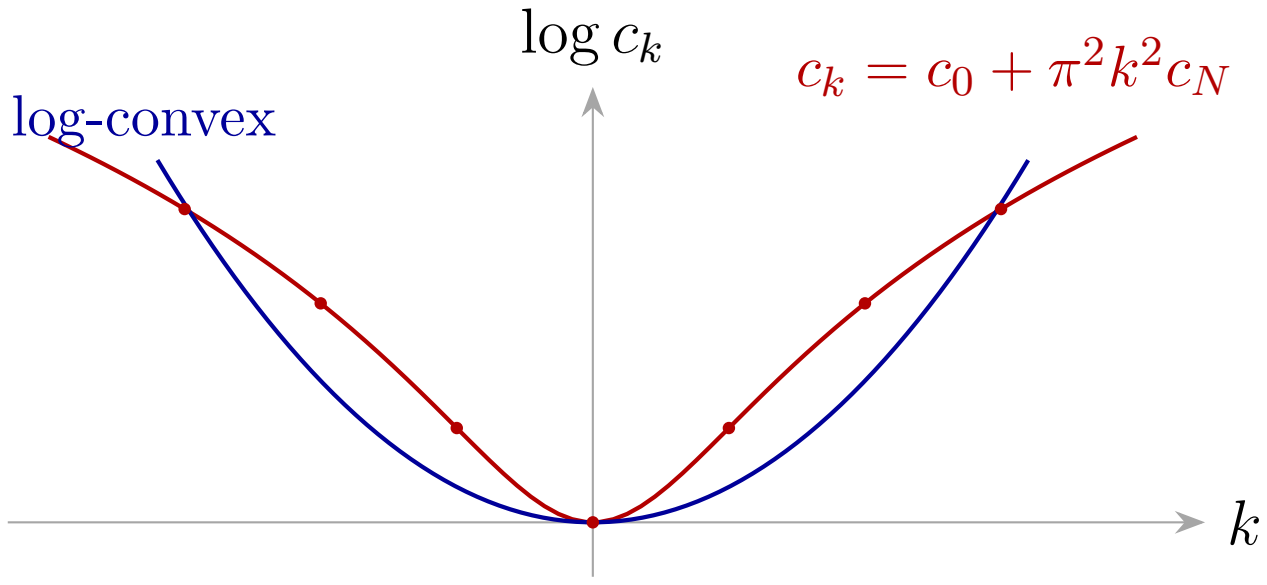
\begin{figure}[htb]
\centering
\begin{tikzpicture}[scale=0.9]
\draw[axisline] (-4.3,0)--(4.5,0) node[right,black] {$k$};
\draw[axisline] (0,-0.4)--(0,3.2) node[above,black] {$\log c_k$};
\draw[thick,red!70!black,domain=-4:4,samples=80] plot (\x,{ln(1+\x*\x)});
\node[red!70!black,font=\small,align=left] at (3.1,3.35) {$c_k=c_0+\pi^2k^2c_N$};
\draw[thick,blue!60!black,domain=-3.2:3.2,samples=80] plot (\x,{0.26*\x*\x});
\node[blue!60!black,font=\small] at (-3.3,2.95) {log-convex};
\foreach \k in {-3,-2,-1,0,1,2,3} {\fill[red!70!black] (\k,{ln(1+\k*\k)}) circle (1.3pt);}
\end{tikzpicture}
\caption{For a non-constant quadratic polynomial $c_k$ the function $\log c_k$ flattens out at large $|k|$, so its slope decreases. The red curve shows $\log c_k$ for the generalized free field \eqref{eq:gffjordan}, with $c_N=\|\varphi_1(h)|\Om\rangle\|^2$. A log-convex sequence, such as the blue one, has a slope that does not decrease.}\label{fig:logconvex}
\end{figure}

Fortunately, this is still enough to exclude a partner, because a quadratic polynomial is not log-convex either, unless it is a constant. To see this, recall that the slope $\log(c_{k+1}/c_k)$ of a convex function does not decrease. Now take the pair with $Na=b$, $Nb=0$ and $O_b(h)|\Om\rangle\neq0$. The quadratic polynomial $c_k$ grows like $\pi^2k^2\,\|O_b(h)|\Om\rangle\|^2$ for large $|k|$, so $\log c_k\simeq2\log|k|$ flattens out at both ends. Its slope tends to zero both for $k\to+\infty$ and for $k\to-\infty$ (Fig.~\ref{fig:logconvex}). But a slope that does not decrease and tends to zero at both ends must vanish everywhere. Hence $c_k$ is constant and $O_b(h)|\Om\rangle=0$, which contradicts our choice of $h$, as in Sec.~\ref{sec:ideal}. We conclude that $N=0$ on every field. Since $\beta$ does not appear in \eqref{eq:logconvex}, this holds for complex dimensions as well.\footnote{The red curve in Fig.~\ref{fig:logconvex} is convex near $k=0$, and it fails to be convex only at larger $|k|$. The three states with $k=0,\pm1$ would therefore not be enough, and this is why we need the whole sequence $|\Psi_k\rangle$.}

Finally, why does this conclusion not apply to the generalized free field \eqref{eq:gffjordan}? There $c_k=\|\varphi_0(h)|\Om\rangle\|^2+\pi^2k^2\|\varphi_1(h)|\Om\rangle\|^2$, which is not log-convex. But there is no contradiction, because without an energy-momentum tensor we have no \eqref{eq:Hg}, on which the argument rests.

\subsection{The imaginary part is an internal charge}\label{sec:charge}
We have excluded the Jordan blocks, and it remains to understand the imaginary part $\beta$ of the dimension. In the ideal case of Sec.~\ref{sec:ideal} we found $\beta=0$. With the cut cones we cannot expect this, since in the CFT whose dilatation has been redefined by a flavor charge we have $\beta\neq0$. What we should expect instead is clear from Sec.~\ref{sec:ideal}. There the generator with respect to which the cone is thermal gave every field a real dimension. Now this generator is $D+Q_0$, so the imaginary part of the dimension under $D$ must be compensated by $Q_0$. Indeed, in the example the field creates a state with $Q_0=-\beta$. We now show that this always holds.

We go back to the two bounds \eqref{eq:twobounds}, which hold without the two simplifications. Now that $N=0$, we have $\Sigma=1$, so all the states $|\Psi_k\rangle$ are the same state $|\Psi_0\rangle=O(h)|\Om\rangle$ and $c_k=c_0$ for every $k$. The ratios $c_1/c_0$ and $c_{-1}/c_0$ in them are both one. By Jensen's inequality the first bound gives $m+\beta\ge0$, and the second gives $m+\beta\le0$, so $m=-\beta$ and Jensen's inequality holds with equality in both bounds. Since the exponential is strictly convex, this happens only if $\mathrm{Prob}_0^+$ and $\mathrm{Prob}_0^-$ are both concentrated at the single value $q=-\beta$. That is, every measurement of $q$ in the state $|\Psi_0\rangle$ gives $-\beta$. In particular, the two probabilities, which we could not compare in Sec.~\ref{sec:logconv}, are now the same, and we no longer have to distinguish the future and the past cones.

Now recall an elementary fact of quantum mechanics. If every measurement of an observable $A$ in a state $|\psi\rangle$ gives the same value $a$, the state is an eigenstate, $A|\psi\rangle=a|\psi\rangle$, because $\|(A-a)|\psi\rangle\|^2=\langle(A-a)^2\rangle=0$. If $q$ were the value of $Q_0$, as in the simplest situation of Sec.~\ref{sec:logconv}, we could conclude at once that $Q_0|\Psi_0\rangle=-\beta|\Psi_0\rangle$.

But in \eqref{eq:twobounds} $q$ is the value of $H^\pm(R)$ in the limit $R\to\infty$ minus the value of $D$, as we defined it in Sec.~\ref{sec:logconv}. By \eqref{eq:HRD} the corresponding operator is $H^\pm(R)-D=Q_0\mp K^+/2R$, which becomes $Q_0$ in the limit. So we expect the same conclusion, and Astra's manuscript confirms it \cite{Astra:2026}.\footnote{More precisely, the manuscript shows $\langle\phi|Q_0|\Psi_0\rangle=-\beta\,\langle\phi|\Psi_0\rangle$ for a dense set of states $|\phi\rangle$, by evaluating the matrix element of $H^\pm(R)$ in two ways.} We thus obtain
\begin{equation}\label{eq:Q0}
\boxed{\ Q_0\,O(h)|\Om\rangle=-\beta\,O(h)|\Om\rangle\qquad\text{for every field of dimension }\Delta=\alpha+\ii\beta\ }
\end{equation}

This is the main result of this lecture. A field of dimension $\alpha+\ii\beta$ creates a state of charge $-\beta$ under $Q_0$, the charge of the part of the virial current that we cannot remove by an improvement. That is, the imaginary part of a scaling dimension is a charge.

We would like to use $Q_0$ as the generator of a symmetry. For this we need \eqref{eq:Q0} for products of fields as well, whereas so far we have it only for states created by a single field. Inside correlation functions with one insertion of $L$, $Q_0$ is the charge of the conserved current $V_\mu-\partial_\mu L$, and the charge of a conserved current acts on a product of fields through its commutators with the individual fields. So each field should contribute its own charge. Astra's manuscript shows that this is indeed the case, using locality to bring the fields one at a time next to the vacuum, where \eqref{eq:Q0} applies \cite{Astra:2026}. The state $O_1(h_1)\cdots O_m(h_m)|\Om\rangle$ is therefore an eigenstate of $Q_0$ with the eigenvalue $-(\beta_1+\dots+\beta_m)$.

In this way $Q_0$, which so far was only a formal expression, becomes a self-adjoint operator $Q$. Indeed, the states created by products of fields are dense, and each of them is an eigenstate with a real eigenvalue. The operator $Q$ commutes with the Poincar\'e group and with $D$, since these transformations map the states of each field to states of the same field, which have the same charge. It annihilates the vacuum and generates an internal symmetry,
\[
\ee^{\ii sQ}O(x)\ee^{-\ii sQ}=\ee^{-\ii s\beta}O(x) .
\]
Since $Q$ annihilates the vacuum, this symmetry gives a selection rule,
\[
\langle\Om|O_1(x_1)\cdots O_n(x_n)|\Om\rangle=\ee^{-\ii s(\beta_1+\dots+\beta_n)}\,\langle\Om|O_1(x_1)\cdots O_n(x_n)|\Om\rangle
\]
for every $s$, so a non-zero correlation function requires $\beta_1+\dots+\beta_n=0$.

The combination $D'=D+Q$ is a dilatation under which every field has the real dimension $\alpha=\mathrm{Re}\,\Delta$. Once $L$ has been constructed as a local field under the condition of Sec.~\ref{sec:conf}, $V_\mu-\partial_\mu L$ is the conserved current of this symmetry, so we are in the harmless case of the Introduction, $V_\mu=\partial_\mu L+J_\mu$ with a conserved $J_\mu$.

\subsection{From the internal charge to conformal invariance}\label{sec:conf}
Under $D'=D+Q$ every field has a real dimension and no Jordan block, so with the dilatation $D'$ the theory is of the kind to which the OpenAI theorem applies, and we would like to conclude that it is conformal. There is one more point to take care of, however. The OpenAI argument works with the operators that the fields generate in each region of spacetime, while we know only how $Q$ acts on the fields themselves. To pass from the fields to these operators we need one more condition, a technical one on the domain of the smeared fields. Most physicists, or at least I, would take it for granted. Following Astra's manuscript \cite{Astra:2026} we call it condition (C).\footnote{By the third assumption below, a field smeared inside a double cone, the intersection of a future and a past light cone, belongs to the operators of that double cone. Strictly speaking, this may be true only after we extend the domain of the field, whereas we know the action of $Q$ only on the states that products of fields create from the vacuum. Condition (C) says that the smeared field defined on these states alone already belongs to the operators of its double cone. It holds, for example, if the fields obey a polynomial bound in the energy. Whether it already follows from the four assumptions listed below is not known.}

We can now state the result of Astra's manuscript \cite{Astra:2026}. Let me first collect the assumptions. They are those of the OpenAI paper \cite{OpenAI:2026}, except for the second one, where the OpenAI paper assumes that the dilatation acts diagonally with real dimensions.
\begin{enumerate}
\item The theory is unitary with positive energy. The states are created from the vacuum by local fields, which transform covariantly under Poincar\'e transformations, and the vacuum is invariant under Poincar\'e transformations and dilatations.
\item Under dilatations the fields mix within finite Jordan blocks, and their scaling dimensions may be complex. The real parts of the dimensions are bounded below, and any bounded range of real parts contains only finitely many dimensions, each of finite multiplicity.
\item The operators localized in a region form an algebra, the algebras of spacelike separated regions commute, and a field smeared with a test function supported in a region is an operator localized in that region.
\item The energy-momentum tensor $T_{\mu\nu}$ and the virial current $V_\mu$ are local fields with $T^\mu{}_\mu=\partial^\mu V_\mu$, and they generate translations, Lorentz transformations and dilatations through Ward--Takahashi identities, including the mixing of the fields under dilatations.\footnote{A field may be a finite sum of components of different dimensions. In the statements below $T_{\mu\nu}$ and $V_\mu$ denote the components of dimension four and three, which carry the Ward--Takahashi charges.}
\end{enumerate}
Condition (C) is not among them, and we need it only for the last of the following three statements.
\begin{itemize}
\item The dilatation acts diagonally, $N=0$ on every field.
\item There is an internal charge $Q$ with the properties found above, and $D'=D+Q$ assigns to every field the real part of its dimension.
\item Under condition (C), there is a local Hermitian scalar $L$ of dimension two such that
\begin{equation}
\Box L=\partial^\mu V_\mu,\qquad t_{\mu\nu}=T_{\mu\nu}+\tfrac13(\partial_\mu\partial_\nu-\eta_{\mu\nu}\Box)L,\qquad t^\mu{}_\mu=0 ,
\end{equation}
and the conformal currents built from $t_{\mu\nu}$ generate local conformal Ward--Takahashi identities on the fields.
\end{itemize}
When all dimensions are real we do not need condition (C). Once the Jordan blocks have been excluded by the first statement, the real-dimensional case follows from the OpenAI theorem \cite{OpenAI:2026}.

\subsection{Back to the examples}\label{sec:examples}
Let us finally see what these results say about the examples that motivated us. We start with the complex generalized free field \eqref{eq:gffc} of the Introduction, in which $\phi$ has the dimension $\alpha+\ii\beta$ and $A$ is the coefficient of $\langle\phi\phi\rangle$. For $A=0$ the charge $Q$ of Sec.~\ref{sec:charge} is $Q=-\beta F$, with the $U(1)$ charge $F$ normalized by $[F,\phi]=\phi$, and $D'=D-\beta F$ is the dilatation of a conformal generalized free field of dimension $\alpha$. For $A\neq0$ there is no internal symmetry that could play the role of $Q$. The selection rule leads to the same conclusion. With $\beta\neq0$ the charges in $\langle\phi\phi^\dagger\rangle$ cancel, but $\langle\phi\phi\rangle$ carries $2\beta$, so the selection rule requires $A=0$. Since we did not need condition (C) to find $Q$, we conclude that a theory of this kind cannot have an energy-momentum tensor, and indeed the example has none.

Let us also return to Polchinski's virial current $V_\mu=Q_{ab}\phi^a\partial_\mu\phi^b-\ii P_{ij}\bar\psi^i\bar\sigma_\mu\psi^j$. We have learned that in a theory with an energy-momentum tensor the rotation can produce complex scaling dimensions only if it is generated by a conserved current, that is, only if it is a symmetry of the theory. In perturbation theory the would-be limit cycles were found to be conformal by a different route, through $B^I=0$ in the trace $T^\mu{}_\mu=B^I\mathcal O_I$ of \eqref{eq:B}. The argument of this lecture is non-perturbative, and it does not use the local renormalization group.

\section{Summary and outlook}\label{sec:summary}
Let me summarize these lectures. Our question was whether $L=\Box^{-1}T^\mu{}_\mu$ is a local operator. The two-point functions cannot answer it, because they give only the norms of states, which are the same as for a local $L$. Throughout the lectures we looked instead at the energy flux of a state through a null plane, measured with the energy-momentum tensor itself.

In Lecture one we followed the OpenAI paper. The only possible non-locality of the state of $L$ is a tail in the distribution of its transverse width, and causality excludes it, through an identity in which the light ray that creates the state and the light ray that measures its energy flux are exchanged. Since the amplitude of this tail is proportional to the rate at which the $a$-anomaly would keep flowing, we also learned that $a$ does not flow at a scale-invariant point. This argument required that $T_{\mu\nu}$ and $V_\mu$ have definite scaling dimensions, and in Lecture two we showed that they do, under two simplifying assumptions whose removal I only sketched. The light-ray operators have no logarithmic partner, but a partner of $T_{\mu\nu}$ would put a logarithm of the invariant mass into the state that $T_{\mu\nu}$ creates, and this would make the second moment of its energy flux infinite, in contradiction with causality.

In Lecture three we used that a dilatation by $\ii\pi$ is a reflection through the origin. If the vacuum in a light cone were thermal with respect to the dilatation, this alone would force every dimension to be real and every Jordan block to be trivial. But in a theory that is only scale invariant, the light cone is thermal with respect to the dilatation shifted by the flux $Q_0$ of the virial current, and only in a limited sense. By combining the future and the past cone we eliminated the unknown $Q_0$, and found that there are no Jordan blocks and that the imaginary part of a dimension is the charge of an internal symmetry. If we remove this charge from the dilatation, we obtain a dilatation with real dimensions, to which the OpenAI theorem applies, so that under a technical condition the theory is conformal.

Let me emphasize what these lectures do not contain. I have not reviewed the second half of the OpenAI paper, the construction of $L$ as a local operator, although Lectures two and three depend on it. In Lecture one we showed only that the state created by $L$ has no tail, and in Lecture two we used one insertion of $L$ in correlation functions. In Lecture three, after we found the dilatation $D'=D+Q$ with real dimensions, we obtained conformal invariance by applying this construction to $D'$, under the condition of Sec.~\ref{sec:conf}.

Note also that the argument uses four dimensions in an essential way. In four dimensions the transverse directions of a null plane form a two-dimensional plane, and the width of the energy flux in this plane is conjugate to the invariant mass through a two-dimensional Fourier transform. We started the Introduction with the three-dimensional Ising model, so whether one can adapt the argument to three dimensions is a natural next question.

Finally, let me comment on the status of the argument. Lectures two and three are my reading, as a physicist, of the manuscript \cite{Astra:2026}, in which the AI system Astra extends the argument of the OpenAI paper \cite{OpenAI:2026}. I have followed the steps presented in the main text at the level of these lectures and checked their algebra, but not the functional-analytic parts, such as domains of unbounded operators, the removal of cutoffs and the order of limits. I have also quoted several inputs without deriving them.\footnote{In Lecture one these are the estimates behind the comparison \eqref{eq:compare} and the sum rule \cite[Secs.~2 and 4]{OpenAI:2026}, and in Appendix~\ref{app:lookback} the normalizations of Hartman, Mathys and Stergiou and the vertex of the Euler term. In Lecture two they are the insertion of $L$ \cite[Sec.~5]{OpenAI:2026} and its adaptation in Astra's manuscript \cite{Astra:2026}, the statement that $\mathcal P_t$ maps the states of $X$ to states of $X$, and, without the assumptions (A) and (B), two long arguments of Astra's manuscript that I only sketched. In Lecture three they are \eqref{eq:Hg} with a Jordan block, \eqref{eq:Kpp} on a dense set of states, the norm identity for the cut cones, the step with $\gamma(D)$, the limit $R\to\infty$ that leads to the bounds \eqref{eq:twobounds}, \eqref{eq:Q0} without the two simplifications and its extension to products of fields, and the construction of $L$ under condition (C) \cite{Astra:2026}.}

I would like to end these lectures with an anecdote. On one occasion I had a job interview, and I told the interviewers that my life-long project was to prove that scale invariance implies conformal invariance. One of the interviewers, a very famous physicist, said, ``It is not a good idea to make it a life-long project. Work on it now and prove it within five years.'' I declined the offer, but maybe he was right. It took more than five years, but now it is time for me to look for another project. Maybe in three dimensions?

\section*{Acknowledgments}
A part of these lecture notes is based on my lecture given at the Yukawa Institute for Theoretical Physics, Kyoto University, which was amended and improved with the help of Anthropic's Claude. Claude also helped me to understand the OpenAI papers (the original one \cite{OpenAI:2026} and the one by Astra \cite{Astra:2026}). I thank Slava Rychkov for the correspondence and discussions. I also thank OpenAI for not responding to my query about the scope of the preprint \cite{OpenAI:2026}, but offering me the ``ChatGPT for Academic Researchers'' program, under which ChatGPT 6.2 Astra generated a proof for the Jordan case. For the record, Anthropic's Claude Fable 5.2 had generated a proof for complex dimensions at that time. This work is supported in part by JSPS KAKENHI Grant Numbers 21K03581 and 26K00699.

\appendix
\section{The sum rule in detail}\label{app:sumrule}
In Sec.~\ref{sec:sumrule} we saw that $\mathcal A$ changes sign when the two light rays are exchanged, and I stated the form \eqref{eq:sumrule0} that the sum rule takes for the weight $|\vy-Y|^2$. In this appendix I explain how the weight leads to \eqref{eq:sumrule0}, and how the cutoff is treated. The computation is that of Sec.~4 of the OpenAI paper \cite{OpenAI:2026}. The paper works with a dispersion relation at fixed spacelike momentum transfer, but here I rewrite its result for the half light rays of Sec.~\ref{sec:key}.

\paragraph{Positions relative to the centroid.} Let $\vec b=\vy-Y$ be the transverse position of a light ray relative to the centroid of the energy flux. For a function $g$ on the transverse plane we define
\[
\mu[g]=\frac1{P^+}\int\dd^2y\,g(\vy-Y)\,E(\vy) ,
\]
the energy flux weighted with $g(\vec b)$, per unit null momentum. By \eqref{eq:moments} and \eqref{eq:kappa},
\[
\mu[1]=1,\qquad\mu[b_i]=0,\qquad\mu[|\vec b|^2]=\kappa .
\]
In $\mathcal A(\vec k_1,\vec k_2)$ the light rays appear through their Fourier transforms, and the bra has a definite transverse momentum. This is the reason why $\vec b$, and not $\vy$, is the relevant position. The centroid $Y$ is conjugate to the transverse momentum, so in a state of definite transverse momentum it is completely undetermined, and only positions relative to it have a meaning. Explicitly,
\[
E(\vec k_1)=\int\dd^2y\,\ee^{\ii\vec k_1\cdot\vy}E(\vy)=P^+\,\ee^{\ii\vec k_1\cdot Y}\,\mu\big[\ee^{\ii\vec k_1\cdot\vec b}\big] .
\]
The factor $\ee^{\ii\vec k_1\cdot Y}$ only shifts the transverse momentum of the state by $\vec k_1$. So if we multiply $\mathcal A$ by the Fourier transform $\hat g(\vec k_1)$ of a weight and integrate over $\vec k_1$, the measuring light ray becomes $P^+\mu[g]$. In the same way, $\vec k_2$ is the transverse momentum of the state that $\bar E_+(\vec k_2)$ creates. Integrating with $\hat g(\vec k_2)$ therefore weights the creating light ray with $g(\vec b)$, where $\vec b$ is now its position relative to the centroid of the state that it creates. We write $|\Phi[g]\rangle$ for this state.

\paragraph{The sum rule for a general weight.} Since $V_v=n^\mu V_\mu$ and $V_u=\bar n^\mu V_\mu$, the bra enters $\mathcal A=p^-A_v-p^+A_u$ only through the component of $V$ along $e=p^-n-p^+\bar n$. This vector is orthogonal to $p$. In the rest frame of $p$ it is $e=M\,e_z+p_i\,e_i$, where $i=1,2$ labels the directions $e_i$ transverse to $z$. Recall from \eqref{eq:spin} that $|l\rangle$ enters $V_\mu|\Om\rangle$ only along $p_\mu$. So $|l\rangle$ drops out, and what is left of the bra is the spin-one state with the two components $M\langle j_z|$ and $p_i\langle j_i|$. Up to an overall factor,
\[
\mathcal A(\vec k_1,\vec k_2)=M\,\langle j_z|E(\vec k_1)\bar E_+(\vec k_2)|\Om\rangle+(k_1+k_2)_i\,\langle j_i|E(\vec k_1)\bar E_+(\vec k_2)|\Om\rangle .
\]
We now take a radial weight $g$, multiply $\mathcal A$ by $\hat g(\vec k_1)\hat g(\vec k_2)$ and integrate. Since $\mathcal A$ changes sign under the exchange, the result is zero. In the first term both light rays are weighted with $g$. In the second term a momentum $k_i$ multiplies one of the two functions $\hat g$, and $k_i\hat g(\vec k)$ is, up to a factor $\ii$, the Fourier transform of $\partial_ig$. So in the second term one of the two light rays is weighted with $\partial_ig$ instead of $g$, and we find
\begin{equation}\label{eq:sumruleg}
M\,\langle j_z|\mu[g]|\Phi[g]\rangle=\ii\,\langle j_i|\mu[\partial_ig]|\Phi[g]\rangle+\ii\,\langle j_i|\mu[g]|\Phi[\partial_ig]\rangle ,
\end{equation}
where the bra has the definite invariant mass $M$. This is eq.~(4.15) of the OpenAI paper.

\paragraph{The state created by a weighted half light ray.} It remains to find $|\Phi[g]\rangle$. In momentum space $T_{uu}$ creates the state $(p^-)^2\big(\tfrac13|l\rangle+|s_{\bar z\bar z}\rangle\big)$. This is \eqref{eq:cVcT} with $u$ and $v$ exchanged, so $\bar z$ is the direction of $\bar n$ in the rest frame. The integral over the half line $u>0$ gives a factor $1/p^-$, up to a constant. At the transverse momentum $\vec r$ and the invariant mass $M$ we have $p^-=(M^2+|\vec r|^2)/2p^+$, so after we absorb $1/2p^+$ into $\zeta$, the half light ray creates the state
\[
(M^2+|\vec r|^2)\big(\tfrac13|l\rangle+|s_{\bar z\bar z}\rangle\big) .
\]
We have to express $|s_{\bar z\bar z}\rangle$ through the components along $z$. In the rest frame $\bar n$ points along the unit vector
\[
\bar e=\frac{|\vec r|^2-M^2}{|\vec r|^2+M^2}\,e_z-\frac{2Mr_i}{|\vec r|^2+M^2}\,e_i .
\]
For $\vec r=0$ this is $-e_z$, as it should be. It depends on $\vec r$ because the boost that brings us to the rest frame does. With $|s_{\bar z\bar z}\rangle=\bar e_A\bar e_B|s_{AB}\rangle$ we find
\begin{equation}\label{eq:chi}
\begin{aligned}
(M^2+|\vec r|^2)\big(\tfrac13|l\rangle+|s_{\bar z\bar z}\rangle\big)={}&\frac{M^2+|\vec r|^2}3\,|l\rangle+\Big(|\vec r|^2-5M^2+\frac{6M^4}{|\vec r|^2+M^2}\Big)|s_{zz}\rangle\\
&+\Big(-4Mr_i+\frac{8M^3r_i}{|\vec r|^2+M^2}\Big)|s_{zi}\rangle+\frac{4M^2r_ir_j}{|\vec r|^2+M^2}\,|s^{\rm TF}_{ij}\rangle ,
\end{aligned}
\end{equation}
where $|s^{\rm TF}_{ij}\rangle=|s_{ij}\rangle+\tfrac12\delta_{ij}|s_{zz}\rangle$ is the traceless part of the transverse components \cite{OpenAI:2026}.

To weight the creating light ray with $g$, we multiply \eqref{eq:chi} by $\hat g(\vec r)$ and integrate over $\vec r$.

First, several terms of \eqref{eq:chi} are polynomials in $\vec r$. A polynomial in $\vec r$ is the Fourier transform of a delta function at $\vec b=0$ and of its derivatives, so these terms depend on the weight only at $\vec b=0$. We remove them by taking a weight that vanishes near $\vec b=0$. In particular, the whole spin-zero part $\tfrac13(M^2+|\vec r|^2)|l\rangle$ drops out. Note that we do not lose the tail in this way, since, as we saw in Sec.~\ref{sec:spin}, it also appears, with the factor $-\tfrac13$, in the spin-two component $|s_{zz}\rangle$.

Second, a radial weight has no direction in the transverse plane, so the terms with $r_i$ and with the traceless part of $r_ir_j$ integrate to zero. Only the component $|s_{zz}\rangle$ of helicity zero remains,
\[
|\Phi[g]\rangle=f_g(M)\,|s_{zz}\rangle,\qquad f_g(M)=6M^4\int\dd^2r\,\frac{\hat g(\vec r)}{|\vec r|^2+M^2}=6M^4\int\dd^2b\,g(\vec b)\,G_M(\vec b) .
\]
Here $G_M$ is the Fourier transform of $1/(|\vec r|^2+M^2)$, that is, the massive propagator in the transverse plane. It depends on $\vec b$ only through $M|\vec b|$, and it decays exponentially for $|\vec b|\gg1/M$. For $g=|\vec b|^2$ we substitute $\vec b=\vec t/M$ and find
\[
f_g(M)=6\int\dd^2t\,|\vec t|^2\,G_1(\vec t)\equiv c_* ,
\]
which does not depend on $M$. The weight $|\vec b|^2$ contains no length scale, and the powers of $M$ cancel. So this weight creates $|s_{zz}\rangle$ with a constant wave function.

Third, the weight $\partial_ig$ is $b_i$ times a radial function. It has a direction in the transverse plane, and now only the term $8M^3r_i/(|\vec r|^2+M^2)$ with the component $|s_{zi}\rangle$ of helicity one survives. For $\partial_i|\vec b|^2=2b_i$ the weight contains one power of length less than $|\vec b|^2$, so the same substitution leaves one power of $M$,
\[
|\Phi[2b_i]\rangle=-\tfrac{2\ii}3\,c_*\,M\,|s_{zi}\rangle .
\]
The wave function is proportional to $M$ and vanishes at $M=0$. With the Fourier convention of the OpenAI paper one finds $c_*=24$ here and the coefficient $-16\ii$ for the weight $2b_i$, which gives the number $\tfrac23$ \cite{OpenAI:2026}.

If we now put $g=|\vec b|^2$ in \eqref{eq:sumruleg}, with $\mu[|\vec b|^2]=\kappa$ and $\mu[2b_i]=2\mathcal D_i$, and divide by $c_*$, we obtain \eqref{eq:sumrule0} with $f=1$. Strictly speaking, the weight $|\vec b|^2$ vanishes at $\vec b=0$ but not near it. The cutoff, to which we now turn, repairs this as well.

\paragraph{The cutoff.} With the weight $|\vec b|^2$ the wave function is a constant, which is not normalizable, and with a tail the individual terms of \eqref{eq:sumrule0} diverge. We therefore replace $|\vec b|^2$ on both light rays with $g_\Lambda(\vec b)=|\vec b|^2\chi_\Lambda(\vec b)$, where $\chi_\Lambda$ is a smooth cutoff to the annulus $1<|\vec b|<\Lambda$ around the centroid. This weight vanishes near $\vec b=0$. In place of $\kappa$ and $\mathcal D_i$ we now have
\[
\kappa_\Lambda=\mu[g_\Lambda],\qquad\mathcal D_i^\Lambda=\tfrac12\,\mu[\partial_ig_\Lambda] .
\]
On the creating light ray the propagator $G_M$ has the range $1/M$. If this range lies inside the annulus, that is, for $1/\Lambda\ll M\ll1$, the cutoff does not matter. So $g_\Lambda$ creates $c_*|s_{zz}(f_\Lambda)\rangle$ with a wave function $f_\Lambda$ that is equal to one for $1/\Lambda\ll M\ll1$ and suppressed outside this range. In the same way $\partial_ig_\Lambda$ creates $|s_{zi}\rangle$ with the wave function $Mf_\Lambda$, up to its form at the two ends of the range, which plays no role below. Finally, we give the bra a wave function as well. We integrate \eqref{eq:sumruleg} over the invariant mass of the bra with the weight $f_\Lambda/M$. The sum rule then reads
\begin{equation}\label{eq:sumruleL}
\underbrace{\langle j_z(f_\Lambda)|\kappa_\Lambda|s_{zz}(f_\Lambda)\rangle}_{\rm (i)}
=\underbrace{2\ii\,\big\langle j_i\big(\tfrac {f_\Lambda}M\big)\big|\mathcal D_i^\Lambda\big|s_{zz}(f_\Lambda)\big\rangle}_{\rm (ii)}+\underbrace{\tfrac23\big\langle j_i\big(\tfrac {f_\Lambda}M\big)\big|\kappa_\Lambda\big|s_{zi}(Mf_\Lambda)\big\rangle}_{\rm (iii)} .
\end{equation}

Let us estimate the three terms for large $\Lambda$ \cite[Lemma~4.5, Proposition~4.6]{OpenAI:2026}. In the term (i) both states carry the wave function $f_\Lambda$. At each finite $\Lambda$ it vanishes at $M=0$, so by itself it has no tail. But it is equal to one down to $M\sim1/\Lambda$, and for widths $\sqrt\kappa$ between $1$ and $\Lambda$ the two states reproduce the tails that we found in Sec.~\ref{sec:spin} for a wave function with $f(0)=1$. Moreover, $\kappa_\Lambda$ acts as $\kappa$ on these widths, since the annulus then contains almost all of the energy. As in \eqref{eq:dd}, every logarithmic interval of widths contributes the same amount, and this range of widths grows with $\Lambda$, so
\[
{\rm (i)}=\tfrac13\,\|\mathbf d_l\|^2\log\Lambda+o(\log\Lambda) .
\]
In the term (ii) the squared norm of the bra is $\int|f_\Lambda/M|^2M\,\dd M\approx\log\Lambda$. The first moment $\mathcal D_i^\Lambda$ does not vanish exactly, because of the edges of the annulus, but on the part of the state whose width lies well inside the annulus it is small. The norm of $\mathcal D_i^\Lambda|s_{zz}(f_\Lambda)\rangle$ therefore grows more slowly than $\sqrt{\log\Lambda}$, and (ii) grows more slowly than $\log\Lambda$. In the term (iii) we have $\kappa_\Lambda\le\kappa$, and the wave function $Mf_\Lambda$ vanishes at $M=0$. The comparison also holds for the individual components of the states of $V$ and $T$, with a value of $\nu$ that depends on the component, and for $|s_{zi}\rangle$ it is $\nu=1$ \cite[eq.~(4.1)]{OpenAI:2026}. It then bounds the norm of $\kappa|s_{zi}(Mf_\Lambda)\rangle$ uniformly in $\Lambda$. With the norm of the bra, (iii) grows at most like $\sqrt{\log\Lambda}$. Dividing \eqref{eq:sumruleL} by $\log\Lambda$ and letting $\Lambda\to\infty$, we obtain $\mathbf d_l=0$.

\section{The $a$-anomaly and dilaton scattering}\label{app:lookback}
The non-perturbative arguments mentioned in the Introduction couple the theory to a background dilaton and ask whether forward dilaton scattering allows the $a$-anomaly to keep flowing at a scale-invariant point. Hartman, Mathys and Stergiou have recently expressed the change of $a$ through the null energy flux \cite{Hartman:2026pos}, which is the quantity that we measured in Lecture one. In this appendix we give the details behind Sec.~\ref{sec:lookback}. We first rewrite the sum rule of Hartman, Mathys and Stergiou in the language of Lecture one (Sec.~\ref{app:HMS}), and show that in a theory that is scale invariant but not conformal the rate at which $a$ keeps flowing is $\pi^3\|\mathbf d_l\|^2$ (Sec.~\ref{app:tailK}). We then compare this rate with the one defined by forward dilaton scattering (Secs.~\ref{app:dil} and \ref{app:BK}), and finally comment on the other non-perturbative arguments (Sec.~\ref{app:others}).

\subsection{The sum rule of Hartman, Mathys and Stergiou}\label{app:HMS}
Let us start with the state that Hartman, Mathys and Stergiou use,
\begin{equation}\label{eq:Thetaomega}
|\Theta_\omega\rangle=\int\dd^2y\int_{-\infty}^{\infty}\dd v\int_{-\infty}^{0}\dd u\ \ee^{-\ii\omega v+\epsilon u}\,\Theta(u,v,\vy)|\Om\rangle ,\qquad \omega>0 ,
\end{equation}
written here in our coordinates, with $\epsilon\to0$ at the end \cite{Hartman:2026pos}. The plane wave in $v$ fixes the null momentum $P^+=\omega$, the integral over $\vy$ sets the transverse momentum to zero, and the integral over $u$ runs over a half line only. The half-line integral produces a factor $-\ii/(p^--\ii\epsilon)$ in momentum space, and with $P^+=\omega$ and $\vec p=0$ we have $M^2=2\omega p^-$, so that
\[
|\Theta_\omega\rangle\ \propto\ \omega\,\frac{\Theta(p)}{p^2}|\Om\rangle\ \propto\ \omega\,L(p)|\Om\rangle .
\]
This is a state of the form of Sec.~\ref{sec:claim}, created by $L=\Box^{-1}\Theta$ with the wave function $f=1$. In particular $f(0)\neq0$, which is the case in which a tail makes the mean squared width diverge.

Decomposing $|\Theta_\omega\rangle$ according to the centroid $Y$ of its energy flux defines the impact-parameter density $\mathcal E(\vec b)$, which Hartman, Mathys and Stergiou call $\rho(\vec b)$ \cite[eqs.~(2.19)--(2.21)]{Hartman:2026pos},
\begin{equation}\label{eq:rhodef}
\langle\Theta_\omega|\,E(\vec 0)\,\delta^{(2)}\big(P^+\vec Y-\omega\vec b\big)\,|\Theta_{\omega'}\rangle=2\pi\delta(\omega-\omega')\,\mathcal E(\vec b) .
\end{equation}
On these states $P^+=\omega$, so the delta function puts the centroid at $\vec b$, and $\mathcal E(\vec b)$ is the energy flux through the origin in the part of the state whose centroid sits at $\vec b$. It is not negative, since $E(\vec 0)\ge0$ commutes with $Y$. Since the state is invariant under transverse translations, $\mathcal E(\vec b)$ is the energy flux at the distance $|\vec b|$ from the centroid, and its second moment is the expectation value of
\[
\int\dd^2y\,|\vy-Y|^2E(\vy)=K^+-P^+|Y|^2=P^+\kappa ,
\]
where we used the moments \eqref{eq:moments} and the definition \eqref{eq:kappa} of $\kappa$. For a renormalization group flow between two CFTs, Hartman, Mathys and Stergiou show that
\begin{equation}\label{eq:HMS}
\boxed{\ \Delta a=\frac1{32}\int\dd^2b\,|\vec b|^2\,\mathcal E(\vec b)\ }
\end{equation}
The integral on the right-hand side is thus $\langle P^+\kappa\rangle$ in the state of $L$, in the normalization of $\mathcal E$ used by Hartman, Mathys and Stergiou. The $a$-theorem $\Delta a\ge0$ follows from $E(\vy)\ge0$, as Hartman and Mathys first showed in a different form \cite{Hartman:2023qdn}. The positivity of $\mathcal E$ at each $\vec b$ is a much stronger statement. If we restrict the integral to $|\vec b|<\mu^{-1}$, we obtain the function
\[
a_{\rm HMS}(\mu)=a_{\rm UV}-\frac1{32}\int_{|\vec b|<1/\mu}\dd^2b\,|\vec b|^2\mathcal E(\vec b) ,
\]
which decreases monotonically along the flow \cite{Hartman:2026pos}.

\subsection{Scale invariance and the tail}\label{app:tailK}
In a theory that is scale invariant but not conformal there is no scale, so $\mathcal E(\vec b)$ must be a pure power of $|\vec b|$. Since the integral in \eqref{eq:HMS} is dimensionless, the power is fixed,
\begin{equation}\label{eq:Kdef}
\mathcal E(\vec b)=\frac{\mathcal K}{|\vec b|^4},\qquad\mathcal K\ge0 .
\end{equation}
Each logarithmic interval of distances then contributes the same amount to \eqref{eq:HMS},
\begin{equation}\label{eq:aslope}
a_{\rm HMS}(\mu_2)-a_{\rm HMS}(\mu_1)=\frac1{32}\int_{1/\mu_2<|\vec b|<1/\mu_1}\dd^2b\,\frac{\mathcal K}{|\vec b|^2}=\frac{\pi\mathcal K}{16}\,\log\frac{\mu_2}{\mu_1} ,
\end{equation}
so $a_{\rm HMS}$ keeps changing at a constant rate. An improvement $T_{\mu\nu}\to T_{\mu\nu}+(\partial_\mu\partial_\nu-\eta_{\mu\nu}\Box)L'$ with a local scalar $L'$ changes $\mathcal E(\vec b)$ only at $\vec b=0$ \cite{Hartman:2026pos}, so it does not change $\mathcal K$. In a CFT, where some improvement makes $\Theta$ vanish, we therefore have $\mathcal K=0$.

The $1/b^4$ profile is the tail of Sec.~\ref{sec:tailL} seen in the energy flux. Let us take $f=1$, as in \eqref{eq:Thetaomega}. By \eqref{eq:tail} the squared norm of the part of $|l\rangle$ with a width between $\lambda$ and $\lambda+\dd\lambda$ is, at large $\lambda$,
\[
-\frac{\dd}{\dd\lambda}\big\|\mathcal P_{\lambda^2}|l\rangle\big\|^2\,\dd\lambda=\|\mathbf d_l\|^2\,\frac{\dd\lambda}{\lambda^3} .
\]
Since $\kappa$ commutes with the energy flux, we can measure the energy in each of these parts separately. By scale invariance the part with width $\lambda$ distributes its energy around the centroid with a shape that depends only on $\vec w=(\vy-Y)/\lambda$, and the second moment of this shape is $\langle|\vec w|^2\rangle=1$ by the definition of $\kappa$. Let us measure the energy with a radial weight $\Phi(b)$ that vanishes near $b=0$. Adding up the parts, we find per unit null momentum
\[
\|\mathbf d_l\|^2\int_0^\infty\frac{\dd\lambda}{\lambda^3}\,\big\langle\Phi(\lambda|\vec w|)\big\rangle=\|\mathbf d_l\|^2\,\big\langle|\vec w|^2\big\rangle\int_0^\infty\frac{\dd b}{b^3}\,\Phi(b)=\int\dd^2b\,\Phi(|\vec b|)\,\frac{\|\mathbf d_l\|^2}{2\pi|\vec b|^4} ,
\]
where we substituted $b=\lambda|\vec w|$. The right-hand side is the energy measured with the weight $\Phi$ in the profile $\|\mathbf d_l\|^2/(2\pi b^4)$, so the energy falls off as $1/b^4$ with $\mathcal K\propto\|\mathbf d_l\|^2$. To fix the constant we need the normalization of \eqref{eq:Thetaomega}. In the conventions of Hartman, Mathys and Stergiou the state \eqref{eq:Thetaomega} is, up to a phase, $2(2\pi)^2\omega$ times the state of $L$ with $f=1$, at $P^+=\omega$ and $\vec p=0$. After the momentum delta function is removed, $\mathcal E$ is $[2(2\pi)^2]^2/(2\pi)=4(2\pi)^3$ times the profile above, and
\begin{equation}\label{eq:adl}
\mathcal K=16\pi^2\|\mathbf d_l\|^2,\qquad \frac{\dd a_{\rm HMS}}{\dd\log\mu}=\frac{\pi\mathcal K}{16}=\pi^3\|\mathbf d_l\|^2 .
\end{equation}
This is the proportionality that we quoted in Sec.~\ref{sec:lookback}, now with its coefficient. Since we have shown $\mathbf d_l=0$, $a$ does not keep flowing.

\subsection{Dilaton scattering}\label{app:dil}
The constant flow of $a$ also appears in dilaton scattering. Following Komargodski and Schwimmer \cite{Komargodski:2011vj}, let us couple the theory to a background metric $g_{\mu\nu}$ and a dilaton $\tau$ through
\[
\hat g_{\mu\nu}=\ee^{-2\tau}g_{\mu\nu} ,
\]
with the decay constant $f_\tau$, and let $A(s,t)$ be the amplitude of dilaton scattering at leading order in $1/f_\tau$. At exact scale invariance the absorptive part of the forward amplitude is fixed up to a constant,
\begin{equation}\label{eq:Bdef}
\mathrm{Im}\,A(s+\ii0,0)=\mathcal B\,\frac{s^2}{f_\tau^4},\qquad s>0 ,
\end{equation}
with $\mathcal B\ge0$ by the optical theorem. The dispersive integral for $a$ \cite{Komargodski:2011vj,Luty:2012ww} runs logarithmically,
\[
a_{\rm KS}(\mu_2)-a_{\rm KS}(\mu_1)=\frac{f_\tau^4}{\pi}\int_{\mu_1^2}^{\mu_2^2}\frac{\dd s}{s^3}\,\mathrm{Im}\,A(s+\ii0,0)=\frac{2\mathcal B}\pi\log\frac{\mu_2}{\mu_1} .
\]
Thus $\mathcal B$ and $\mathcal K$ both measure how fast $a$ keeps flowing, one with dilaton scattering and one with the null energy. In the next subsection we relate them.

\subsection{Relating the two rates}\label{app:BK}
We now show that, under three assumptions stated below, the two rates agree, $\mathcal B=\pi^2\mathcal K/32$. The strategy is to write both numbers as the response of a dimensionless function to a change of scale, and then to observe that the two responses are two vertices of one local functional, of which only the Euler term of Komargodski and Schwimmer contributes. Let us first write both numbers as coefficients of logarithms. For the dilaton we divide the forward amplitude by its natural scaling,
\[
\mathcal F(z)=\frac{f_\tau^4\,A(z,0)}{z^2} ,
\]
where $z$ is the complex Mandelstam variable. By \eqref{eq:Bdef} its discontinuity on the cut $z=s>0$ is $\mathrm{Im}\,\mathcal F(s+\ii0)=\mathcal B$, and by crossing the cut at $z<0$ has the opposite discontinuity. Since $\log[-(s+\ii0)^2]=\log s^2-\ii\pi$, this gives
\[
\mathcal F(z)=-\frac{\mathcal B}{\pi}\,\log\frac{-z^2}{\mu_0^4}+{\rm const}
\]
with a reference mass $\mu_0$. For the energy flux we use the representation of $\mathcal E$ by a time-ordered three-point function \cite[eqs.~(2.27)--(2.31)]{Hartman:2026pos}. Let
\begin{equation}\label{eq:GTTT}
G_{\Theta\Theta T}(\vec q)=\lim_{k_1^2,\,k_2^2\to0^-}\ \frac1{\omega^2}\,\big\langle\!\big\langle\Theta(k_1)\,\Theta(k_2)\,E(k_3)\big\rangle\!\big\rangle_{\rm T},\qquad
k_1^+=-k_2^+=\omega,\quad \vec k_1+\vec k_2=\vec q ,
\end{equation}
where $\langle\!\langle\cdots\rangle\!\rangle_{\rm T}$ is the time-ordered correlation function in momentum space with the delta function $(2\pi)^4\delta^{(4)}(k_1+k_2+k_3)$ removed, and $E(k_3)=\int\dd^4x\,\ee^{\ii k_3\cdot x}\,T_{vv}(x)$ at $k_3^+=0$ is the light-ray operator with the transverse momentum $\vec k_3=-\vec q$. The two momenta $k_1$ and $k_2$ of $\Theta$ carry the null momentum of the states in \eqref{eq:rhodef}. The limit makes them null from spacelike values. Since $k_1^++k_2^+=0$, we have $(k_1+k_2)^2=2(k_1+k_2)^+(k_1+k_2)^--|\vec q|^2=-|\vec q|^2$. The transverse Fourier transform of $G_{\Theta\Theta T}$ gives the density,
\[
\int\frac{\dd^2q}{(2\pi)^2}\,\ee^{-\ii\vec q\cdot\vec b}\,G_{\Theta\Theta T}(\vec q)=-\mathcal E(\vec b)=-\frac{\mathcal K}{|\vec b|^4}\qquad(\vec b\neq0) .
\]
Away from the origin the Fourier transform of $|\vec q|^2\log|\vec q|^2$ is $4/(\pi b^4)$.\footnote{The Fourier transform of $|\vec q|^{2\alpha}$ is $4^\alpha\Gamma(1+\alpha)/\big(\pi\Gamma(-\alpha)\,b^{2+2\alpha}\big)$ away from the origin. It vanishes at $\alpha=1$, and its derivative there is $4/(\pi b^4)$.} The dimensionless function $\mathcal H=G_{\Theta\Theta T}/|\vec q|^2$ is therefore
\[
\mathcal H(|\vec q|^2)=-\frac{\pi\mathcal K}{4}\,\log\frac{|\vec q|^2}{\mu_0^2}+{\rm const} .
\]
If we rescale all momenta by a factor $\Lambda$, the constants drop out, and
\begin{equation}\label{eq:FHslopes}
\Lambda\frac{\partial\mathcal F}{\partial\Lambda}=-\frac{4\mathcal B}{\pi},\qquad\Lambda\frac{\partial\mathcal H}{\partial\Lambda}=-\frac{\pi\mathcal K}{2} .
\end{equation}
The two rates are thus the responses of $\mathcal F$ and $\mathcal H$ to a rigid change of scale.

These two responses are coefficients of a single functional. Let $\mathcal W[g,\tau]$ be the generating functional of the theory coupled to $g_{\mu\nu}=\eta_{\mu\nu}+h_{\mu\nu}$ and $\tau$. At linear order $\tau$ couples to $\Theta$ and $h_{\mu\nu}$ to $\tfrac12T^{\mu\nu}$, so the variations of $\mathcal W$ at $h_{\mu\nu}=\tau=0$ are the time-ordered correlation functions of $\Theta$ and $T_{\mu\nu}$, including contact terms. Up to constant factors, the fourth variation in $\tau$ is the dilaton amplitude, which in forward kinematics is $s^2\mathcal F/f_\tau^4$, and the variation in $\tau$, $\tau$ and $h_{\mu\nu}=n_\mu n_\nu h$ is the correlation function \eqref{eq:GTTT}, which is $\omega^2|\vec q|^2\mathcal H$. Let us now rescale the metric by a constant at fixed $\tau$ and define
\[
\mathcal R[g,\tau]=\frac{\dd}{\dd \ell}\,\mathcal W\big[\ee^{-2\ell}g,\tau\big]\Big|_{\ell=0} ,
\]
of which we keep the terms with four derivatives. This rescales $\hat g_{\mu\nu}$ by the same factor. A constant rescaling of the metric is a rescaling of all lengths by $\ee^{-\ell}$, and hence of all momenta by $\Lambda=\ee^{\ell}$. The variations of $\mathcal R$ are therefore the scale derivatives $\Lambda\partial_\Lambda$ of the correlation functions. A term with four derivatives and no logarithm does not change under this rescaling, since $\int\dd^4x$ with four derivatives is dimensionless, so only the logarithms contribute. The four-dilaton variation of $\mathcal R$ is thus $(s^2/f_\tau^4)\,\Lambda\partial_\Lambda\mathcal F$, and its variation in $\tau$, $\tau$ and $h$ is $\omega^2|\vec q|^2\,\Lambda\partial_\Lambda\mathcal H$. We call the coefficients $\Lambda\partial_\Lambda\mathcal F$ and $\Lambda\partial_\Lambda\mathcal H$ the two vertices of $\mathcal R$. The correlation functions themselves are not local, since they contain logarithms of the momenta, but their scale derivatives can be, since the scale derivative of a logarithm is a constant. To relate the two vertices we make three assumptions.
\begin{itemize}
\item[1.] $\mathcal R$ is a local functional of $g_{\mu\nu}$ and $\tau$, and its variations, including their contact terms, are the scale derivatives of the correlation functions and of the dilaton amplitude.
\item[2.] Under the simultaneous change $\delta_\sigma g_{\mu\nu}=2\sigma g_{\mu\nu}$, $\delta_\sigma\tau=\sigma$, which leaves $\hat g_{\mu\nu}$ unchanged,
\begin{equation}\label{eq:redund}
\frac{\delta}{\delta\tau(x)}\,\delta_\sigma\mathcal R=0\qquad\text{for every smooth }\sigma\text{ of compact support} .
\end{equation}
This expresses that the simultaneous change is a redundancy of the sources, and it does not require the theory to be Weyl invariant.
\item[3.] The scale derivative commutes with the null limit in \eqref{eq:GTTT} and with the forward limit of the amplitude, in which the Mandelstam variable $t$ goes to zero.
\end{itemize}
By the first two assumptions, $\delta\mathcal R/\delta\tau(x)$ is a local density built from $\hat g_{\mu\nu}$ alone. Among the parity-even densities with four derivatives this leaves, up to boundary terms,
\[
\mathcal R=r_E\,S_E[g,\tau]+r_C\int\sqrt g\,\tau\,C^2+r_R\int\sqrt{\hat g}\,\hat R^2+\mathcal R_0[g] ,
\]
where $C$ is the Weyl tensor of $g_{\mu\nu}$, $\hat R$ the scalar curvature of $\hat g_{\mu\nu}$, $\mathcal R_0[g]$ contains no dilaton, and $S_E$ is the Euler term of Komargodski and Schwimmer,
\[
S_E=-\int\sqrt g\,\Big(\tau E_4+4G^{\mu\nu}\partial_\mu\tau\,\partial_\nu\tau-4(\partial\tau)^2\Box\tau+2(\partial\tau)^4\Big) ,
\]
with the Euler density $E_4$ and the Einstein tensor $G^{\mu\nu}$ of $g_{\mu\nu}$. Its change under the simultaneous rescaling,
\[
\delta_\sigma S_E=-\int\sqrt g\,\sigma\,E_4(g) ,
\]
does not depend on $\tau$, so \eqref{eq:redund} allows $S_E$ with an arbitrary coefficient $r_E$.

Only the Euler term contributes to the two vertices. For on-shell dilatons in flat space, $\Box\,\ee^{-\tau}=0$, the curvature $\hat R=-6\,\ee^{3\tau}\Box\,\ee^{-\tau}$ vanishes, so neither $\int\sqrt{\hat g}\,\hat R^2$ nor its first variation in $h_{\mu\nu}$ contributes. The term $\tau C^2$ is of second order in $h_{\mu\nu}$, and $\mathcal R_0[g]$ contains no dilaton. In the Euler term we use $\Box\tau=(\partial\tau)^2$, which is the on-shell condition, and $G^{\mu\nu}=-\tfrac12\Box h^{\mu\nu}$ at first order in a transverse traceless $h_{\mu\nu}$. To the order that we need,
\[
r_ES_E\ \supset\ 2r_E\int(\partial\tau)^4+2r_E\int(\Box h^{\mu\nu})\,\partial_\mu\tau\,\partial_\nu\tau .
\]
For the canonical dilaton $\sqrt2f_\tau\tau$ the first term gives the amplitude $r_E(s^2+t^2+u^2)/f_\tau^4$, as in Komargodski and Schwimmer \cite{Komargodski:2011vj}, which at $t=0$ is $2r_Es^2/f_\tau^4$. Since $\mathcal R$ is the scale derivative of $\mathcal W$, this is the scale derivative of the forward amplitude $s^2\mathcal F/f_\tau^4$, and hence $\Lambda\partial_\Lambda\mathcal F=2r_E$. With the normalization of $G_{\Theta\Theta T}$ the second term gives $\Lambda\partial_\Lambda\mathcal H=8r_E$. As a check, in a CFT the generating functional contains $-aS_E$, and the same projection gives the contact term $\mathcal H=-8a$ found by Hartman and Mathys \cite{Hartman:2023qdn}. The coefficient $r_E$ drops out of the ratio, $\Lambda\partial_\Lambda\mathcal H=4\,\Lambda\partial_\Lambda\mathcal F$, and with \eqref{eq:FHslopes} we obtain
\begin{equation}\label{eq:BK}
\mathcal B=\frac{\pi^2}{32}\,\mathcal K,\qquad\frac{\dd a_{\rm KS}}{\dd\log\mu}=\frac{\dd a_{\rm HMS}}{\dd\log\mu}=\frac{\pi\mathcal K}{16} .
\end{equation}

The absorptive part of forward dilaton scattering, the $1/b^4$ profile of the null energy and the tail amplitude $\|\mathbf d_l\|^2$ of the squared width are thus proportional to one another. The assumptions do not determine whether this number vanishes, since $r_E$ is arbitrary. What would a non-zero $r_E$ mean? Since $\mathcal R$ is the scale derivative of $\mathcal W$, a term $r_ES_E$ in $\mathcal R$ means that the coefficient of the Euler term in $\mathcal W$, which in a CFT is the constant $-a$, depends logarithmically on the scale. Comparing \eqref{eq:FHslopes} with the dispersive integral of Sec.~\ref{app:dil}, we find $r_E=-\dd a_{\rm KS}/\dd\log\mu$. On shell and in flat space $S_E$ reduces to $2\int(\partial\tau)^4$, so in the amplitude this is a logarithm $(\partial\tau)^4\log\mu$.

Luty, Polchinski and Rattazzi exclude such a logarithm by a stronger requirement \cite{Luty:2012ww}. Since $\Theta=\partial^\mu V_\mu$, a local rescaling of the metric can be compensated by a shift of a source $C_\mu$ of the virial current, $C_\mu\to C_\mu+\partial_\mu\sigma$. They require that under this combined transformation the generating functional changes only by a local anomaly that satisfies the Wess--Zumino consistency condition. With the dilaton in the metric $\hat g_{\mu\nu}$ and $C_\mu=0$, a constant rescaling then gives a local functional of $\hat g_{\mu\nu}$ alone,
\[
\mathcal R=\int\sqrt{\hat g}\,\big(c_1\,\hat C^2+c_2\,\hat R^2+c_3\,\hat E_4\big)
\]
with constants $c_i$, where the last term is a total derivative. So $\delta_\sigma\mathcal R=0$, the Euler term $S_E$ is excluded, and for on-shell dilatons in flat space, where $\hat C=0$ and $\hat R=0$, nothing contributes to the two vertices. Hence $\mathcal B=\mathcal K=0$. That is, the dilaton enters only as the conformal factor of a metric, and in a curved background scale invariance is violated only by the usual geometric anomaly. In a logarithm $(\partial\tau)^4\log\mu$, on the other hand, $\partial_\mu\tau$ would appear as an independent source of the virial current, and not as part of a metric.

Note that the logic here is the reverse of that in Lecture one. Luty, Polchinski and Rattazzi put in the geometric form of the anomaly and obtain $\mathcal B=0$. In Lecture one we obtained $\mathbf d_l=0$, and hence $\mathcal K=0$, from causality and the positivity of energy, without a dilaton. Under our three assumptions this gives $\mathcal B=0$ and $r_E=0$, so what they require comes out as a result, as far as the two vertices are concerned. Once $L$ has been constructed as a local field, $\Theta=\Box L$, the improved theory is conformal, and the requirement holds without these assumptions.

Of the three assumptions the third is the most delicate. The vertices of $\mathcal R$ are scale derivatives taken before the limits, while $\mathcal K$ and $\mathcal B$ are read off from $\mathcal H$ and $\mathcal F$ after the limits, and the two orders can differ. Let $k_1$ be the momentum of one dilaton, and suppose that away from the null limit $\mathcal H$ contains a term $\log(|\vec q|^2/|k_1^2|)$. This term depends only on a ratio of momenta, so its scale derivative vanishes, and it does not contribute to the vertex of $\mathcal R$. In the null limit $k_1^2\to0$, however, it diverges, and to define $\mathcal H$ we have to subtract the divergence at a fixed scale $\mu_0$,
\[
\log\frac{|\vec q|^2}{|k_1^2|}=\log\frac{|\vec q|^2}{\mu_0^2}-\log\frac{|k_1^2|}{\mu_0^2} ,
\]
where we drop the last term. The remaining term $\log(|\vec q|^2/\mu_0^2)$ has the scale derivative $2$, so it contributes to $\mathcal K$. In this example the scale derivative and the null limit do not commute, so $\mathcal K$ differs from the value that the vertex of $\mathcal R$ predicts. The third assumption excludes such terms. The same question arises for the forward limit of $\mathcal F$.

\subsection{The other arguments}\label{app:others}
We can read the other non-perturbative arguments in the same way. Dymarsky, Komargodski, Schwimmer and Theisen argued from unitarity that the forward dilaton amplitudes, and with them infinitely many matrix elements of $\Theta$, vanish \cite{Dymarsky:2013pqa}. From this vanishing Yonekura constructed a local $L$ with $\Theta=\Box L$ \cite{Yonekura:2014tha}. Since at linear order the dilaton couples only to $\Theta$, with these arguments we probe only correlation functions of $\Theta$ itself, and we learn nothing if $\Theta$ is a generalized free field. Dymarsky, Farnsworth, Komargodski, Luty and Prilepina closed this loophole with the anomaly of global scale transformations \cite{Dymarsky:2014zja}. They showed that the anomaly cannot be reproduced if $\Theta$ is a generalized free field, unless the theory has a scalar of dimension two. If it has one, we can use it to improve the energy-momentum tensor, and at least one of the improved tensors is not a generalized free field. We met scalars of dimension two and their improvement terms in Lecture two, where the partner $X$ of the energy-momentum tensor is such a scalar.

With the OpenAI argument of Lecture one we showed $\mathbf d_l=0$, and hence $\mathcal K=0$, with the causality of the energy-momentum tensor, $[T_{\mu\nu}(x),T_{\rho\sigma}(y)]=0$ at spacelike separation, and the positivity of energy as the main physical inputs. We needed no dilaton, no flow and no hypothesis on the forward limits of dilaton amplitudes. The loophole does not arise either, since we used three-point functions with one insertion of $T$, whose lowest integrated moments the Ward--Takahashi identities fix in terms of two-point functions. For a state $|\Theta\rangle$ that $\Theta$ creates from the vacuum, for example,
\[
\int\dd^2y\,\langle\Theta|E(\vy)|\Theta\rangle=\langle\Theta|P^+|\Theta\rangle ,
\]
which is positive unless $|\Theta\rangle=0$. So $\Theta$ cannot decouple from the energy flux, whatever its own correlation functions are.

\section{The single insertion of $L$}\label{sec:insert}
In Step 3 of Lecture two we used one input that I did not explain there. To obtain \eqref{eq:Kpp}, $\dd^2K^+/\dd u_0^2=4P^-$, we wrote $\Theta=\Box L$ inside matrix elements between the states $|X\rangle$, so that $\Theta$ integrates to zero over the null plane. In momentum space this integral sets the momentum $k$ that the operator transfers to $k^+=0$ and $\vec k=0$, so that $k^2=0$, and the matrix element of $\Theta=-k^2L$ vanishes there only if that of $L$ has no pole at $k^2=0$. A non-local operator such as $\Box^{-1}\phi^4$ of the free-field example has such a pole between excited states. In this appendix I explain why the two terms of $\Box L$ integrate to zero in \eqref{eq:Kpp}.

First, we recall from the OpenAI paper that correlation functions with one insertion of $L$ exist and are local as soon as we can separate the vector $L(h)|\Om\rangle$ from the other fields by a wedge, in a sense that we define below. Second, when $T_{\mu\nu}$ and $V_\mu$ have definite dimensions, this localization follows from the absence of the tail, $\mathbf d_l=0$. Third, in Lecture two we cannot assume definite dimensions, and $|l\rangle$ may contain a logarithm. There we obtain the localization of $|l\rangle$ from that of the state of $T_{vv}$ and of its spin-two part $|s_{zz}\rangle$, and for the latter we show that its tail vanishes by the sum rule of Lecture one.

\paragraph{One insertion of $L$ from a localized vector.} We say that a state is localized in a region if it can be approximated by states $A|\Om\rangle$ with Hermitian operators $A$ localized in this region. A Hermitian local field smeared with a real test function supported in the region creates such a state. Because of the restriction to Hermitian operators, the localized states form only a closed real subspace, the standard subspace of the region, which is not closed under multiplication by $\ii$. By the Reeh--Schlieder theorem its complex span is dense, but the subspace itself is not, so localization in this sense is a non-trivial property of a vector.

Suppose first that $L$ were a local field. Then it would be determined by its vector $|l\rangle$. Indeed, if we place another operator $A$ in a region spacelike to the support of $h$, then $L(h)$ commutes with $A$, and $L(h)A|\Om\rangle=A\,L(h)|\Om\rangle$. By the Reeh--Schlieder theorem the states $A|\Om\rangle$ with $A$ in such a region are dense, so the vector $L(h)|\Om\rangle$ fixes the action of $L(h)$. 

Conversely, suppose that $L(h)|\Om\rangle$ is localized in a wedge $W_g$ that separates it from the other fields in a correlation function. Here $W_g$ is the wedge $u<0<v$ with its edge moved to $v=g(\vy)$, which we will meet in Sec.~\ref{sec:edge}. Then the same rule defines the correlation function with $L(h)$ next to the vacuum, at either end of the product, and the two definitions agree. The wave equation $\Box L=\Theta$ then determines the correlation functions with $L$ in the middle of the product. The OpenAI paper carries this out with the retarded and the advanced fundamental solutions of the wave equation \cite[Proposition~5.3]{OpenAI:2026}, and Astra's manuscript adapts it to the case with a Jordan block \cite[Proposition~9.1]{Astra:2026}. So to construct one insertion of $L$, it is enough to separate the state $L(h)|\Om\rangle$ from the other insertions by a wedge.

\paragraph{Definite dimensions.} We therefore have to show that $L(h)|\Om\rangle$ is localized in such wedges. When $T$ and $V$ have definite dimensions, as in Lecture one, this follows from the absence of the tail. With $\mathbf d_l=0$ the state of $L$ has the property of the Claim: for a wave function whose Fourier transform is supported in a disk of radius $R$, its width is never larger than $R$, just as for the state $L_{\rm ref}(h)|\Om\rangle$ of the generalized free field $L_{\rm ref}$ of dimension two with the same two-point function. The latter state has the same inner products as $L(h)|\Om\rangle$, and it is localized in every region that contains the support of $h$. Lemma~5.1 of the OpenAI paper turns the bound on the width into a statement about localization \cite[Lemma~5.1]{OpenAI:2026}. Roughly speaking, the width enters the modular Hamiltonian of a region $W_g$ with a parabolic edge as in Appendix~\ref{sec:entered}, and replacing the width of the reference state by the smaller physical one amounts to translating the state in the positive $v$ direction. Since such translations map these regions into themselves, $L(h)|\Om\rangle$ is localized in the wedges $W_g$ that we need to separate it from the other fields. If the dilatation is diagonal, this is all we need. With a Jordan block, $|l\rangle$ may contain $\log M\,|X\rangle$, the Claim is not available for it, and we cannot proceed in the same way.

\paragraph{With a Jordan block.} In this case we obtain the localization of $|l\rangle$ from that of the other components. Let us take one of the wedges $W_g$ that we use to separate the state from the other fields. The field $T_{vv}$ is local, so the state $|c_T\rangle=\tfrac13|l\rangle+|s_{zz}\rangle$ that it creates is localized in $W_g$. If $|s_{zz}\rangle$ is localized in the same region, then so is $|l\rangle=3(|c_T\rangle-|s_{zz}\rangle)$, since the coefficients are real. By \eqref{eq:NV} the partner of $T$ has no spin-two part, so $|s_{zz}\rangle$ has no logarithm, and once we know that its tail vanishes, it is localized in $W_g$ \cite{Astra:2026}. Astra's manuscript shows that this tail vanishes without the assumptions (A) and (B), by a long computation with the three-point function of $T_{\mu\nu}$. Under these assumptions we can use the sum rule of Lecture one instead, as follows.

A local field that creates only the spin-two component is the $vv$ component of
\[
\mathcal G_{\mu\nu}=-\Box T_{\mu\nu}-\tfrac13(\partial_\mu\partial_\nu-\eta_{\mu\nu}\Box)\Theta .
\]
By \eqref{eq:NV}, and since the partner of $\Theta$ is $-3\Box X$, its partner is
\[
N\mathcal G_{\mu\nu}=-\Box(\partial_\mu\partial_\nu-\eta_{\mu\nu}\Box)X+(\partial_\mu\partial_\nu-\eta_{\mu\nu}\Box)\Box X=0 ,
\]
so $\mathcal G_{\mu\nu}$ has a definite dimension. Its component $\mathcal G_{vv}$ creates the state $(p^+)^2M^2|s_{zz}\rangle$, and the comparison \eqref{eq:compare} with $\nu=2$ applies to it \cite{Astra:2026}. The argument of Sec.~\ref{sec:tailL} then goes through as for $|l\rangle$ (Fig.~\ref{fig:tail}). For $t>R^2$ the vector $\mathcal P_t|s_{zz}(f)\rangle$ depends on $f$ only through $f(0)$, and
\[
\|\mathcal P_t|s_{zz}(f)\rangle\|^2=\frac{|f(0)|^2\,\|\mathbf d_s\|^2}{2t}\qquad(t>R^2)
\]
with a constant $\|\mathbf d_s\|^2$, the squared tail amplitude of $|s_{zz}\rangle$. We have to show $\mathbf d_s=0$.

To show this, let us first look at the field $-\partial_vV_v-3T_{vv}$. By \eqref{eq:NV} and $\eta_{vv}=0$, its partner is
\[
N\big(-\partial_vV_v-3T_{vv}\big)=-\partial_v\big(-3\partial_vX\big)-3\,\partial_v^2X=0 .
\]
The field $-\partial_vV_v-3T_{vv}$ thus has a definite dimension. On the vacuum $-\partial_v$ acts as $-\ii p^+$, so the states created by $V_v$ and $T_{vv}$ combine into
\[
(p^+)^2\big(|l\rangle+|j_z\rangle\big)-3\,(p^+)^2\big(\tfrac13|l\rangle+|s_{zz}\rangle\big)=(p^+)^2\big(|j_z\rangle-3|s_{zz}\rangle\big).
\]
The spin-zero component $|l\rangle$, which has the logarithm, cancels. Since the field has a definite dimension, we may apply the Claim \eqref{eq:claim} to it, so its state has no tail. Therefore
\[
\mathcal P_t|j_z(f)\rangle=3\,\mathcal P_t|s_{zz}(f)\rangle\qquad(t>R^2) .
\]
At large widths the two states are thus multiples of one vector, and their interference is a squared norm, $\langle j_z(f)|\mathcal P_t|s_{zz}(f)\rangle=3\|\mathcal P_t|s_{zz}(f)\rangle\|^2$.

Now we use the sum rule with a cutoff, \eqref{eq:sumruleL} of Appendix~\ref{app:sumrule}. In Lecture one its term (i) grew like $\log\Lambda$ times a positive multiple of $\|\mathbf d_l\|^2$, while the terms (ii) and (iii) grew more slowly, (ii) because $\mathcal D_i^\Lambda$ would vanish without the cutoff, and (iii) because the wave function $Mf_\Lambda$ vanishes at $M=0$.

The derivation of the sum rule in Appendix~\ref{app:sumrule} used the key identity \eqref{eq:key}, the decomposition \eqref{eq:spin} and the form \eqref{eq:chi} of the state created by $\bar E_+$. None of these depends on how $|l\rangle$ depends on $M$, so the sum rule holds here as well. Moreover, the component $|l\rangle$ with the logarithm does not appear in it, since the vector $p^-n-p^+\bar n$ removes the spin-zero part of the bra and only $|s_{zz}\rangle$ and $|s_{zi}\rangle$ survive in the ket. So \eqref{eq:sumruleL} involves only the spin-one and spin-two states, and neither of them has a logarithm. For $|s_{\mu\nu}\rangle$ this follows from \eqref{eq:NV}. In $|j_\mu\rangle$ of \eqref{eq:spin} the logarithms of $V_\mu|\Om\rangle$ and of $|l\rangle$ cancel, since $NV_\mu=-3\partial_\mu X$ and the partner of $|l\rangle$ is $-3|X\rangle$. Since we know the tails of $|j_z\rangle$ and $|s_{zz}\rangle$, we can estimate the three terms of \eqref{eq:sumruleL} as before.

The last term involves the other polarization $|s_{zi}\rangle$, and its estimate in the OpenAI paper uses the comparison with $\nu=1$ for this component \cite[Lemma~4.5]{OpenAI:2026}. The local field $\mathcal S_i=\partial_vT_{vi}-\partial_iT_{vv}$ creates only this component. In momentum space it contracts $T_{\mu\nu}$ with $n^\mu$ and with $w^\nu=p^+\hat y_i^\nu-p_in^\nu$, where $\hat y_i$ is the unit vector along $y^i$. The vector $w$ is orthogonal to $p$ and to $n$. The spin-zero part therefore drops out, and in the rest frame of $p$ the vector $w$ is spatial and orthogonal to $e_z$, so that only the components $|s_{zi}\rangle$ remain. By \eqref{eq:NV} the partner of $\mathcal S_i$ is $\partial_v^2\partial_iX-\partial_i\partial_v^2X=0$, so the comparison applies to it. The term (i) now grows like $3\|\mathbf d_s\|^2\log\Lambda$, up to a positive normalization, since both states carry $f_\Lambda$, which is constant down to $M\sim1/\Lambda$. The terms (ii) and (iii) again grow more slowly. The coefficient of $\log\Lambda$ must therefore vanish, and $\mathbf d_s=0$. With the spin-two tail gone we can subtract the spin-two part from the state created by $T_{vv}$, as explained above, and we obtain the localization of the scalar part that we need for the single insertion of $L$ in \eqref{eq:Kpp}.

\section{Two inputs from the thermal vacuum}\label{sec:entered}
In Lectures one and two we used two facts without proof. The first is the finiteness \eqref{eq:Kfinite} of the second moment of the energy flux,
\[
\langle\psi|K^+|\psi\rangle<\infty ,\qquad K^+=\int\dd^2y\,|\vy|^2E(\vy) ,
\]
for a state $|\psi\rangle$ that $T_{vv}$ creates with a test function supported in a bounded region. The second is the comparison \eqref{eq:compare} with the CFT,
\[
\mathrm{Prob}(\kappa>t)\ \le\ \mathrm{Prob}_{\rm CFT}(\kappa>t) ,
\]
for the states created by a local field of definite dimension. We will explain it for a scalar of dimension two. Neither of the two facts is obvious. Consider first the finiteness. Light from a bounded source reaches the null plane at every transverse position, so the energy can arrive arbitrarily far away, and it is not clear that the second moment converges. Moreover, in Lecture two we needed the finiteness for $T_{vv}$, which might have a logarithmic partner, so we cannot use any formula that relies on its scaling dimension. For the comparison the difficulty is that $K^+$ is not a symmetry, so we cannot compute the width of the energy distribution. In this appendix we see that both facts follow from the thermal property of the vacuum of Sec.~\ref{sec:thermal}, as does the positivity $E(\vy)\ge0$ that we also used (Sec.~\ref{sec:edge}). In both we consider a light cone cut by a null plane, whose modular Hamiltonian contains the second moment $K^+$, and use two general properties of modular Hamiltonians, their positivity on states created in the region and their monotonicity.

\paragraph{The cut cone.} Both arguments use one region, which we obtain by bending the edge of the wedge into a parabola. Let us recall from Sec.~\ref{sec:edge} that we can write \eqref{eq:Hg} as
\[
K_{W_g}=2\pi\int\dd v\,\dd^2y\,\big(v-g(\vy)\big)\,T_{vv}(u=0,v,\vy) .
\]
On each light ray the modular Hamiltonian of $W_g$ is the boost about the point $v=g(\vy)$ where the light ray crosses the edge, in the same way as the modular Hamiltonian $2\pi B$ of the wedge is the boost about $v=0$. In this form we see that the distance $g(\vy)$ by which we move the edge is the weight with which the null energy on the light ray at $\vy$ enters the modular Hamiltonian. We are interested in the second moment $K^+=\int\dd^2y\,|\vy|^2E(\vy)$, so let us move the edge on each light ray by an amount proportional to $|\vy|^2$, that is, let us bend it into the parabola $g(\vy)=c\,|\vy|^2$ with a constant $c>0$. Then $W_g$ is the future light cone of the point $u=-1/2c$, $v=0$, $\vy=0$, cut by the null plane $u=0$, as we saw in Sec.~\ref{sec:cones}. Let us call it $C_c$. For a small $c$ its tip is far away, and near a bounded region it looks like the wedge. If we could use the formula above for this $g$, the modular Hamiltonian of $C_c$ would be $2\pi(B-cK^+)$. Since $g$ is not bounded, we cannot use it directly. In Sec.~\ref{sec:cones} we quoted from Astra's manuscript that the Tomita relation with $B-cK^+$ still holds for the states prepared in $C_c$, but for this appendix we need less. Let us write
\[
K_{C_c}=2\pi\big(B-\hat K_c\big)
\]
with an operator $\hat K_c$ that we do not know in general. It plays the role of $cK^+$.

First, $\hat K_c\ge cK^+$. The cut cone is contained in $W_g$ for every bounded $g\le c\,|\vy|^2$, so by monotonicity $K_{C_c}\le K_{W_g}$, which by the formula above means $\hat K_c\ge\int\dd^2y\,g(\vy)E(\vy)$ for all such $g$, and hence $\hat K_c\ge cK^+$. Second, in a CFT $\hat K_c=cK^+$. There $K^+$ is a component of the special conformal generator (Sec.~\ref{sec:claim}), and it generates the conformal transformation that moves the points of the light ray at $\vy$ by $|\vy|^2$ in the $v$ direction. This transformation maps the wedge onto $C_c$, and the modular Hamiltonian $2\pi B$ of the wedge to $2\pi(B-cK^+)$.

\paragraph{Finiteness of the second moment.} Let us start with the finiteness, for which a classical argument is simple. Let $h$ be a real smooth test function supported in a bounded region on the side $u<0$. Since $K^+$ commutes with the translations in $v$, we may move the region into $C_c$ for a small enough $c$. By the definition of $W_g$, no light signal from $C_c$ reaches the part $v<c\,|\vy|^2$ of the null plane. So the energy that the source sends through the plane arrives on each light ray after the edge, and if the energy density were a positive classical quantity, we would find
\[
\langle B\rangle-c\,\langle K^+\rangle=\int\dd v\,\dd^2y\,\big(v-c\,|\vy|^2\big)\,\langle T_{vv}\rangle\ \ge\ 0 .
\]
Since $\langle B\rangle$ is finite, so would be $\langle K^+\rangle$. In quantum field theory this argument does not work as it stands, since the energy density can be negative and the state is not exactly empty before the edge. Modular theory provides the quantum version of the argument.

The quantum version rests on a positivity property of the modular Hamiltonian. Let us first see it in the system of two copies of Sec.~\ref{sec:thermal}, with $|\Om\rangle=Z^{-1/2}\sum_n\ee^{-E_n/2}|n\rangle\otimes|n\rangle$ and $K=H\otimes1-1\otimes H$. The modular energy $K$ is the energy of the first copy minus that of the second, and its mean in the thermal state vanishes. If we act on the first copy with a Hermitian operator, the mean of $K$ becomes non-negative. For a Hermitian operator $A$ acting on the first copy, $A|\Om\rangle=Z^{-1/2}\sum_{m,n}\ee^{-E_n/2}A_{mn}|m\rangle\otimes|n\rangle$, so
\[
\langle\Om|AKA|\Om\rangle=\frac1Z\sum_{m,n}\ee^{-E_n}|A_{mn}|^2(E_m-E_n)=\frac1{2Z}\sum_{m,n}|A_{mn}|^2(E_m-E_n)\big(\ee^{-E_n}-\ee^{-E_m}\big) ,
\]
where we used $|A_{mn}|=|A_{nm}|$ and symmetrized in $m$ and $n$. The two factors in the last expression have the same sign, so it is not negative. A Hermitian operator causes transitions up and down with equal amplitudes, and the lower states are more populated, so on average the energy of the first copy rises relative to that of the second. The positivity concerns only the difference $K$. If $A$ projects onto the ground state of the first copy, for example, the energy of the first copy decreases, but the second copy is projected onto its ground state as well, and $K$ vanishes. In general the same follows from the Tomita relation \eqref{eq:tomita}. For a Hermitian $A$ in $\mathcal M(\mathcal U)$ and $|\psi\rangle=A|\Om\rangle$ it says $\langle\psi|\ee^{-K_{\mathcal U}}|\psi\rangle=\langle\psi|\psi\rangle$, and with $\ee^{-x}\ge1-x$ we obtain
\begin{equation}\label{eq:passive}
\langle\psi|K_{\mathcal U}|\psi\rangle\ \ge\ 0 .
\end{equation}

Now let us apply \eqref{eq:passive} to $A=T_{vv}(h)$ and the region $C_c$. Since $T_{vv}$ is local, $A$ belongs to $\mathcal M(C_c)$, and \eqref{eq:passive} gives $\langle\psi|\hat K_c|\psi\rangle\le\langle\psi|B|\psi\rangle$. With $\hat K_c\ge cK^+$ we obtain
\begin{equation}\label{eq:Kbound}
\langle\psi|K^+|\psi\rangle\ \le\ \frac1c\,\langle\psi|B|\psi\rangle\ <\infty ,
\end{equation}
which is the quantum version of the classical inequality above. The right-hand side is finite, since the boost acts on $T_{vv}(h)$ only by deforming the smooth test function $h$ and through the tensor indices. No scaling dimension enters here, and this is why we could use \eqref{eq:Kfinite} for $T_{vv}$ in Lecture two. The same derivation applies to any Hermitian local field, and hence, by splitting a field into its Hermitian and anti-Hermitian parts, to any local field.

\paragraph{The comparison with the CFT.} Let us now turn to the comparison. We take $c=1$ and write $C=C_1$ and $\hat K=\hat K_1$, so that $K^+\le\hat K$, with equality in a CFT. It is then enough to show that in the states created by $O$ the operator $\hat K$ is not larger than in a CFT. Let $O$ be a Hermitian scalar of dimension two, and let $O_{\rm ref}$ be the generalized free field with the same two-point function \eqref{eq:wightman}, which is a CFT. Since the two-point functions agree, the states $O_{\rm ref}(h)|\Om\rangle$ and $O(h)|\Om\rangle$ have the same inner products, so we can identify them by the map
\[
\iota:\ O_{\rm ref}(h)|\Om\rangle\ \mapsto\ O(h)|\Om\rangle .
\]
This map commutes with the Poincar\'e transformations, and in particular with the boost $B$. Since $O$ is local, a state $O(h)|\Om\rangle$ with $h$ supported in $C$ is prepared in $C$. In the physical theory we can therefore prepare in $C$ all the states that $\iota$ brings from the reference theory, and possibly more, with the other fields of the theory. Monotonicity holds for the real subspaces of localized states as well as for regions, and a larger real subspace has a larger modular Hamiltonian. Let us denote by $\hat K_{\rm ref}$ and $K^+_{\rm ref}$ the operators $\hat K$ and $K^+$ of the reference theory, carried over to the physical Hilbert space by $\iota$. On these states we therefore have $B-\hat K\ge B-\hat K_{\rm ref}$, and since $\iota$ does not change $B$, we find $\hat K\le\hat K_{\rm ref}=K^+_{\rm ref}$, where the last equality holds because the reference theory is a CFT. Together with $K^+\le\hat K$, in the states created by $O$ we have
\[
K^+\ \le\ \hat K\ \le\ K^+_{\rm ref} .
\]

The OpenAI paper proves this chain in the stronger form that we need, for the probabilities of large values instead of mean values \cite[Lemma~2.3, Proposition~2.7]{OpenAI:2026}. On the null plane $P^+$ generates the translations in $v$ and $P^+Y^i$ the null rotations, so $\iota$ commutes with $P^+$ and with the centroid $Y$. We can therefore fix them, and with $K^+=P^+(|Y|^2+\kappa)$ we obtain the comparison \eqref{eq:compare}. We also see why the comparison is only an inequality. The other fields of the physical theory can make the modular Hamiltonian of $C$ larger than in the reference theory, and $\hat K$ correspondingly smaller. We needed the definite dimension of $O$ only to have a CFT with the same two-point function. In Lecture two and Appendix~\ref{sec:insert} we used the comparison for $X$, $\mathcal G_{vv}$, $-\partial_vV_v-3T_{vv}$ and $\mathcal S_i$, which have definite dimensions, but for $T_{vv}$ itself only \eqref{eq:Kbound}.

\bibliographystyle{JHEP}
\bibliography{main}

\end{document}